\documentclass[floats,twocolumn,pra]{revtex4}%
\usepackage{graphics}
\usepackage{amsmath}
\usepackage{graphicx}
\usepackage{amsfonts}
\usepackage{amssymb}
\usepackage{eurosym}
\usepackage{graphics}
\usepackage{float}
\usepackage{longtable}
\usepackage{epsfig}
\usepackage{latexsym}
\usepackage{theorem}
\usepackage{bbm}
\usepackage{bm}
\usepackage{psfrag}
\usepackage{color}%
\begin{document}
\title{Satellite Quantum Communications: \\Fundamental Bounds and Practical Security}
\author{Stefano Pirandola}
\affiliation{Department of Computer Science, University of York, York YO10 5GH, United Kingdom}

\begin{abstract}
Satellite quantum communications are emerging within the panorama of quantum
technologies as a more effective strategy to distribute completely-secure keys
at very long distances, therefore playing an important role in the
architecture of a large-scale quantum network. In this work, we apply and
extend recent results in free-space quantum communications to determine the
ultimate limits at which secret (and entanglement) bits can be distributed via
satellites. Our study is comprehensive of the various practical scenarios,
encompassing both downlink and uplink configurations, with satellites at
different altitudes and zenith angles. It includes effects of diffraction,
extinction, background noise and fading, due to pointing errors and
atmospheric turbulence (appropriately developed for slant distances). Besides
identifying upper bounds, we also discuss lower bounds, i.e., achievable rates
for key generation and entanglement distribution. In particular, we study the
composable finite-size secret key rates that are achievable by protocols of
continuous variable quantum key distribution, for both downlink and uplink,
showing the feasibility of this approach for all configurations. Finally, we
present a study with a sun-synchronous satellite, showing that its key
distribution rate is able to outperform a ground chain of ideal quantum repeaters.
\end{abstract}
\maketitle
\affiliation{Department of Computer Science, University of York, York YO10 5GH, United Kingdom}

\section{Introduction}

Satellite quantum communications~\cite[Sec.~VI]{QKDreview} represent a new
collective endeavour of the scientific community, with pioneering experiments
already demonstrated. A number of quantum protocols have been successfully
realized, including satellite-to-ground quantum key distribution
(QKD)~\cite{satexp1,satexp1b,satexp1c}, entanglement
distribution~\cite{satexp2}, entanglement-based QKD~\cite{satexp2b,satexp2c},
and ground-to-satellite quantum teleportation~\cite{satexp3}. Further
experiments have considered a space lab (Tiangong-2~\cite{satexp4}), and
microsatellites, such as SOCRATES~\cite{satexp4b} and CubeSats~\cite{satexp4c}.

An important driving reason behind the development of free-space quantum
communications with satellites is the possibility to by-pass fundamental
limitations that restrict rates and distances achievable by ground-based fiber
communications. It is in fact well known that the amount of secret bits or
entanglement bits (ebits) that can be distributed through a lossy
communication channel with transmissivity $\eta$ cannot exceed its secret key
capacity $-\log_{2}(1-\eta)$ bits/use, also known as the repeaterless PLOB
bound~\cite{QKDpaper} (see also Ref.~\cite{RCI} for the very first
investigation of the fundamental limits of quantum communication). In a
ground-based fiber link, the transmissivity decays exponentially with the
distance and so does the communication rate of any protocol for QKD or
entanglement distribution.

One strategy to mitigate such a problem is the introduction of quantum
repeaters or relays~\cite[Sec.~XII]{QKDreview}. In QKD the cheapest solution
is the use of a chain of trusted nodes between the two end-users. These nodes
distribute pairs of keys with their neighbors, whose composition via one-time
pad generates a final secret key for the remote users. Here a non-trivial
issue is the fact that all the nodes need to be trusted, so that the longer is
the chain, the higher is the probability that security could be compromised.
An alternative strategy relies in the adoption of nodes able to distribute
entanglement, which is then swapped to the remote users. However, this
solution is rather expensive because it involves the development of quantum
repeaters with long coherence times and distillation capabilities.

In this scenario, satellites open the way for new opportunities. Free-space
connection with a satellite may have far less decibels of loss than a long
ground-based fiber connection. Furthermore, most satellites are fast-moving
objects, therefore able to physically travel between two far locations over
the globe. These features have the potential to drastically reduce the
complexity of a ground-based quantum network. In fact, a chain of nodes could
just be replaced by a single satellite acting as a trusted QKD\ node or as a
distributor of entanglement. In such an exciting new setting, it is crucial to
understand the optimal performances allowed by quantum mechanics, and also
what practical performances may be achieved with current technology. This work
serves for this purpose.

Here we establish the information-theoretic limits of satellite quantum
communications and also show their practical security on the basis of
state-of-the-art technology. Our study extends the free-space analysis of
Ref.~\cite{FreeSpacePaper}, there developed for ground-to-ground free-space
communications, to the more general setting of ground-satellite
communications, where the optical signals travel slant distances with variable
altitudes and zenith angles (in uplink or downlink). This scenario involves
more general models for the underlying physical processes occurring within the
atmosphere (refraction, extinction and turbulence), and different descriptions
for the background noise (planetary albedos besides sky brightness).
Accounting for these accurate models, we study the ultimate rates for secret
key generation and entanglement distribution with a satellite in all scenarios
(uplink/downlink, night/day-time operations).

Once we established the ultimate converse rates for secret key and
entanglement distribution, we also study lower bounds. In particular, we focus
our investigation on the practical rates that are achievable by a
coherent-state QKD\ protocol, suitably modified to account for the fading
channel between satellite and ground station, and including the orbital
dynamics of the satellite. Our security analysis considers finite-size effects
and composable aspects. We show that high-rate ground-satellite QKD\ with
continuous variable (CV) systems~\cite{RMP} is feasible for both downlink and
uplink, during night and day.

Finally, we show that the number of secret key bits per day that can be
distributed between two stations by a sun-synchronous satellite can be much
larger than what achievable by a standard fiber connection between these
stations, even when a substantial number of repeaters are employed in the
middle and assumed to operate at their capacity level~\cite{netpaper}. This
analysis proves the potential advantages of satellite links over ground
networks and strongly corroborates their role for near-future realization of
large-scale quantum communications.

\subsection{Structure of the paper}

The paper has two main parts. In the first part, we investigate the
fundamental bounds for satellite quantum communications (Sec.~\ref{SECbounds}%
). We start with some basic geometric considerations (Sec.~\ref{subsecGEO})
and then we present general upper bounds based on free-space diffraction
(Sec.~\ref{diffsubsec}). We then increase the complexity of the description by
introducing atmospheric extinction and setup inefficiencies
(Sec.~\ref{atmextsubsec}). Next we describe the fading process induced by
pointing errors and turbulence (Sec.~\ref{fadingsubsec}), followed by the
corresponding expressions of the loss-limited upper bounds for key and
entanglement distribution (Sec.~\ref{lossboundsubsec}). In our next
generalization (Sec.~\ref{thermalsubsec}), we consider the effect of
background thermal noise and derive more advanced thermal-loss upper and lower
bounds for key and entanglement distribution with the satellite.

Once we have established the ultimate performances, we then study the secret
key rates that are achievable in satellite CV-QKD\ accounting for composable
finite-size aspects, fading and orbital dynamics. This is the second part of
the paper (Sec.~\ref{SECpractical}). We start with an overview of the problem
(Sec.~\ref{SECqkd1}) and a discussion on the composable security at fixed
transmissivity (Sec.~\ref{SECqkd2}). Then, we delve into the problem of
free-space fading by discussing the use of pilots, post-selection, and a
suitable defading technique (Sec.~\ref{SECqkd3}). Next we discuss how to
perform parameter estimation and we provide the general forms of the
composable key rates (Sec.~\ref{SECqkd4}). After important observations on the
setup noise (Sec.~\ref{SECqkd5}), we analyze the performances of the key rates
accounting for the orbital dynamics (Secs.~\ref{orbitalSEC} and~\ref{SECqkd7}%
). Finally, we show how the satellite-based key rates can overcome the
performance of a chain of quantum repeaters on the ground
(Sec.~\ref{satcompsec}). Sec.~\ref{SECconclusions} is for conclusions.

\section{Fundamental bounds for satellite quantum
communications\label{SECbounds}}

\subsection{Geometric considerations\label{subsecGEO}}

Consider a ground station (G), at some relatively-low altitude $h_{0}\simeq0$
above the sea level, and a satellite (S), that is\ orbiting at some variable
altitude $h$ beyond the K\'{a}rm\'{a}n line ($h$ $\geq100~$km)\ with a
variable zenith angle $\theta$. The latter is the angle between the zenith
point at the ground station and the direction of observation pointing at the
satellite. It takes positive values between $0$ (satellite at the zenith) and
$\pi/2$ (satellite at the horizon). For a zenith-crossing orbit (studied later
in Sec.~\ref{SECqkd7}), it may be useful to associate a sign to $\theta$, so
that $-\pi/2$ represents the \textquotedblleft front\textquotedblright%
\ horizon and $+\pi/2$ is the \textquotedblleft back\textquotedblright%
\ horizon. (Including the sign does not change the main geometric formulas
since $\theta$ appears in cosine functions).

Calling $R_{\text{E}}\simeq6371$~km the approximate radius of the Earth, the
slant distance $z$ between the ground station and the satellite can be written
as
\begin{equation}
z(h,\theta)=\sqrt{h^{2}+2hR_{\text{E}}+R_{\text{E}}^{2}\cos^{2}\theta
}-R_{\text{E}}\cos\theta. \label{slantANALYTICAL}%
\end{equation}
Equivalently, the altitude $h$ of the satellite reads
\begin{equation}
h(z,\theta)=\sqrt{R_{\text{E}}^{2}+z^{2}+2zR_{\text{E}}\cos\theta}%
-R_{\text{E}}. \label{altitudeANALYTICAL}%
\end{equation}
See Appendix~\ref{SecGEOMETRY} for more details on this geometry, which can be
easily extended to the case of non-negligible atmospheric altitudes $h_{0}$
for the ground station. (We remark that, while this extension may be useful,
in our main text we investigate the basic scenario of a low-altitude ground
station for which $h_{0}$ can be considered to be negligible with respect to
the typical satellite altitudes.)

The formulas above are very good approximations for angles $\theta\lesssim1$
(i.e., within about $60^{\circ}$ from the zenith). For larger zenith angles,
one needs to consider the apparent angle and the optical-path elongation
induced by atmospheric refraction, which become more and more prominent close
to the horizon. In such a case, the formulas above undergo some modifications
as discussed in Appendix~\ref{SECrefraction}. In our main text below, we omit
this technicality for two reasons: (i) formulas above can still be used to
provide (larger) upper-bounds in the proximity of the horizon; (ii) when we
treat achievable rates (lower bounds), we will restrict our study to the good
window $\theta\lesssim1$, an assumption which is also justified by the
analysis of turbulence carried out later on in the manuscript.

In terms of configurations, we consider both uplink and downlink. In uplink,
the ground station is the transmitter (Alice) and the satellite is the
receiver (Bob); in downlink, it is the satellite to be the transmitter and the
ground station to operate as a receiver. In these two configurations, the
effects of free-space diffraction and atmospheric extinction are the same.
Different is the case for the fading induced by turbulence (more relevant in
uplink) and the thermal noise induced by background sources (with further
differences between day- and night-time operations). For the sake of
simplicity, we start by accounting for diffraction and extinction only; then,
we will introduce the other effects, which need to be treated quite
differently with respect to the models that are valid for ground-to-ground
free-space communications.

\subsection{Free-space diffraction\label{diffsubsec}}

We assume that free-space quantum communication is based on a
quasi-monochromatic optical mode with temporal duration $\Delta t$\ and narrow
bandwidth $\Delta\lambda$ around a carrier wavelength $\lambda$ (so that the
angular frequency is $\omega=2\pi c/\lambda$ and the wavenumber is
$k=\omega/c=2\pi/\lambda$). This model is represented by a Gaussian beam with
field spot size $w_{0}$ and curvature $R_{0}$%
~\cite{svelto,Siegman,Andrews93,Andrews94}. Spot size is sufficiently smaller
than the transmitter's aperture so that the latter does not induce
relevant\ diffraction. After free-space propagation for a distance $z$, the
beam is detected by a receiver whose telescope has a circular aperture with
radius $a_{R}$. To fix the ideas, one can assume that the propagation
direction is uplink so that the ground station is the transmitter, but the
model is completely symmetric and applies to downlink in exactly the same way.

Because of the inevitable free-space diffraction, the waist of the beam will
broaden during propagation. After travelling for a distance $z$, the beam is
intercepted by the receiver that will see an increased spot size
\begin{equation}
w_{\text{d}}(z)=w_{0}\sqrt{\left(  1-z/R_{0}\right)  ^{2}+\left(
z/z_{R}\right)  ^{2}},\label{diffWAIST}%
\end{equation}
where $z_{R}:=\pi w_{0}^{2}\lambda^{-1}$ is the Rayleigh range. Due to the
finite aperture $a_{R}$ of the receiving telescope, only a fraction of the
initial beam will be detected, and this fraction is given by the
diffraction-induced transmissivity
\begin{equation}
\eta_{\text{d}}(z)=1-e^{-2a_{R}^{2}/w_{\text{d}}^{2}}.\label{diffINDUCED}%
\end{equation}
In the far field $z\gg z_{R}$, this can be approximated as%
\begin{equation}
\eta_{\text{d}}\simeq\eta_{\text{d}}^{\text{far}}:=\frac{2a_{R}^{2}%
}{w_{\text{d}}^{2}}\ll1.
\end{equation}

Using $\eta_{\text{d}}$ with the PLOB bound~\cite{QKDpaper}, one finds that
the maximum number of secret bits that can be distributed by the most general
(adaptive) QKD protocols over the free-space communication channel is
upper-bounded by~\cite{FreeSpacePaper}%
\begin{equation}
\mathcal{U}(z)=\frac{2}{\ln2}\frac{a_{R}^{2}}{w_{\text{d}}^{2}}~~\text{bits
per use.} \label{diffBOUND}%
\end{equation}
In other words, the secret key capacity $K$ of the free-space channel must
satisfy $K\leq\mathcal{U}$ and, similarly, this bound also holds for the
channel's entanglement distribution capacity $E\leq K$ (which is the number of
ebits per use of the channel that can be distributed by the most general
adaptive protocols of entanglement distribution; see Ref.~\cite{QKDpaper} for
exact mathematical definitions).

Note that a focused beam ($R_{0}=z$) optimizes the bound in
Eq.~(\ref{diffBOUND}) but, at long distances, optical focusing becomes a very
challenging task. For this reason, a better strategy is to just generate a
collimated beam ($R_{0}=\infty$) so that Eq.~(\ref{diffBOUND}) is computed by
assuming
\begin{equation}
w_{\text{d}}(z)=w_{0}\sqrt{1+z^{2}/z_{R}^{2}}.
\end{equation}
We will therefore use the specific case of a collimated Gaussian beam in our
numerical investigations.

\begin{figure}[t]
\vspace{0.2cm}
\par
\begin{center}
\includegraphics[width=0.48\textwidth] {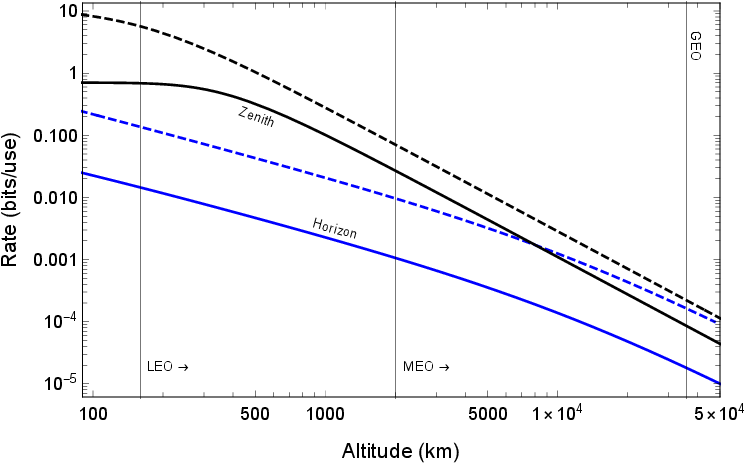}
\end{center}
\par
\vspace{-0.5cm}\caption{Key rate between a ground station and a satellite at
various altitudes $h$. In dashed we show the diffraction-based bound
$\mathcal{U}(h,\theta)$ of Eq.~(\ref{diffBOUND}) while, in solid, we show the
upper-bound $\mathcal{V}(h,\theta)$ of Eq.~(\ref{PLOBder1}) which includes the
combined effects of diffraction, extinction and quantum efficiency. The upper
black curves refer to the satellite at the zenith position ($\theta=0$), while
the lower blue curves refer to the horizon position ($\theta=\pi/2$). We
assume a collimated Gaussian beam with $\lambda=800~$nm and $w_{0}=20~$cm, so
that $z_{\text{R}}\simeq160~$km (LEO\ boundary). Then, we assume $a_{R}=40$~cm
and $\eta_{\text{eff}}=0.4$.}%
\label{RatePic2}%
\end{figure}

It is also clear that $\mathcal{U}(z)$ can be expressed in terms of the
altitude $h$ and zenith angle $\theta$ of the satellite. In fact, we may
replace the function $z=z(h,\theta)$ of Eq.~(\ref{slantANALYTICAL}) in
Eq.~(\ref{diffBOUND}) to get the expression for $\mathcal{U}(h,\theta
)=\mathcal{U}[z(h,\theta)]$. The diffraction-bound $\mathcal{U}(h,\theta)$ is
numerically investigated in Fig.~\ref{RatePic2}, for a collimated Gaussian
beam and a typical choice of parameters. In particular, we show this ultimate
bound in two extreme angles for the satellite, i.e., zenith position
($\theta=0$) and horizon ($\theta=\pi/2$). The reduction of the rate at the
horizon is due to the greater slant distance to be travelled by the beam. As
mentioned before, in this case the bound is optimistic because the
refraction-induced elongation of the optical path is here neglected.

\subsection{Atmospheric extinction\label{atmextsubsec}}

Another important physical process that causes loss in the free-space
propagation of an optical beam is atmospheric extinction; this is induced by
both aerosol absorption and Rayleigh/Mie\ scattering. For a free-space
communication at fixed altitude $h$, this effect is described by the simple
Beer-Lambert equation
\begin{equation}
\eta_{\text{atm}}(h)=\exp[-\alpha(h)z], \label{BLextinction}%
\end{equation}
where $\alpha(h)$ is the extinction factor~\cite[Ch.~11]{Huffman}. This is
given by $\alpha(h)=\alpha_{0}\exp(-h/\tilde{h})$, where $h$ is expressed in
meters, $\tilde{h}=6600~$m, and the sea-level value $\alpha_{0}$ takes the
value $\simeq5\times10^{-6}$ m$^{-1}$ at $\lambda=800$~nm~\cite[Sec.~III.C]%
{Vasy19}.

It is clear that the model in Eq.~(\ref{BLextinction}) needs to be suitably
modified in order to describe free-space optical communications at variable
altitudes $h$. First suppose that the satellite is exactly at the zenith, so
that its slant range $z$\ is equal to its altitude $h$. Then, we can easily
compute
\begin{align}
\eta_{\text{atm}}^{\text{zen}}(h)  &  =\exp\left[  -\int_{0}^{h}dh^{\prime
}\alpha(h^{\prime})\right] \nonumber\\
&  =\exp\left[  \alpha_{0}\tilde{h}(e^{-h/\tilde{h}}-1)\right] \nonumber\\
&  \geq e^{-\alpha_{0}\tilde{h}}\simeq0.967~(\simeq0.14~\text{dB}).
\label{etaATMzenith}%
\end{align}
The value in Eq.~(\ref{etaATMzenith}) is valid for any altitude and is already
approximated at $h=30$~km, in the middle of the stratosphere, after which the
atmospheric density is negligible. For this reason, for any satellite at the
zenith position, we can use the estimate $\eta_{\text{atm}}^{\text{zen}%
}(\infty)\simeq0.967$.

Consider now a generic zenith angle $\theta$. Neglecting refraction, we can
therefore use the expressions in Eqs.~(\ref{slantANALYTICAL})
and~(\ref{altitudeANALYTICAL}), and write the following expression for the
atmospheric transmissivity
\begin{equation}
\eta_{\text{atm}}(h,\theta)=\exp\left\{  -\int_{0}^{z(h,\theta)}%
dy~\alpha\lbrack h(y,\theta)]\right\}  =e^{-\alpha_{0}g(h,\theta)},
\label{atmEQ2}%
\end{equation}
where we have introduced the integral function%
\begin{equation}
g(h,\theta):=\int_{0}^{z(h,\theta)}dy\exp\left[  -\frac{h(y,\theta)}{\tilde
{h}}\right]  . \label{IntegralFUN}%
\end{equation}
For zenith angles $\theta\lesssim1$, one may check that Eq.~(\ref{atmEQ2}) can
be approximated as follows%
\begin{equation}
\eta_{\text{atm}}(h,\theta)\simeq\lbrack\eta_{\text{atm}}^{\text{zen}%
}(h)]^{\sec\theta}\simeq\left[  \eta_{\text{atm}}^{\text{zen}}(\infty)\right]
^{\sec\theta}. \label{approxALTT}%
\end{equation}
This approximation is already good at $30$~km of altitude and becomes an
almost exact formula beyond $100$~km.

Combining atmospheric extinction with free-space diffraction and the
inevitable internal loss $\eta_{\text{eff}}$ affecting the setup of the
receiver (due to non-unit quantum efficiency of the detector and other optical
imperfections), we can write the total amount of fixed loss of the free-space
channel from the generation of the Gaussian beam to its final detection. This
is given by
\begin{equation}
\eta_{\text{tot}}(h,\theta):=\eta_{\text{eff}}\eta_{\text{atm}}(h,\theta
)\eta_{\text{d}}(h,\theta),\label{combinedETA}%
\end{equation}
where $\eta_{\text{d}}(h,\theta)=\eta_{\text{d}}[z(h,\theta)]$ and we can
assume $\eta_{\text{eff}}\simeq0.4$, i.e., about $4~$dB (e.g., as in
Ref.~\cite{BrussSAT}). Using this transmissivity in the PLOB bound, we get an
immediate extension of a result in Ref.~\cite{FreeSpacePaper}, i.e.,%
\begin{align}
&  K\leq\mathcal{V}(h,\theta):=-\log_{2}[1-\eta_{\text{tot}}(h,\theta)]\\
&  =-\log_{2}\left[  1-\eta_{\text{eff}}e^{-\alpha_{0}g(h,\theta)}\left(
1-e^{-\frac{2a_{R}^{2}}{w_{\text{d}}[z(h,\theta)]^{2}}}\right)  \right]
\label{PLOBder1}\\
&  \simeq\frac{2}{\ln2}\frac{a_{R}^{2}\eta_{\text{eff}}e^{-\alpha
_{0}g(h,\theta)}}{w_{\text{d}}[z(h,\theta)]^{2}},
\end{align}
where the last approximation is valid for $\eta_{\text{tot}}(h,\theta)\ll1$
which is certainly true in the far field regime $z\gg z_{R}$.

From Fig.~\ref{RatePic2}, we see that the combined effects of diffraction,
extinction and non-ideal quantum efficiency decrease (by about one order of
magnitude) the ultimate communication bounds that are only based on free-space
diffraction. As for $\mathcal{U}(h,\theta)$, also the value of the upper bound
$\mathcal{V}(h,\theta)$ is over-estimated at the horizon due to the fact that
refraction has been neglected (see Appendix~\ref{SECrefraction} for an
extension of the bound which includes refraction). The ultimate performances
discussed so far\ will further decrease when we include fading
(turbulence/pointing errors) and then background noise.

\subsection{Fading process induced by beam wandering: Turbulence and pointing
errors\label{fadingsubsec}}

The combined transmissivity $\eta_{\text{tot}}$ in Eq.~(\ref{combinedETA}) is
constant for a fixed geometry, $h$ and $\theta$, between ground station and
satellite. At each time instant, it corresponds to the maximum transmissivity
which can be reached by a beam that is perfectly aligned between transmitter
and receiver. In a realistic scenario, such alignment is however not
maintained and we need to consider a process of beam wandering; this
inevitably induces a fading process for the communication channel whose
instantaneous transmissivity will
fluctuate~\cite{Esposito,Fried73,Titterton,Vasy12}.

Beam wandering is due to random errors in the pointing mechanism of the
transmitter and also to the action of atmospheric turbulence on a section of
the optical path. These two effects are independent and they sum up. In
practice, they have a different weights depending on the configuration. In
downlink, pointing error is quite relevant, since on-board optics is limited,
while turbulence can be neglected, because it occurs in the final section of
the optical path where the beam has been already spread by diffraction. In
uplink, pointing error can be reduced, because ground stations may adopt more
extensive and sophisticated optics; by contrast, turbulence represents a major
effect in this case due to the fact that it affects the beam right after its generation.

In order to treat beam wandering and the corresponding fading process, we
assume the regime of weak turbulence, which is appropriate for
relatively-small zenith angles $\theta\lesssim1$. In this regime, we may
separate effects occurring on fast and slow time-scales. Turbulent eddies
smaller than the beam waist act with a fast dynamics; these tend to broaden
the beam, so that the diffraction-limited spot size $w_{\text{d}}$ is replaced
by a larger \textquotedblleft short-term\textquotedblright\ spot size
$w_{\text{st}}=w_{\text{st}}(z,\theta)$, also known as \textquotedblleft hot
spot\textquotedblright. On the other hand, turbulent eddies that are larger
than the beam waist act on a much slower time scale~\cite{Fante75} (of the
order of $10-100~$ms~\cite{Burgoin}); these tend to deflect the beam, whose
centroid will then wander according to a Gaussian distribution with variance
$\sigma_{\text{TB}}^{2}=\sigma_{\text{TB}}^{2}(z,\theta)$. This slow dynamics
can be fully resolved and closely followed by a fast detector, e.g., with a
realistic bandwidth of the order of $100~$MHz. On top of this process, there
are pointing errors whose dynamics is also slow and causes an additional
Gaussian random walk with variance $\sigma_{\text{P}}^{2}\simeq\left(
10^{-6}z\right)  ^{2}$ for a typical $1~\mu$rad error at the transmitter.
Overall, the wandering of the beam centroid has variance $\sigma^{2}%
=\sigma_{\text{TB}}^{2}+\sigma_{\text{P}}^{2}$.

A crucial theoretical step in our treatment is the explicit derivation of
$w_{\text{st}}$ and $\sigma_{\text{TB}}^{2}$ according to turbulence models
that are appropriate for satellite communications. As discussed in detail in
Appendix~\ref{TurbSECTION}, we start from the Hufnagel-Valley (H-V)
model~\cite{Stanley,Valley}, which provides the atmospheric profile for the
refraction-index structure constant $C_{n}^{2}(h)$~\cite[Sec.~12.2.1]%
{AndrewsBook}. This altitude-dependent constant measures the strength of the
fluctuations in the refraction index caused by spatial variations of
temperature and pressure.\ Assuming this model, we then compute the
scintillation index and the Rytov variance. The latter allows us to verify
that the angular window $\theta\lesssim1$ is compatible with the regime of
weak turbulence, which is why we choose this angular window for quantum
communication in both uplink and downlink.

From the structure constant $C_{n}^{2}(h)$, the wave-number $k$ of the beam,
and the geometry (slant distance $z$ and zenith angle $\theta$)\ one can
define the spherical-wave coherence length $\rho_{0}=\rho_{0}(z,\theta)$ for
uplink $\mathrm{(up}$\textrm{)} and downlink ($\mathrm{down}$). Using the
expression for generic $z$-long propagation~\cite{Fante75,Fante80} and
accounting for the altitude function $h=h(z,\theta)$ in
Eq.~(\ref{altitudeANALYTICAL}), this length takes the form%
\begin{align}
&  \rho_{0}^{\text{up/down}}=\left[  1.46k^{2}\int_{0}^{z}d\xi\left(
1-\tfrac{\xi}{z}\right)  ^{\frac{5}{3}}\gamma^{\text{up/down}}(\xi)\right]
^{-\frac{3}{5}},\nonumber\\
&  \gamma^{\text{up}}(\xi)=C_{n}^{2}[h(\xi,\theta)],~\gamma^{\text{down}}%
(\xi)=C_{n}^{2}[h(z-\xi,\theta)].
\end{align}

An analysis of $\rho_{0}^{\text{down}}$ confirms that, within the good angular
window and not-too large receiver apertures, downlink communication can be
considered to be free of turbulence, so we can set $\sigma_{\text{TB}%
}^{2}\simeq0$ and $w_{\text{st}}\simeq w_{\text{d}}$. This assumption for
downlink is well justified by examining the long-term spot-size of the beam
\begin{equation}
w_{\text{lt}}^{2}=w_{\text{st}}^{2}+\sigma_{\text{TB}}^{2},\label{ltspot}%
\end{equation}
which takes the following form (valid in general conditions of
turbulence)~\cite{Fante75}
\begin{equation}
w_{\text{lt}}^{2}\simeq w_{\text{d}}^{2}+2\left(  \frac{\lambda z}{\pi\rho
_{0}}\right)  ^{2}.
\end{equation}
A quick calculation of $\rho_{0}^{\text{down}}$ shows that one has
$w_{\text{lt}}\simeq w_{\text{d}}$ in downlink at satellite altitudes (e.g.,
at the LEO lower border, the difference in these two standard deviations is
basically in the third significative digit). This is equivalent to say that
diffraction is the only relevant effect, i.e., we have the collapse
$\sigma_{\text{TB}}^{2}\simeq0$ and $w_{\text{lt}}\simeq w_{\text{st}}\simeq
w_{\text{d}}$.

By contrast, for uplink communication, the value of $\rho_{0}^{\text{up}}$
becomes rather small (of the order of 1 cm at $\lambda=800~$nm), meaning that
turbulence is relevant in this scenario. In this case, we can resort to
Refs.~\cite{Yura73,Fante75} and write analytical formulas for the short-term
spot size and the variance of the centroid wandering.

For satellite distances one can easily check the validity of Yura's condition
for uplink $\phi:=0.33\left(  \rho_{0}^{\text{up}}/w_{0}\right)  ^{1/3}\ll1$,
which allows us to write~\cite{Yura73}\
\begin{equation}
w_{\text{st}}^{2}\simeq w_{\text{d}}^{2}+2\left(  \frac{\lambda z}{\pi\rho
_{0}^{\text{up}}}\right)  ^{2}\Psi,~~\sigma_{\text{TB}}^{2}\simeq2\left(
\frac{\lambda z}{\pi\rho_{0}^{\text{up}}}\right)  ^{2}(1-\Psi),
\end{equation}
where $\Psi:=(1-\phi)^{2}\simeq1-2\phi$. For satellites in the LEO region and
beyond, we can adopt the asymptotic planar approximation%
\begin{align}
\rho_{0}^{\text{up}} &  \simeq\rho_{p}^{\text{up}}\simeq\left[  1.46k^{2}%
(\sec\theta)I_{\infty}\right]  ^{-3/5},\\
I_{\infty} &  :=\int_{0}^{\infty}d\xi C_{n}^{2}(\xi),
\end{align}
where $\rho_{p}^{\text{up}}$\ bounds the value of $\rho_{0}^{\text{up}}$\ from
below at any relevant altitude and any $\theta\lesssim1$ (e.g., see the
comparison in Fig.~\ref{compareCLpic} of Appendix~\ref{TurbSECTION}). This
leads to the simpler expressions%
\begin{align}
w_{\text{st}}^{2} &  \simeq w_{\text{d}}^{2}+z^{2}\Delta(\theta
),\label{wSTuplink}\\
\sigma_{\text{TB}}^{2} &  \simeq\frac{7.71I_{\infty}}{w_{0}^{1/3}}z^{2}%
\sec\theta,\label{TBuplink}%
\end{align}
where we have set%
\begin{equation}
\Delta(\theta):=\frac{26.28(I_{\infty}\sec\theta)^{6/5}}{\lambda^{2/5}}%
-\frac{7.71I_{\infty}\sec\theta}{w_{0}^{1/3}}.
\end{equation}

In these formulas, $I_{\infty}$ takes different values depending on the
parameters chosen for the H-V model. In particular, we compute $I_{\infty
}\simeq2.2354\times10^{-12}$~m$^{1/3}$ for the standard H-V$_{5/7}$
model~\cite[Sec.~12.2.1]{AndrewsBook}, which is good for describing night-time
operation. During the day, turbulence on the ground is higher and we consider
a typical day-time version of the H-V model, for which $I_{\infty}%
\simeq3.2854\times10^{-12}$~m$^{1/3}$ (see Appendix~\ref{TurbSECTION}\ for
more details). Numerical investigations at $\lambda=800~$nm show that
$w_{\text{st}}$ exceeds $w_{\text{d}}$ by one order of magnitude in uplink,
with an almost constant gap in the far field (e.g., see Fig.~\ref{stPic}\ in
Appendix~\ref{TurbSECTION}). Finally, note that we can re-obtain the downlink
diffraction-limited values by setting $I_{\infty}=0$ in Eqs.~(\ref{wSTuplink})
and~(\ref{TBuplink}).

\subsection{Bounds for the satellite fading channels in uplink and
downlink\label{lossboundsubsec}}

Following the theory of the previous section, it follows that we can adopt a
unified approach to treat fading in uplink and downlink. In fact, we may
consider the general parameters $w_{\text{st}}$ and $\sigma^{2}=\sigma
_{\text{TB}}^{2}+\sigma_{\text{P}}^{2}$, which can then be simplified for the
specific case of downlink, for which we may set $w_{\text{st}}\simeq
w_{\text{d}}$ and $\sigma^{2}\simeq\sigma_{\text{P}}^{2}$.

In general, the broader short-term spot size $w_{\text{st}}$ decreases the
maximum value of the transmissivity. In fact, the diffraction-induced
transmissivity $\eta_{\text{d}}$ has to be replaced by the (lower) short-term
transmissivity
\begin{equation}
\eta_{\text{st}}(z,\theta)=1-e^{-2a_{R}^{2}/w_{\text{st}}^{2}},
\label{shortTERMtransmissivity}%
\end{equation}
with far-field approximation
\begin{equation}
\eta_{\text{st}}\simeq\eta_{\text{st}}^{\text{far}}:=\frac{2a_{R}^{2}%
}{w_{\text{st}}^{2}}.
\end{equation}
As a result the combined expression $\eta_{\text{tot}}$ of
Eq.~(\ref{combinedETA}) has to be replaced by the more general parameter
\begin{equation}
\eta(h,\theta):=\eta_{\text{eff}}\eta_{\text{atm}}(h,\theta)\eta_{\text{st}%
}(h,\theta), \label{etaTTT}%
\end{equation}
where $\eta_{\text{st}}(h,\theta):=\eta_{\text{st}}[z(h,\theta),\theta]$ using
Eq.~(\ref{slantANALYTICAL}).

At any fixed geometry $h$ and $\theta$, the loss parameter $\eta(h,\theta)$
describes the maximum transmissivity that is achievable in the communication
through the generally-turbulent free-space channel, which corresponds to the
case where the incoming beam is perfectly aligned with the receiver's
aperture. Note that, more generally, one may assume the case of a constant
deflection for the beam; here we omit this technicality for two reasons: we
are interested in the optimal rate performance of the communication and such a
deflection can anyway be compensated by using adaptive optics.

As a consequence of beam wandering, the actual instantaneous value of the
transmissivity will be $\tau\leq\eta$ and this value will depend on how far
the beam is deflected from the center of the receiver's aperture. The Gaussian
random walk of the beam centroid~\cite{Dowling} results in a Weibull
distribution for the instantaneous deflection which, in turn, leads to a
probability distribution $P_{\sigma}(\tau)$ for the instantaneous
transmissivity. Let us introduce the two functions
\begin{align}
f_{0}(x)  &  :=\left[  1-\exp\left(  -2x\right)  I_{0}\left(  2x\right)
\right]  ^{-1},\\
f_{1}(x)  &  :=\exp\left(  -2x\right)  I_{1}\left(  2x\right)  ,
\end{align}
in terms of the modified Bessel function $I_{n}$\ of the first kind with order
$n=0,1$. These are useful to introduce the geometry-dependent positive
parameters~\cite{Vasy12}
\begin{align}
\gamma(z,\theta)  &  =\frac{4\eta_{\text{st}}^{\text{far}}f_{0}(\eta
_{\text{st}}^{\text{far}})f_{1}(\eta_{\text{st}}^{\text{far}})}{\ln\left[
2\eta_{\text{st}}f_{0}(\eta_{\text{st}}^{\text{far}})\right]  },\label{expGG2}%
\\
r_{0}(z,\theta)  &  =\frac{a_{R}}{\left\{  \ln\left[  2\eta_{\text{st}}%
f_{0}(\eta_{\text{st}}^{\text{far}})\right]  \right\}  ^{1/\gamma}}.
\end{align}
Using these parameters, we may then write
\begin{equation}
P_{\sigma}(\tau)=\frac{r_{0}^{2}}{\gamma\sigma^{2}\tau}\left(  \ln\frac{\eta
}{\tau}\right)  ^{\frac{2}{\gamma}-1}\exp\left[  -\frac{r_{0}^{2}}{2\sigma
^{2}}\left(  \ln\frac{\eta}{\tau}\right)  ^{\frac{2}{\gamma}}\right]  .
\end{equation}

The free-space fading channel $\mathcal{E}_{\text{fad}}$ can therefore be
described by an ensemble $\{P_{\sigma}(\tau),\mathcal{E}_{\tau}\}$ of
pure-loss channels $\mathcal{E}_{\tau}$ whose transmissivity $\tau$ is chosen
with probability $P_{\sigma}(\tau)$. From the PLOB\ bound and the convexity of
the relative entropy of entanglement~\cite{QKDpaper}, one has that the
secret-key capacity of the fading channel $\mathcal{E}_{\text{fad}}$\ is
bounded by the average
\begin{equation}
K\leq-\int_{0}^{\eta}d\tau~P_{\sigma}(\tau)\log_{2}(1-\tau).
\end{equation}
Repeating the steps of Ref.~\cite{FreeSpacePaper}, we get
\begin{equation}
K\leq\mathcal{B}(\eta,\sigma):=-\Delta(\eta,\sigma)\log_{2}(1-\eta),
\label{mainOTHER}%
\end{equation}
where $\Delta(\eta,\sigma)$ is defined by the expression
\begin{equation}
\Delta(\eta,\sigma):=1+\frac{\eta}{\ln(1-\eta)}\int_{0}^{+\infty}dx\frac
{\exp\left(  -\frac{r_{0}^{2}}{2\sigma^{2}}x^{2/\gamma}\right)  }{e^{x}-\eta}.
\end{equation}

Note that, while Eq.~(\ref{mainOTHER}) has exactly the same analytical form of
the free-space bound in Ref.~\cite{FreeSpacePaper}, it is here implicitly
extended from the setting of ground-based communications to that of satellite
communications. In $\mathcal{B}(\eta,\sigma)$, the component $-\log_{2}%
(1-\eta)\simeq\eta/\ln2$ upper-bounds the key rate achievable with a
perfectly-aligned link between ground station and receiver, while $\Delta
(\eta,\sigma)$ is a correction factor accounting for beam wandering, induced
by turbulence and/or pointing errors. In order to investigate this bound for
satellite communications, we need to include the necessary geometry and
distinguish between downlink and uplink.

\begin{figure}[t]
\vspace{0.2cm}
\par
\begin{center}
\includegraphics[width=0.48\textwidth] {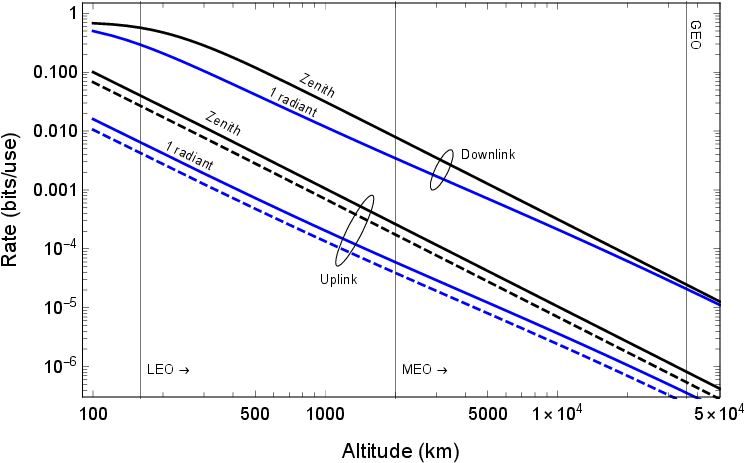}
\end{center}
\par
\vspace{-0.5cm}\caption{Key rate between a ground station and a satellite at
various altitudes $h$. We consider the upper-bound in Eq.~(\ref{mainOTHER})
specified for downlink ($\mathcal{B}_{\text{down}}$) and uplink ($\mathcal{B}%
_{\text{up}}$), the latter being presented for night-time (solid) and day-time
(dashed). In each case, we show the performance both at the zenith position
($\theta=0$) and at $\theta=1$ radiant. In the various configurations, the
lines upper-bound the maximum number of secret (and entanglement) bits that
can be distributed per use of the channel, considering the combined effects of
diffraction, atmospheric extinction, quantum efficiency, pointing error
($1$~$\mu$rad) and atmospheric turbulence (night-/day-time H-V model, only for
uplink). As in Fig.~\ref{RatePic2},\ we assume a collimated beam with
$\lambda=800~$nm and $w_{0}=20~$cm. Receiver has aperture $a_{R}=40~$cm and
total efficiency $\eta_{\text{eff}}=0.4$.}%
\label{TurbulenceBounds}%
\end{figure}

By replacing $z=z(h,\theta)$ in the formulas of $\gamma$, $r_{0}$ and the
total variance $\sigma^{2}$, we can express all these parameters in terms of
$h$ and $\theta$. We may then use these functionals together with $\eta
=\eta(h,\theta)$\ of Eq.~(\ref{etaTTT}) in Eq.~(\ref{mainOTHER}), so as to
obtain a geometry-dependent expression for the bound $\mathcal{B}%
=\mathcal{B}(h,\theta)$. In Fig.~\ref{TurbulenceBounds},\ we investigate the
behavior of $\mathcal{B}(h,\theta)$ in uplink and downlink for a satellite at
various altitudes $h$ and for zenith angles equal to zero or $1$. Besides the
effects of free-space diffraction, atmospheric extinction and limited quantum
efficiency, the downlink bound $\mathcal{B}_{\text{down}}(h,\theta)$ includes
the centroid wandering due to pointing error $\sigma_{\text{P}}^{2}$, while
the uplink bound $\mathcal{B}_{\text{up}}(h,\theta)$ also includes
turbulence-induced beam spreading ($w_{\text{st}}$) and wandering (so that
$\sigma^{2}=\sigma_{\text{P}}^{2}+\sigma_{\text{TB}}^{2}$).

By comparing the zenith-performance of $\mathcal{B}_{\text{down}}$ in
Fig.~\ref{TurbulenceBounds} and that of $\mathcal{V}$ in Fig.~\ref{RatePic2},
we can see how the pointing error decreases the rate already from the
beginning of the LEO region. Then, by comparing the downlink bound
$\mathcal{B}_{\text{down}}$ with the uplink bound $\mathcal{B}_{\text{up}}$ in Fig.~\ref{TurbulenceBounds}, we see how turbulence induces a further
non-trivial decrease, which is about one-two orders of magnitude. Also note
that $\mathcal{B}_{\text{up}}$ is additionally decreased during day time
(while $\mathcal{B}_{\text{down}}$ does not depend on the operation time).

Several important considerations are in order about Eq.~(\ref{mainOTHER}). The
first is that, as long as the free-space fading channel $\mathcal{E}%
_{\text{fad}}$ can indeed be described by an ensemble of instantaneous
pure-loss channels $\mathcal{E}_{\tau}$, i.e., $\mathcal{E}_{\text{fad}%
}=\{P_{\sigma}(\tau),\mathcal{E}_{\tau}\}$, then the bound $\mathcal{B}$ in
Eq.~(\ref{mainOTHER}) is also achievable. As discussed in
Ref.~\cite{FreeSpacePaper}, this bound is achieved by optimal protocols of
CV-QKD, either based on the use of quantum memories or employing
largely-squeezed states to be transmitted in an extremely biased manner. The bound can also be
achieved by optimal protocols of entanglement distribution, where distillation
is assisted by one-way backward classical communication. The performance of
these protocols is equal to the (bosonic) reverse coherent
information~\cite{RCI}, which achieves $-\log_{2}(1-\tau)$ for each
$\mathcal{E}_{\tau}$. As a result, we may write $E=K=\mathcal{B}$ for both the
entanglement distribution ($E$) and the secret key ($K$) capacities of the
fading channel $\mathcal{E}_{\text{fad}}$.

The pure-loss assumption $\mathcal{E}_{\text{fad}}=\{P_{\sigma}(\tau
),\mathcal{E}_{\tau}\}$, which implies the achievability of the bound
$\mathcal{B}$, is appropriate for night-time operation at typical satellite
altitudes. Different is the case for day-time operation, where the background
noise becomes non-trivial. In this setting, the quantity $\mathcal{B}$ in
Eq.~(\ref{mainOTHER}) is still an upper bound but no longer guaranteed to be
achievable. In the following (Sec.~\ref{thermalsubsec}), we will therefore
consider a more refined upper bound and a corresponding lower bound in the
presence of noise. These two bounds are useful for the study of day-time
operation and also clarify the validity of the pure-loss assumption for
night-time operation.

A final observation regards the potential use of slow detection strategies, so
that the wandering of the centroid is not time-resolved but averaged over long
acquisition times of the order of $100$~ms or more, i.e., beyond the typical
time-scales associated with turbulence and pointing error. In such a case, the
transmissivity of the ground-satellite link has to be averaged over the fading
process and is given by%
\begin{equation}
\eta_{\text{slow}}=\eta_{\text{eff}}\eta_{\text{atm}}\left[  1-e^{-2a_{R}%
^{2}/(w_{\text{lt}}^{2}+\sigma_{\text{P}}^{2})}\right]  ,
\end{equation}
where $w_{\text{lt}}$ is the long-term spot size of the beam given in
Eq.~(\ref{ltspot}). As a result, the upper bound takes the simple
form~\cite{FreeSpacePaper}%
\begin{equation}
K_{\text{slow}}\leq-\log_{2}(1-\eta_{\text{slow}})\leq\frac{2}{\ln2}%
\frac{a_{R}^{2}}{w_{\text{lt}}^{2}+\sigma_{\text{P}}^{2}}. \label{ineq2}%
\end{equation}
This detection strategy may be helpful to increase the detection efficiency,
but it has the downside to reduce the clock rate and also involves a larger
amount of noise to be collected by the receiver.

\subsection{Satellite bounds with background noise\label{thermalsubsec}}

Let us account for the presence of background noise. First of all, it is
important to adopt an appropriate model for the input-output number of
photons. Call $\bar{n}_{T}$ the mean number of photons in the beam at the
transmitter. For an instantaneous transmissivity $\tau$, the mean number of
photons reaching the receiver can be written as
\begin{equation}
\bar{n}_{R}=\tau\bar{n}_{T}+\bar{n}, \label{TLchannel}%
\end{equation}
where $\bar{n}$ is the mean number of thermal photons describing the overall
noise affecting the propagation.

It is natural to decompose the thermal number $\bar{n}$ as follows (see also
Ref.~\cite[Fig.~1]{FreeSpacePaper})%
\begin{equation}
\bar{n}:=\eta_{\text{eff}}\bar{n}_{B}+\bar{n}_{\text{ex}},
\label{thermalCnoise}%
\end{equation}
where $\bar{n}_{B}$ is the background thermal noise collected by the
receiver's aperture, whose detector has quantum efficiency $\eta_{\text{eff}}$
and extra setup noise $\bar{n}_{\text{ex}}$. The noise contribution from the
setup $\bar{n}_{\text{ex}}$ is sometimes considered to be trusted. It is
assumed to be negligible ($\bar{n}_{\text{ex}}\simeq0$) in our numerical
investigation of the ultimate bounds since we aim at analyzing
optimal/almost-optimal performances. (A realistic estimate of $\bar
{n}_{\text{ex}}$ for a practical receiver setup will be explicitly taken into
account in our subsequent analysis of the composable QKD rates.)

The value of the background noise $\bar{n}_{B}$ depends on the operational
setting (time of the day/direction of the link), besides features of the
receiver, such as its aperture $a_{R}$, field of view $\Omega_{\text{fov}}$,
detection time $\Delta t$, carrier frequency $\lambda$\ and spectral filter
$\Delta\lambda$. Let us evaluate $\bar{n}_{B}$ in the various settings. It is
convenient to define the parameter
\begin{equation}
\Gamma_{R}:=\Delta\lambda\Delta t\Omega_{\text{fov}}a_{R}^{2}.
\label{gammaERRE}%
\end{equation}
Assuming $\Delta\lambda=1~$nm and $\Delta t=10~$ns for the detector, and
$\Omega_{\text{fov}}=10^{-10}~$sr and $a_{R}=40~$cm for the receiving
telescope, we compute $\Gamma_{R}=1.6\times10^{-19}~$m$^{2}$ s nm sr.

Then, for uplink we may write%
\begin{equation}
\bar{n}_{B}^{\text{up}}=\kappa H_{\lambda}^{\text{sun}}\Gamma_{R},
\label{UPthermalNOISE}%
\end{equation}
where $H_{\lambda}^{\text{sun}}$ is the solar spectral irradiance at the
relevant wavelength $\lambda$, e.g., $H_{\lambda}^{\text{sun}}=4.61\times
10^{18}$ photons m$^{-2}$ s$^{-1}$ nm$^{-1}$ sr$^{-1}$\ at $\lambda=800~$nm.
In the formula above, the dimensionless parameter $\kappa$ depends on the
geometry and albedos of the Earth and the Moon. Its value is $\kappa
_{\text{day}}\simeq0.3$ for day-time and $\kappa_{\text{night}}\simeq
7.36\times10^{-7}$ for full-Moon night-time (see Appendix~\ref{APPnoiseSAT}
for more details). For downlink, we instead have
\begin{equation}
\bar{n}_{B}^{\text{down}}=H_{\lambda}^{\text{sky}}\Gamma_{R},
\label{DOWNthermalNOISE}%
\end{equation}
where $H_{\lambda}^{\text{sky}}$ is the spectral irradiance of the sky in
units of photons m$^{-2}$ s$^{-1}$ nm$^{-1}$ sr$^{-1}$. At $\lambda=800~$nm,
its value ranges from $1.9\times10^{13}$ (full-Moon clear night)\ to
$1.9\times10^{16}$\ (clear day-time) and $1.9\times10^{18}$ (cloudy day-time).

A summary of the resulting values for the mean number of photons is provided
in Table~\ref{TableDDDD}. These values confirm that background noise is
practically negligible at night time, while it plays an important role for day
time. Under general sky conditions, day-time operations need to be described
by thermal-loss channels, where loss and fading effects are combined with the
thermal bath collected by the field of view of the receiver.

\begin{table}[h]
\vspace{-0.3cm}
\[%
\begin{tabular}
[c]{c|c|c}
& Day & Night\\\hline
Downlink & $%
\begin{array}
[c]{c}%
\simeq0.3\text{ (cloudy)}\\
\simeq3\times10^{-3}\text{ (clear)}%
\end{array}
$ & $\simeq3\times10^{-6}$\\\hline
Uplink & $\simeq0.22$ & $\simeq5.4\times10^{-7}$%
\end{tabular}
\]
$\ \vspace{-0.4cm} $\caption{Environmental noise in satellite communications
(mean number of thermal photons $\bar{n}_{B}$ per mode). This is shown for
uplink and downlink in various conditions, considering a typical receiver
($\Gamma_{R}=1.6\times10^{-19}~$m$^{2}$ s nm sr).}%
\label{TableDDDD}%
\end{table}

\subsubsection{Noise filtering\label{NoiseFILTERsection}}

It is important to observe that these values strongly depend on the filter
$\Delta\lambda$. At $800~$nm, the filter $\Delta\lambda=1~$nm corresponds to a
bandwidth of $\Delta\nu=c\lambda^{-2}\Delta\lambda\simeq470~$GHz (here $c$ is
the speed of light). This value for the filter is certainly appropriate for
GEO\ satellites, while it may become a bit more challenging for fast-moving
satellites in the LEO region where the Doppler shift need to be compensated by
adaptive optics~\cite{Gruneisen} (so that the central frequency of the beam is
suitably tracked during the flyby of the satellite, with generally-different
speeds at different zenith angles). However, due to other observations, the
value of $\Delta\lambda=1~$nm can also be considered to be relatively large,
since it can be reduced by employing specific detection techniques at the
receiver. In other words, the effective value of $\Delta\nu$ can be greatly
reduced by suitable interferometric measurements at the receiver.

In fact, an important ingredient of CV\ quantum communications is the local
oscillator (LO). In the method of \textquotedblleft transmitted
LO\textquotedblright\ (TLO), each signal is multiplexed in polarization\ with
a bright classical LO pulse, carrying phase information. Signal and LO pulse
are de-multiplexed at the receiver via a polarizing beam splitter and made to
interfere on balanced beam-splitter(s) in a homodyne or heterodyne setup.
Alternatively, one can use the method of the \textquotedblleft local
LO\textquotedblright\ (LLO)~\cite{LLO}. Here there is the transmission of
bright reference pulses regularly interleaved with the signal pulses (time
multiplexing). The reference pulses are detected and used to digitally
reconstruct the LO at the receiver. Each signal is homodyned/heterodyned by
using an independent LO which is then rotated according to the phase reconstruction.

In both cases, the output of homodyne is proportional to $\sqrt{\bar
{n}_{\text{LO}}}\hat{x}$, where $\hat{x}$ is signal's generic quadrature and
$\bar{n}_{\text{LO}}\gg1$ is the number of photons of the LO pulse interfering
with the signal. Only thermal noise mode-matching with the LO will be detected
(together with the signal), while all the other noise will be filtered out.
For this reason, the actual filter of the receiver will be limited by the
bandwidth of the LO\ pulses. Compatibly with the time-bandwidth product (which
is $\Delta t\Delta\nu\geq0.44$ for Gaussian pulses), the bandwidth of the LO
can be made very narrow so that very small values of $\Delta\lambda$ are
indeed accessible. For $\Delta t=10~$ns, one can consider $\Delta\nu=50~$MHz,
which corresponds to $\Delta\lambda=0.1~$pm around $800$~nm. Such small
bandwidths can certainly be generated by current lasers, whose line-widths can
easily go down to $1$~KHz (e.g., for continuous-wave lasers).

From this point of view, the value of $\Delta\nu\simeq470~$GHz is pessimistic
in the setting of LO-based CV\ quantum communications. For this reason, we
also account for the possibility of receiver designs that may have such
ultra-narrow filters. An effective filter of $\Delta\lambda=0.1~$pm is
$10^{-4}$ narrower than the value considered above in Table~\ref{TableDDDD}.
With such a filter, we have a corresponding $10^{-4}$ suppression of thermal
noise $\bar{n}_{B}$, so that its day-time values become almost negligible as
reported in Table~\ref{Filter_Table} (where we do not report the night-time
values since they become $\lesssim10^{-10}$). In our following analysis we
therefore include this better regime.

\begin{table}[h]
\vspace{-0.3cm}
\[%
\begin{tabular}
[c]{c|c}%
Day-downlink & $%
\begin{array}
[c]{c}%
\simeq3\times10^{-5}\text{ (cloudy)}\\
\simeq3\times10^{-7}\text{ (clear)}%
\end{array}
$\\\hline
Day-uplink & $\simeq2.2\times10^{-5}$%
\end{tabular}
\
\]
$\ \vspace{-0.4cm} $\caption{Day-time noise $\bar{n}_{B}$ with a narrow filter
$\Delta\lambda=0.1~$pm (corresponding to $\Gamma_{R}=1.6\times10^{-23}~$%
m$^{2}$ s nm sr).}%
\label{Filter_Table}%
\end{table}

As also noted in Ref.~\cite{FreeSpacePaper}, when thermal noise is
non-negligible, there is a general trade-off associated with the receiver
aperture $a_{R}$. If we increase $a_{R}$\ we certainly increase the
transmissivity of the link according to Eqs.~(\ref{diffINDUCED})
and~(\ref{shortTERMtransmissivity}). However, larger values of $a_{R}$ also
lead to higher values of background noise collected by the receiver as is also
clear from Eqs.~(\ref{gammaERRE})-(\ref{DOWNthermalNOISE}). As a result, an
optimal value for $a_{R}$ can be determined by optimizing over the rate [or
the thermal bound in Eq.~(\ref{ThermalLOSSub}) of the next sub-section]. In
this regard, it is clear that the use of an ultra-narrow filter able to
suppress the noise would mitigate this problem and give access to large
apertures for the receiver. However, even in total absence of thermal noise,
too large apertures would encounter other problems related to enhanced
turbulence, in terms of increasing off-axis scintillation and number of
short-term speckles (see Appendix~\ref{TurbSECTION} for more details on these effects).

\subsubsection{Accounting for the untrusted noise}

It is important to observe that the total input-output relation of
Eq.~(\ref{TLchannel}) from transmitter to receiver is equivalent to the action
of a thermal-loss channel $\mathcal{E}_{\tau,\bar{n}}$ with transmissivity
$\tau$ and environmental number of photons
\begin{equation}
\bar{n}_{e}=\bar{n}/(1-\tau).
\end{equation}
From the point of view of the generic quadrature $\hat{x}$ of the mode (i.e.,
position $\hat{q}$ or momentum $\hat{p}$), Eq.~(\ref{TLchannel}) corresponds
to the following transformation
\begin{equation}
\hat{x}_{T}\rightarrow\hat{x}_{R}=\sqrt{\tau}\hat{x}_{T}+\xi_{\text{add}%
},\label{IOdopo}%
\end{equation}
where the additive-noise variable $\xi_{\text{add}}$ can be written as
$\xi_{\text{add}}=\sqrt{1-\tau}\hat{e}$, and $\hat{e}$ is the quadrature of an
environmental mode with $\bar{n}_{e}$ mean thermal photons. Therefore, the
overall process can be represented as an effective beam-splitter of
transmissivity $\tau$ mixing the input mode of the transmitter with an
environmental thermal mode. At the output of the beam splitter, one mode is
detected by the receiver, while the other mode goes back into the environment.

In the worst-case scenario, one assumes that the eavesdropper (Eve) controls
the input environmental mode, so that the latter is part of a TMSV state
in her hands (which realizes a purification of the channel, unique up to local
isometries on the environment). One also assumes that all environmental modes
after interaction are stored by Eve in a quantum memory to be subject to an
optimal joint measurement. This active strategy is known as collective
entangling-cloner attack and represents the most typical collective Gaussian
attack~\cite{CollectiveATT}. According to Ref.~\cite{QKDpaper}, the relative
entropy of entanglement suitably computed over the asymptotic Choi matrix of
the thermal-loss channel $\mathcal{E}_{\tau,\bar{n}}$ provides an upper bound
for its secret key capacity $K$. For each instantaneous channel $\mathcal{E}%
_{\tau,\bar{n}}$ describing the satellite link, we have $K(\mathcal{E}%
_{\tau,\bar{n}})\leq\Phi_{\tau,\bar{n}}$, where~\cite{QKDpaper}%
\begin{equation}
\Phi_{\tau,\bar{n}}=\left\{
\begin{array}
[c]{l}%
-\log_{2}\left[  (1-\tau)\tau^{\bar{n}_{e}}\right]  -h\left(  \bar{n}%
_{e}\right)  ,~\text{for~}\bar{n}\leq\tau\text{,}\\
\\
0\text{~~for~}\bar{n}>\tau\text{,}%
\end{array}
\right.  \label{FIplobTHERMAL}%
\end{equation}
and we have set $h\left(  x\right)  :=(x+1)\log_{2}(x+1)-x\log_{2}x$.

For the analysis of the ultimate bounds, we may assume that $\bar{n}$ does not
depend on $\tau$. Indeed, the external background noise $\bar{n}_{B}$ does not
depend on the transmissivity affecting the signals, since it is noise
collected by the field of view of the receiver. Then, in the presence of
non-negligible values for the setup noise $\bar{n}_{\text{ex}}$ that may
depend on $\tau$, we can always optimize $\bar{n}_{\text{ex}}$ over $\tau$
(and, in particular, minimize its value for the study of upper bounds, and
maximize it for that of the lower bounds).

Thus, we can represent the satellite fading channel in the presence of
background noise by means of the channel ensemble $\mathcal{E}_{\text{fad}%
}^{\text{noi}}=\{P_{\sigma}(\tau),\mathcal{E}_{\tau,\bar{n}}\}$. By averaging
$\Phi_{\tau,\bar{n}}$ over the fading process in $\tau$,
Ref.~\cite{FreeSpacePaper} computed the general free-space upper bound
\begin{equation}
K\leq\int_{\bar{n}}^{\eta}d\tau~P_{\sigma}(\tau)\Phi_{\tau,\bar{n}}%
\leq\mathcal{B}(\eta,\sigma)-\mathcal{T}(\bar{n},\eta,\sigma),
\label{ThermalLOSSub}%
\end{equation}
for $\bar{n}\leq\eta$\ and where the thermal correction $\mathcal{T}$ is given
by\
\begin{align}
\mathcal{T}(\bar{n},\eta,\sigma)  &  =\left\{  1-e^{-\frac{r_{0}^{2}}%
{2\sigma^{2}}\left[  \ln\left(  \eta/\bar{n}\right)  \right]  ^{2/\gamma}%
}\right\} \nonumber\label{ThermalCORRECTION}\\
&  \times\left[  \frac{\bar{n}\log_{2}\bar{n}}{1-\bar{n}}+h\left(  \bar
{n}\right)  \right]  +\mathcal{B}(\bar{n},\sigma).
\end{align}
Similarly, one may write the lower bound~\cite{FreeSpacePaper}
\begin{align}
K  &  \geq E\geq\mathcal{B}(\eta,\sigma)-\int_{0}^{\eta}d\tau~P_{\sigma}%
(\tau)h\left(  \frac{\bar{n}}{1-\tau}\right) \\
&  \geq\mathcal{B}(\eta,\sigma)-h\left(  \frac{\bar{n}}{1-\eta}\right)  ,
\label{ThermalLOSSlb}%
\end{align}
which is based on the RCI and can be approximated by ideal implementations of
CV-QKD protocols~\cite{FreeSpacePaper,NoteTRUSTED}.

The application of these formulas to the specific satellite models developed
above allows us to bound the ultimate rates for key (and entanglement)
distribution that are achievable in the presence of background noise.

\subsubsection{Maximum ranges}

The presence of thermal noise restricts the maximum slant distance for secure
key generation to some finite value $z_{\max}$. From Eq.~(\ref{ThermalLOSSub}%
), we see that no key (or entanglement) distribution is possible in
correspondence to the entanglement-breaking condition $\bar{n}=\eta$, which
automatically leads to an upper bound for $z_{\max}$. If we assume ideal
conditions, where all the effects are negligible with the exception of
diffraction and thermal noise, the over-optimistic threshold condition
$\bar{n}=\eta_{\text{d}}$ would still restricts the range to some finite
value. In fact, the latter leads to
\begin{equation}
\bar{n}\leq\sqrt{f_{0R}(z_{\max})},~~f_{0R}(z):=\left[  \pi w_{0}%
a_{R}/(\lambda z)\right]  ^{2},
\end{equation}
where $f_{0R}(z)$ is the Fresnel number product of the beam and the receiver.
Then, using Eqs.~(\ref{UPthermalNOISE}) and~(\ref{DOWNthermalNOISE}), we get
the following bounds for the maximum ranges in uplink and downlink%
\begin{equation}
z_{\max}^{\text{up}}\leq\frac{\Sigma}{\kappa H_{\lambda}^{\text{sun}}%
},~z_{\max}^{\text{down}}\leq\frac{\Sigma}{H_{\lambda}^{\text{sky}}%
},\label{simpleRANGE}%
\end{equation}
where we have set $\Sigma:=\pi w_{0}/(\lambda\Delta\lambda\Delta
t\Omega_{\text{fov}}a_{R})$. It is also clear that $\bar{n}\geq1$ leads to
entanglement breaking, so that no key or entanglement can be distributed when
$\Gamma_{R}\geq(\kappa H_{\lambda}^{\text{sun}})^{-1}$ in uplink or
$\Gamma_{R}\geq1/H_{\lambda}^{\text{sky}}$ in downlink.

\begin{figure*}[t]
\vspace{-0.0cm}
\par
\begin{center}
\includegraphics[width=0.99\textwidth] {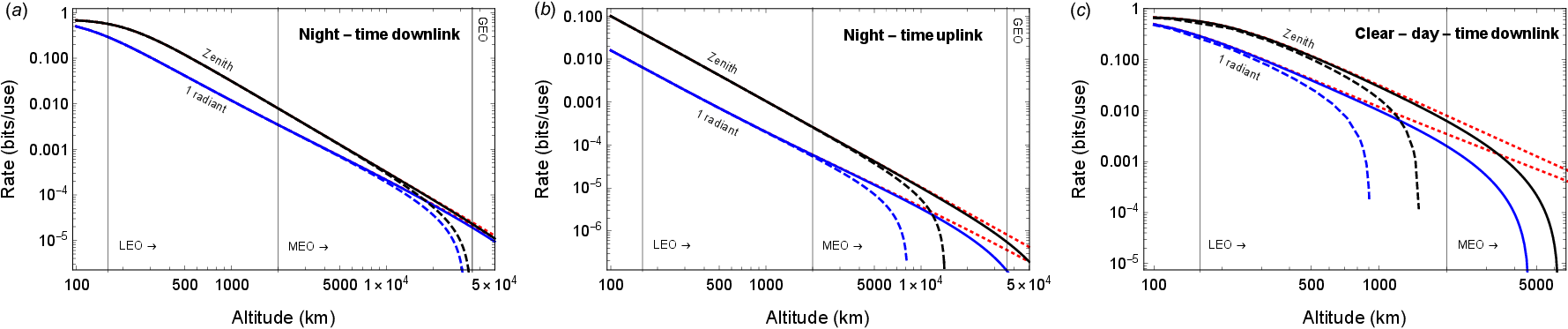}
\end{center}
\par
\vspace{-0.2cm}\caption{Optimal key rate between a ground station and a
satellite at altitude $h$, in low-noise configurations of: (a)~night-time
downlink, (b)~night-time uplink, and (c)~clear-day-time downlink. In each
configuration we consider the thermal-loss upper bound $\mathcal{B}%
-\mathcal{T}$ of Eq.~(\ref{ThermalLOSSub}) (solid lines) and the thermal-loss
lower bound of Eq.~(\ref{ThermalLOSSlb}) (lower dashed lines), specified for
the zenith position (black lines) and $1$ radiant (blue lines). We also plot
the loss-limited upper bound $\mathcal{B}$ of Eq.~(\ref{mainOTHER}) (red
dashed lines). We account for diffraction, extinction, quantum efficiency,
pointing error ($1$~$\mu$rad) and atmospheric turbulence (H-V model).
Collimated beam has $\lambda=800~$nm and $w_{0}=20~$cm. Receiver has a
telescope with $a_{R}=40~$cm and $\Omega_{\text{fov}}=10^{-10}~$sr, and a
detector with filter $\Delta\lambda=1~$nm, $\Delta t=10~$ns, $\eta
_{\text{eff}}=0.4$ and $\bar{n}_{\text{ex}}=0$ (no extra setup noise).
Therefore, the receiver has $\Gamma_{R}=1.6\times10^{-19}~$m$^{2}$ s nm sr and
the background thermal photons $\bar{n}_{B}$ are those specified in
Table~\ref{TableDDDD}. For a receiver implementing a narrow filter
$\Delta\lambda=0.1~$pm, the thermal-loss upper and lower bounds in (a), (b)
and (c) collapse and coincide with the corresponding loss-limited bounds at
the zenith and $1$ radiant (red dashed lines).}%
\label{Trittico}%
\end{figure*}

\begin{figure}[t]
\vspace{0.2cm}
\par
\begin{center}
\includegraphics[width=0.39\textwidth] {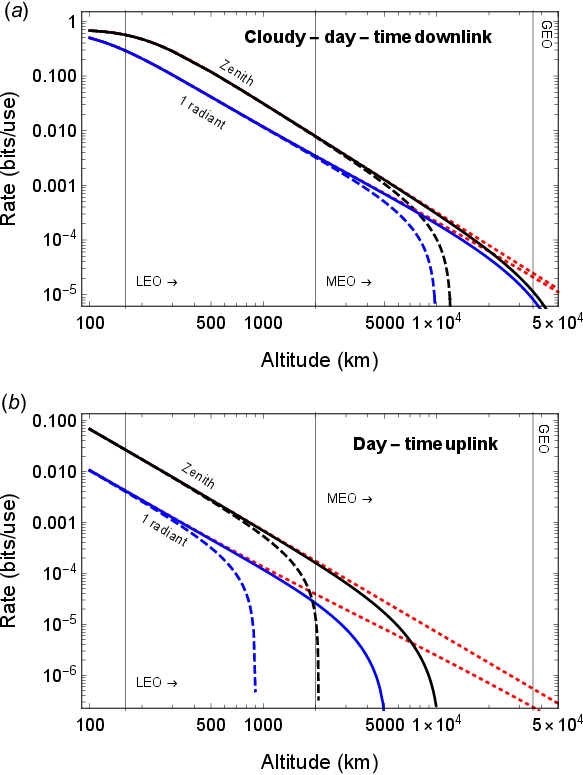}
\end{center}
\par
\vspace{-0.3cm}\caption{Optimal performances in day-light conditions with a
narrow filter. As in Fig.~\ref{Trittico}, we consider the key rate between a
ground station and a satellite at various altitudes $h$. In each configuration
we consider the thermal-loss upper bound $\mathcal{B}-\mathcal{T}$ of
Eq.~(\ref{ThermalLOSSub}) (solid lines) and the thermal-loss lower bound of
Eq.~(\ref{ThermalLOSSlb}) (lower dashed lines), specified for the satellite at
the zenith position (black lines) or at $1$ radiant (blue lines). We also plot
the loss-limited upper bound $\mathcal{B}$ of Eq.~(\ref{mainOTHER}) (red
dashed lines). In particular, we consider cloudy-day-time downlink in (a), and
day-time uplink in (b). We account for diffraction, extinction, quantum
efficiency, pointing error ($1$~$\mu$rad) and atmospheric turbulence (H-V
model). Collimated beam has $\lambda=800~$nm and $w_{0}=20~$cm. Receiver has a
telescope with $a_{R}=40~$cm and $\Omega_{\text{fov}}=10^{-10}~$sr, and a
detector with narrow filter $\Delta\lambda=0.1~$pm, $\Delta t=10~$ns,
$\eta_{\text{eff}}=0.4$ and $\bar{n}_{\text{ex}}=0$ (no extra setup noise).
Therefore, the receiver has $\Gamma_{R}=1.6\times10^{-23}~$m$^{2}$ s nm sr and
the background thermal photons $\bar{n}_{B}$ are strongly suppressed as in
Table~\ref{Filter_Table}.}%
\label{Duetto}%
\end{figure}

While the conditions in Eq.~(\ref{simpleRANGE}) are particularly simple,
tighter bounds on $z_{\max}$ can be obtained by directly imposing
$\mathcal{B}(\eta,\sigma)=\mathcal{T}(\bar{n},\eta,\sigma)$ in
Eq.~(\ref{ThermalLOSSub}),\ and using Eqs.~(\ref{UPthermalNOISE})
and~(\ref{DOWNthermalNOISE}). For typical parameters, the situation is the one
depicted in Table~\ref{MaxRanges}, where we observe that thermal noise
represents quite an important limitation for day-time operation. As a matter
of fact, day-time uplink does seem to be particularly challenging for
satellite QKD. For the regime considered, the security range is roughly
limited to the K\'{a}rm\'{a}n line. In downlink, day-time limitations appear
to be less severe. For the considered regime, secret key generation is
confined to the LEO region in a cloudy day, but may access MEO altitudes in a
clear-sky day.

\begin{table}[h]
\vspace{-0.3cm}
\[%
\begin{tabular}
[c]{c|c|c}
& Day & Night\\\hline
Downlink & $%
\begin{array}
[c]{c}%
\lesssim650~\text{km (cloudy)}\\
\lesssim6300~\text{km (clear)}%
\end{array}
$ & $\lesssim2\times10^{5}~$km\\\hline
Uplink & $\lesssim110~$km & $\lesssim9\times10^{4}~$km
\end{tabular}
\]
$\ \vspace{-0.4cm} $\caption{Bounds on maximum ranges for secure key
distribution in uplink and downlink, considering a typical receiver with
$\Gamma_{R}=1.6\times10^{-19}~$m$^{2}$ s nm sr (with filter $\Delta\lambda
=1~$nm).}%
\label{MaxRanges}%
\end{table}

In order to reach higher altitudes we need to consider better setup
parameters. For instance, a faster detector working at $1~$GHz (instead of
$100~$MHz) will reduce $\Gamma_{R}$ and $n_{B}^{\text{up}}$\ by a factor of
$10$. As a result, we get a larger bound for the maximum range in day-uplink,
i.e., $z_{\max}\lesssim340~$km, so that LEO altitudes are reached. It is clear
that a strong mitigation for this problem comes from the use of ultra-narrow
filters $\Delta\lambda$, so that the amount of background thermal noise
entering the detector becomes negligible, even for day-time operation. With an
effective filter of $\Delta\lambda=0.1~$pm around $800$~nm, day-time ranges
are sensibly increased. Night-time values of the bounds go well beyond
satellite applications, while the bounds for day-time ranges become large as
in Table~\ref{filterGGG}.

\begin{table}[t]
\vspace{-0.3cm}
\[%
\begin{tabular}
[c]{c|c}%
Day-downlink & $%
\begin{array}
[c]{c}%
\lesssim6.2\times10^{4}\text{ km~(cloudy)}\\
\lesssim6.2\times10^{5}\text{ km (clear)}%
\end{array}
$\\\hline
Day-uplink & $\lesssim10^{4}~$km
\end{tabular}
\ \
\]
$\ \vspace{-0.4cm} $\caption{Bounds on maximum ranges for day-time secure key
distribution in uplink and downlink, with a narrow filter $\Delta\lambda
=0.1$~pm, so that $\Gamma_{R}=1.6\times10^{-23}~$m$^{2}$ s nm sr.}%
\label{filterGGG}%
\end{table}

\subsubsection{Analysis of the thermal-loss bounds}

Following our preliminary analysis on the security ranges, we now explicitly
study the thermal-loss upper bound in Eq.~(\ref{ThermalLOSSub}) and the
corresponding lower bound in Eq.~(\ref{ThermalLOSSlb}). Because we are here
interested in investigating optimal performances, we assume the ideal case of
negligible setup noise ($\bar{n}_{\text{ex}}\simeq0$, noiseless receiver),
even though we allow for non-unit quantum efficiency $\eta_{\text{eff}}$. We
start with the investigation of the optimal rates that are achievable by using
a relatively-large filter ($\Delta\lambda=1~$nm) in low-noise conditions of
night-time downlink/uplink and day-time downlink with clear sky (see
Fig.~\ref{Trittico}).

For night-time downlink/uplink, one can numerically check that the thermal
correction $\mathcal{T}(\bar{n},\eta,\sigma)$ is practically negligible, so the thermal-loss upper bound coincides with the loss-limited bound
$\mathcal{B}(\eta,\sigma)$ of Eq.~(\ref{mainOTHER}), with only small
deviations at GEO altitudes. Furthermore, as we can see from
Figs.~\ref{Trittico}(a,b), the thermal-loss upper bound of
Eq.~(\ref{ThermalLOSSub}) numerically coincides with the lower bound of
Eq.~(\ref{ThermalLOSSlb}) at LEO altitudes. This means that, for LEO
satellites, the two bounds collapse into the loss-limited bound $\mathcal{B}%
(\eta,\sigma)$ which therefore represents the secret key capacity $K$ (and
entanglement-distribution capacity $E$) of the satellite
channel~\cite{NoteCapacity}. Said in other words, for night-time
downlink/uplink and LEO altitudes, the free-space fading channel can certainly
be approximated by an ensemble of pure-loss channels and we can write the
achievability result $E\simeq K\simeq\mathcal{B}$. At higher altitudes in the
MEO region, the presence of a gap does not allow us to enforce the pure-loss
assumption and claim achievability with respect to the chosen parameters (in
particular, for the value $\Delta\lambda=1~$nm).

For clear-day-time downlink in Fig.~\ref{Trittico}(c), the gap between the two
thermal-loss bounds of Eqs.~(\ref{ThermalLOSSub}) and~(\ref{ThermalLOSSlb})
already appears in the LEO region, even though at low altitudes (of the order
of $160$~km) there is a substantial coincidence. We can appreciate how, for
increasing altitudes, not only the gap increases but the thermal-loss upper
bound also substantially departs from the loss-limited bound $\mathcal{B}%
(\eta,\sigma)$, confirming the relevant role of the background thermal noise
in limiting the rate performance for day-time operation.

As expected, the role of the thermal background noise can be strongly
mitigated by the use of a narrow filter, e.g., $\Delta\lambda=0.1~$pm (as in
homodyne-like setups at the receiver). In conditions of low background noise,
our numerical investigation shows that the use of such a narrow filter enables
the parties to achieve the loss-limited bound $\mathcal{B}$ at the relevant
altitude and zenith angle, expressed by the red dashed lines in
Fig.~\ref{Trittico}. In other words, the thermal-loss upper and lower bounds
collapse into the loss-limited bound $\mathcal{B}$ at all satellite altitudes,
so that the remote parties can distribute ebits and secret bits at the
capacity value $E\simeq K\simeq\mathcal{B}$. Such a collapse is more limited
in conditions of stronger background noise, as typical in cloudy-day-time
downlink and day-time uplink, but yet the use of a narrow filter would allow
to reach good communication rates in such scenarios as shown in
Fig.~\ref{Duetto}. As we can see from the figure, we can match the
loss-limited bound $\mathcal{B}$ in all the LEO\ region for day-time downlink
with cloudy sky, and for the first part of LEO in day-time uplink. Extending
the achievability of $\mathcal{B}$ to all altitudes would require an even
narrower filter.

Before concluding this section, it is important to remark that, in the various
scenarios studied above, the achievability of the upper bound refers to a
satellite \textit{at a fixed} geometry, i.e., with fixed altitude and zenith
angle. This means that we ignore the orbital dynamics or we assume that such a
dynamics is much slower than the time-scale of the quantum communication, so
that the channel is used many times before the satellite has substantially
moved (fast-clock limit/large repetition rate). However, apart from GEO
satellites, the orbital dynamics is relevant and, for realistic clocks, we
need to modify the lower bound in order to account for this process. In fact,
suppose that the rate in Eq.~(\ref{ThermalLOSSlb}) can approximately be
achieved after $N$ pulses. For some realistic clock, such a block of $N$
pulses corresponds to a slice of the orbit over which we identify the
worst-case values for $h$ and $\theta$. An achievable rate would certainly be
given by Eq.~(\ref{ThermalLOSSlb}) computed over such values, but this rate
will no longer match the upper bound in the slice (to be computed on the
best-case values for $h$ and $\theta$). In the following, we will consider the
problem of orbital slicing more carefully for the derivation of a composable
secret key rate.


\section{Composable finite-size security for satellite
CV-QKD\label{SECpractical}}

Once we have clarified the ultimate limits for distributing keys (and
entanglement) with satellites, we study the rates that are achievable by
practical CV-QKD\ protocols, where we explicitly account for finite-size and
composable aspects. We adopt the pilot-based post-selected protocol studied in
Ref.~\cite{FreeSpacePaper}, here suitably applied and extended to considering
the underlying physical models valid for satellite communications.

\subsection{Overview\label{SECqkd1}}

The idea is to use a coherent-state protocol, where Gaussian-modulated signal
pulses (used to encode the information) are randomly interleaved with more
energetic pilot pulses, that are used to monitor the instantaneous
transmissivity of the satellite link in real-time. In this way, the parties
are able to allocate the distributed data to slots of transmissivity and to
perform classical post-processing slot-by-slot. In general, one may divide the
interval $[0,\eta]$\ by introducing a lattice with step $\delta\tau$, so that
we have $M=\eta/\delta\tau$ slots $[\tau_{k},\tau_{k+1}]$ with $\tau
_{k}:=(k-1)\delta\tau$ and $k=1,\ldots,M$. When the $k$th slot is selected,
with probability
\begin{equation}
p_{k}=\int_{\tau_{k}}^{\tau_{k+1}}d\tau~P_{\sigma}(\tau), \label{pkappaFAD}%
\end{equation}
the corresponding data points are associated to its minimum transmissivity
$\tau_{k}$.

A more practical post-selection strategy consists of introducing a threshold
value $\eta_{\text{th}}$ and only accepting data points with $\tau
>\eta_{\text{th}}$, i.e., within the post-selection interval $\Delta
=[\eta_{\text{th}},\eta]$. This happens with success probability
\begin{equation}
p_{\text{th}}=\int_{\eta_{\text{th}}}^{\eta}d\tau~P_{\sigma}(\tau).
\label{pthresholdEX}%
\end{equation}
The data points are then associated with the value of the threshold
transmissivity $\eta_{\text{th}}$. In our work, we adopt this post-selection
strategy from Ref.~\cite{FreeSpacePaper}, which is more robust, especially in
terms of collecting enough statistics for parameter estimation. This approach
is different from the clusterization method of Ref.~\cite{RuppertCluster}%
\ which may instead be affected by poor statistics.

In the next subsections, we first discuss the security of the link for a fixed
value of the transmissivity, as if there were no fading and with a satellite
in a fixed geometry. Then, we introduce the fading process, for which we use
the post-selection strategy above, and we subsequently account for orbital
dynamics in the key rate. These derivations will provide realistic rates for
satellite quantum communications and will also be used for our comparison with
a ground network.

\subsection{Composable key rate at fixed transmissivity\label{SECqkd2}}

Let us start by considering the link to be a thermal-loss channel
$\mathcal{E}_{\tau,\bar{n}}$ with fixed transmissivity $\tau$\ and thermal
noise\ $\bar{n}_{e}=\bar{n}/(1-\tau)$, where $\bar{n}$ may take the form in
Eq.~(\ref{thermalCnoise}). In a coherent-state protocol~\cite{GG02,Noswitch},
the transmitter generates an input mode with generic quadrature $\hat{x}%
_{T}=x+\hat{v}$, where $\hat{v}$ is the vacuum quadrature and $x$ is a
Gaussianly modulated variable with variance $\sigma_{x}^{2}=\mu-1$ (with $\mu
$\ being the variance of the average thermal state generated by the
transmitter). At the output of the thermal-loss channel $\mathcal{E}%
_{\tau,\bar{n}}$, the receiver gets a mode with generic quadrature $\hat
{x}_{R}$ as given in Eq.~(\ref{IOdopo}). Assuming that the output is homodyned
(randomly in $\hat{q}$ or $\hat{p}$), then the receiver's classical outcome
takes the form
\begin{equation}
y=\sqrt{\tau}x+z, \label{IOrel}%
\end{equation}
where $z$ is a random noise variable, distributed according to a Gaussian with
zero mean and variance $\sigma_{z}^{2}=2\bar{n}+1$. If heterodyne is used,
then the outcome $y$ (for each of the two quadratures) is affected by thermal
noise with larger variance $\sigma_{z}^{2}=2\bar{n}+2$. Compactly, we may
therefore write
\begin{equation}
\sigma_{z}^{2}=2\bar{n}+\nu_{\text{det}},
\end{equation}
where $\nu_{\text{det}}=1$ holds for homodyne, and $\nu_{\text{det}}=2$ for
heterodyne detection.

For fixed $\tau$ and $\bar{n}$, one computes the mutual information
$I(x:y|\tau,\bar{n})$ which takes simple expressions for the two types of
output detections, i.e.,%
\begin{align}
I^{\text{hom}}(x &  :y|\tau,\bar{n})=\frac{1}{2}\log_{2}\left(  1+\frac
{\tau\sigma_{x}^{2}}{2\bar{n}+1}\right)  ,\label{Ihom}\\
I^{\text{het}}(x &  :y|\tau,\bar{n})=\log_{2}\left(  1+\frac{\tau\sigma
_{x}^{2}}{2\bar{n}+2}\right)  .\label{Ihet}%
\end{align}
In reverse reconciliation, it is easy to compute Eve's Holevo bound
$\chi(E:y|\tau,\bar{n})$ with corresponding expressions for $\chi^{\text{hom}%
}$ and $\chi^{\text{het}}$ (see Ref.~\cite[Sec.~III.C]{FreeSpacePaper} for
details). Therefore, one may write the asymptotic key rate against collective
Gaussian attacks, which takes the form%
\begin{equation}
R_{\text{asy}}(\tau,\bar{n})=\beta I(x:y|\tau,\bar{n})-\chi(E:y|\tau,\bar
{n}),\label{AsyRATE}%
\end{equation}
where $\beta\in\lbrack0,1]$ is the reconciliation parameter (here $\beta I$
corresponds to the effective rate of the code which is employed in the step of
error correction).

In an actual practical implementation, there is a number $N$ of transmitted
signals, whose $m$ are randomly selected by the parties and used for the
estimation of the channel parameters $\tau$ and $\bar{n}$. From $m_{p}%
:=m\nu_{\text{det}}$ pairs of data points $\{x_{i},y_{i}\}$ related by
Eq.~(\ref{IOrel}), the parties build the following unbiased estimators
$\widehat{\tau}=\widehat{T}^{2}$ and $\widehat{\bar{n}}=\left(  \widehat
{\sigma_{z}^{2}}-\nu_{\text{det}}\right)  /2$, where~\cite{FreeSpacePaper}%
\begin{equation}
\widehat{T}=\frac{\sum_{i=1}^{m_{p}}x_{i}y_{i}}{m_{p}\sigma_{x}^{2}}%
,~\widehat{\sigma_{z}^{2}}:=\frac{1}{m_{p}}\sum_{i=1}^{m_{p}}(y_{i}-\widehat{T}x_{i})^{2}.
\end{equation}
Up to $\mathcal{O}(m_{p}^{-2})$, these have variances%
\begin{align}
\sigma_{\tau}^{2}  &  :=\mathrm{var}(\widehat{\tau})\simeq\frac{4\tau^{2}%
}{m_{p}}\left(  2+\frac{\sigma_{z}^{2}}{\tau\sigma_{x}^{2}}\right)
,\label{sigmatVAL}\\
\sigma_{\bar{n}}^{2}  &  :=\mathrm{var}(\widehat{\bar{n}})\simeq\frac
{\sigma_{z}^{4}}{2m_{p}}.
\end{align}

We can therefore build the worst-case parameters
\begin{align}
\tau^{\prime} &  :=\widehat{\tau}-w\sigma_{\tau}\simeq\tau-2w\sqrt{\frac
{2\tau^{2}+\tau\sigma_{z}^{2}/\sigma_{x}^{2}}{m_{p}}},\label{wscase}\\
\bar{n}^{\prime} &  :=\widehat{\bar{n}}+w\sigma_{\bar{n}}\simeq\bar{n}%
+\frac{w\sigma_{z}^{2}}{\sqrt{2m_{p}}}.
\end{align}
Each of them bounds the corresponding actual value up to an error probability
$\varepsilon_{\text{pe}}$, with
\begin{equation}
w=\sqrt{2}\operatorname{erf}^{-1}(1-2\varepsilon_{\text{pe}}%
).\label{wexpression1}%
\end{equation}
Alternatively, one may write a more robust tail bound which leads to the same
expressions as above but with
\begin{equation}
w=\sqrt{2\ln(1/\varepsilon_{\text{pe}})},\label{wexpression2}%
\end{equation}
which can be used in the cases where $\varepsilon_{\text{pe}}$ is very small,
e.g., less than $10^{-17}$~\cite[Sec.~III.D]{FreeSpacePaper}.

It is therefore clear that the effect of parameter estimation is to modify the
asymptotic rate in Eq.~(\ref{AsyRATE}) as follows%
\begin{equation}
R_{\text{asy}}\rightarrow\frac{n}{N}R_{\text{pe}},~R_{\text{pe}}%
=R_{\text{asy}}(\tau^{\prime},\bar{n}^{\prime})
\end{equation}
where $n=N-m$ is the number of signals remaining for key generation. This is
valid up to a total error $\simeq2\varepsilon_{\text{pe}}$. The rate
$R_{\text{pe}}$ takes specific formulas for the homodyne ($R_{\text{pe}%
}^{\text{hom}}$) and the heterodyne ($R_{\text{pe}}^{\text{het}}$) protocol by
using Eqs.~(\ref{Ihom}) and~(\ref{Ihet}) together with corresponding
expressions for the Holevo information $\chi^{\text{hom}}$ and $\chi
^{\text{het}}$.

Parameter estimation provides the parties with crucial information about the
amount of loss and noise present in the channel, so they can apply a
suitable code to correct their data. The procedure of error correction is
successful with a probability $p_{\text{ec}}$ which depends on the rate $\beta
I$ of the code and a pre-established $\varepsilon$-correctness $\varepsilon
_{\text{cor}}$, which bounds the residual probability that the local strings
are different after passing a hashing test. Thus, only $np_{\text{ec}}$ pulses
are successfully error-corrected and further processed into a key. This means
that secret-key rate $R_{\text{pe}}$ needs to be rescaled by the pre-factor
$np_{\text{ec}}/N$.

The next step is that of privacy amplification which is also not perfect. This
means that there will be some non-zero distance between the final strings
(composing the shared key) and an ideal classical-classical-quantum state
where the eavesdropper is completely de-correlated from the parties. Such a
distance is bounded by some target value of the $\varepsilon$-secrecy of the
protocol, which is further decomposed as $\varepsilon_{\text{sec}}%
=\varepsilon_{\text{s}}+\varepsilon_{\text{h}}$, where $\varepsilon_{\text{s}%
}$ is a smoothing parameter $\varepsilon_{\text{s}}$ and $\varepsilon
_{\text{h}}$ is a hashing parameter. Overall, all these imperfections are composed
into a single global parameter which is the $\varepsilon$-security of the
protocol, and given by $\varepsilon=2p_{\text{ec}}\varepsilon_{\text{pe}%
}+\varepsilon_{\text{cor}}+\varepsilon_{\text{sec}}$~\cite{FreeSpacePaper}.

For a Gaussian-modulated coherent-state protocol~\cite{GG02,Noswitch} with
success probability $p_{\text{ec}}$ and $\varepsilon$-security against
collective (Gaussian) attacks~\cite{CollectiveATT}, we may write the
composable finite-size key rate~\cite{FreeSpacePaper}%
\begin{equation}
R\geq\frac{np_{\text{ec}}}{N}\left(  R_{\text{pe}}-\frac{\Delta_{\text{aep}}%
}{\sqrt{n}}+\frac{\Theta}{n}\right)  , \label{sckeee}%
\end{equation}
where%
\begin{align}
&  \Delta_{\text{aep}}:=4\log_{2}\left(  2\sqrt{d}+1\right)  \sqrt{\log
_{2}\frac{18}{p_{\text{ec}}^{2}\varepsilon_{\text{s}}^{4}}},
\label{deltaAEPPP}\\
&  \Theta:=\log_{2}[p_{\text{ec}}(1-\varepsilon_{\text{s}}^{2}/3)]+2\log
_{2}\sqrt{2}\varepsilon_{\text{h}}, \label{bigOMEGA}%
\end{align}
and $d$ is the size of the alphabet after analog-to-digital conversion of the
continuous variables of the parties (typically $d=2^{5}$ for a $5$-bit digitalization).

For the specific case of the heterodyne protocol, one can extend the security
to general coherent attacks by combining Eq.~(\ref{sckeee}) with the approach
of Ref.~\cite{Lev2017}. This involves a suitable symmetrization of the data
and performing $m_{\text{et}}=f_{\text{et}}n$ local energy tests, where they
check if their local mean number of photons is $\lesssim\bar{n}_{T}%
+\mathcal{O}(m_{\mathrm{et}}^{-1/2})$. This test is certainly passed for large
enough $m_{\text{et}}$ and for typical communication conditions with small
values of excess noise (as is the case here, so that we certainly have
$\bar{n}_{R}=\tau\bar{n}_{T}+\bar{n}\lesssim\bar{n}_{T}$ at the receiver).

Thus, the parties achieve the rate~\cite{FreeSpacePaper}
\begin{equation}
R_{\text{gen}}^{\text{het}}\geq\frac{np_{\text{ec}}}{N}\left[  R_{\text{pe}%
}^{\text{het}}-\frac{\Delta_{\text{aep}}}{\sqrt{n}}+\frac{\Theta-2\left\lceil
\log_{2}\binom{K_{n}+4}{4}\right\rceil }{n}\right]  ,\label{RhetgenFIN}%
\end{equation}
where the number of key-generation pulses is now%
\begin{equation}
n=N-(m+m_{\text{et}})=\frac{N-m}{1+f_{\text{et}}},\label{keyP1}%
\end{equation}
and we have set%
\begin{align}
K_{n} &  =\max\left\{  1,2n\bar{n}_{T}\Sigma_{n}\right\}  ,\label{keyP2}\\
\Sigma_{n} &  :=\frac{1+2\sqrt{\frac{\ln(8/\varepsilon)}{2n}}+\frac
{\ln(8/\varepsilon)}{n}}{1-2\sqrt{\frac{\ln(8/\varepsilon)}{2f_{\text{et}}n}}%
}.\label{keyP3}%
\end{align}
The final epsilon security here is $\varepsilon^{\prime}=K_{n}^{4}%
\varepsilon/50$.

In a practical implementation, the parties exchange many pulses~\cite{LeoEXP},
which are split into $n_{\text{b}}\gg1$ blocks of suitable size $N$\ for data
processing; typically, $N=\mathcal{O}(10^{6})$ or more. Assuming that the
channel parameters are stable, one can write Eq.~(\ref{sckeee}) for the
generic block, i.e., for $N=\mathcal{O}(10^{6})$. The success probability
$p_{\text{ec}}$ provides the fraction of blocks that are successfully
processed into key generation. Under conditions of stability, parameter
estimation can be performed in one-go over the entire set of sacrificed pulses
$mn_{\text{b}}$, so that the worst-case estimators are expected to be quite
close to the actual values. However, if the channel is not stable, but slowing
varying with respect to the time-frame associated with a single block, then
parameter estimation has to be done block-by-block and the final rate will be
averaged over the blocks. In the following, we will assume that parameter
estimation is based over the single-block statistics, so that we encompass the
possibility of variability.

\subsection{Pilots, post-selection, lattice allocation, and
defading\label{SECqkd3}}

In a fading scenario where the value of the transmissivity $\tau$ fluctuates,
as is the case of a free-space quantum communication, it is useful to use
bright pilot pulses to track $\tau$ so as to create a lattice where signals
with almost-equal transmissivity are grouped together. Assume that Alice
randomly interleaves her signals with $m_{\text{PL}}$ pilots. These are
prepared in an energetic coherent-state $\left\vert \sqrt{\bar{n}_{\text{PL}}%
}e^{i\pi/4}\right\rangle $, with $\bar{n}_{\text{PL}}$ mean photons and fixed
$\pi/4$-phase, so as to provide components for both $q$- and $p$-measurements.
The first moments of the pilots follow Eq.~(\ref{IOrel}), where the magnitude
of the noise $z$ is negligible with respect to that of $\sqrt{\tau}x$. As
discussed in Ref.~\cite{FreeSpacePaper}, they allow to achieve an almost
perfect estimation of the instantaneous value of $\tau$.

The high energy regime is easily accessible using $10~$ns-long pulses from a
$100$~mW laser source. At $\lambda=800~$nm, each $1$~nJ pulse contains an
average of $x^{2}\simeq4\times10^{9}$ photons. Such energy can be used for the
TLO multiplexed with each signal or pilot. A fraction of this energy, say
$x^{2}\simeq10^{6}$ photons can be used for the pilots. Considering $\simeq
30$dB loss, i.e., $\tau\simeq10^{-3}$, which is the worst-case scenario in the
LEO region (reached in uplink with a $2000$-km-high satellite at $1$ radiant),
we have that $>10^{6}$ photons are collected for the LO, and about $10^{3}$
for the pilots. The LO is bright enough to allow for shot-noise limited
measurements, and the pilots are also bright enough to provide an almost
perfect estimation of $\tau$. It is clear that, in the case of an LLO, the LO
is even brighter since it is not even attenuated by the transmission.

On the basis of the pilots, Alice and Bob determine a post-selection interval
$\Delta=[\eta_{\text{th}},\eta]$ where $\eta$ is the maximum transmissivity
and $\eta_{\text{th}}=f_{\text{th}}\eta$ is computed for some fixed
$f_{\text{th}}<1$. As a result, only a fraction $p_{\text{th}}$ of the pulses
will be post-selected [cf. Eq.~(\ref{pthresholdEX})]. Within $\Delta$, they
introduce a regular lattice of $M$ bins/slots with step $\delta\tau=(\eta
-\eta_{\text{th}})/M$ and generic bin/slot $\Delta_{k}=[\tau_{k},\tau_{k+1}]$,
with $\tau_{k}:=\eta_{\text{th}}+(k-1)\delta\tau$ and $k=1,\ldots,M$. The
generic slot $\Delta_{k}$ is populated with probability $p_{k}$ computed
according to Eq.~(\ref{pkappaFAD}), i.e., it is associated with (an integer
approximation of) $S_{k}:=(N-m_{\text{PL}})p_{k}$ signals and corresponding
$\nu_{\text{det}}S_{k}$ pairs of data points $\{x_{i},y_{i}^{k}\}$. For a
sufficiently small step $\delta\tau$, these points satisfy the input-output
relation
\begin{equation}
y^{k}\simeq\sqrt{\tau_{k}}x+z^{k},
\end{equation}
for a Gaussian noise variable $z^{k}$ whose variance $\sigma_{z}^{2}(\tau
_{k})=2\bar{n}(\tau_{k})+\nu_{\text{det}}$ generally depends on $\tau_{k}$.

Instead of processing each slot independently from the others (with limited
statistics), the parties may adopt a de-fading procedure mapping all the slots
into the first one, with minimum transmissivity $\eta_{\text{th}}$. To the
points in slot $\Delta_{k}$, Bob applies the classical channel%
\begin{equation}
y^{k}\rightarrow\tilde{y}^{k}:=\sqrt{\frac{\eta_{\text{th}}}{\tau_{k}}}%
y^{k}+\sqrt{1-\frac{\eta_{\text{th}}}{\tau_{k}}}\xi_{\text{add}},
\end{equation}
where $\xi_{\text{add}}$ is a Gaussian variable with variance equal to
$\nu_{\text{det}}$. By repeating this mapping for all the slots, Bob creates a
new variable%
\begin{equation}
\tilde{y}=\sqrt{\eta_{\text{th}}}x+\tilde{z},
\end{equation}
where $\tilde{z}$ is non-Gaussian with variance%
\begin{equation}
\sigma_{\tilde{z}}^{2}=2\bar{n}_{\ast}+\nu_{\text{det}},~\bar{n}_{\ast}%
:=\frac{\eta_{\text{th}}}{p_{\text{th}}}\sum_{k}\frac{p_{k}}{\tau_{k}}\bar
{n}(\tau_{k}).
\end{equation}

Using the optimality of Gaussian attacks, Alice and Bob assume that $\tilde
{z}$ is Gaussian, thus overestimating Eve's strategy and under-estimating
their performance. In this way, the entire de-fading procedure reduces the
initial non-Gaussian fading channel into a worst-case thermal-loss Gaussian
channel $\mathcal{E}_{\eta_{\text{th}},\bar{n}_{\ast}}$ with fixed minimum
transmissivity $\eta_{\text{th}}$ and thermal number equal to $\bar{n}_{\ast}$.

\subsection{Parameter estimation and composable key rate\label{SECqkd4}}

Once they have reduced the problem to a worst-case thermal loss channel
$\mathcal{E}_{\eta_{\text{th}},\bar{n}_{\ast}}$, Alice and Bob build the
worst-case estimators, $\eta_{\text{th}}^{\prime}$ and $\bar{n}_{\ast}%
^{\prime}$, for the parameters $\eta_{\text{th}}$ and $\bar{n}_{\ast}$
associated with the channel, by using $m_{p}p_{\text{th}}$ pairs of data
points (coming from $mp_{\text{th}}$ sacrificed signals). Each of these
worst-case estimators is correct up to an error $\varepsilon_{\text{pe}}$,
related to the confidence parameter $w$ according to Eq.~(\ref{wexpression1})
or~(\ref{wexpression2}). See also Ref.~\cite[Sec.~IV.F]{FreeSpacePaper}.

Note that also the threshold transmissivity $\eta_{\text{th}}$ needs to be
estimated, even though this is agreed by the parties via the pilots. In this
way, in fact, Alice and Bob can detect and account for Eve's potential
strategies that are aimed at discriminating signals and pilots, and then
applying different interactions. As a matter of fact, any potential error
associated with the slot allocation of the signals (due to Eve's action or
coarse-graining imperfections) will be detected and accounted by directly
estimating $\eta_{\text{th}}$ over the signals~\cite[Sec.~IV.E]%
{FreeSpacePaper}.

Assume the realistic case where signals and pilots undergo the same
interaction with the environment, so that the estimation of $\eta_{\text{th}}$
via the signals confirms the threshold value agreed by the parties using the
pilots. (As said above, in the potential presence of signal-pilot
discrepancies, our general formalism still applies by generating the
corresponding worst-estimators from the signals.) Then, the worst-case
estimators satisfy the bounds~\cite{FreeSpacePaper}%
\begin{align}
\eta_{\text{th}}^{\prime} &  \gtrsim\eta_{\text{LB}}:=\eta_{\text{th}}%
-2w\sqrt{\frac{2\eta_{\text{th}}^{2}+\eta_{\text{th}}\sigma_{\text{wc}}%
^{2}/\sigma_{x}^{2}}{m_{p}p_{\text{th}}}},\label{LBpara}\\
\bar{n}_{\ast}^{\prime} &  \lesssim\bar{n}_{\text{UB}}:=\bar{n}_{\text{wc}%
}+w\frac{\sigma_{\text{wc}}^{2}}{\sqrt{2m_{p}p_{\text{th}}}}%
,\label{nGestimator}%
\end{align}
where $\sigma_{\text{wc}}^{2}=2\bar{n}_{\text{wc}}+\nu_{\text{det}}$ is the
worst-case value for the total thermal noise. This value upper-bounds
$\sigma_{\tilde{z}}^{2}$ and is computed by taking $\bar{n}_{\text{wc}}%
\geq\bar{n}(\tau)$ for any $\tau\in\Delta$. In order to compute $\bar
{n}_{\text{wc}}$, i.e., maximize the thermal noise over the transmissivity, we
need to introduce a practical expression for the setup noise (see the next subsection).

Using $\eta_{\text{LB}}$ and $\bar{n}_{\text{UB}}$, we can lower-bound Alice
and Bob's key rate affected by parameter estimation%
\begin{equation}
R_{\text{pe}}=R_{\text{asy}}(\eta_{\text{th}}^{\prime},\bar{n}_{\ast}^{\prime
})\geq R_{\text{LB}}:=R_{\text{asy}}(\eta_{\text{LB}},\bar{n}_{\text{UB}}).
\end{equation}
Then, the composable key rate is a direct modification of Eq.~(\ref{sckeee})
and takes the form~\cite{FreeSpacePaper}%
\begin{equation}
R\geq\frac{np_{\text{th}}p_{\text{ec}}}{N}\left(  R_{\text{LB}}-\frac
{\Delta_{\text{aep}}}{\sqrt{np_{\text{th}}}}+\frac{\Theta}{np_{\text{th}}%
}\right)  , \label{ratePOSTS}%
\end{equation}
with $n=N-(m+m_{\text{PL}})$ and security $\varepsilon=2p_{\text{ec}%
}\varepsilon_{\text{pe}}+\varepsilon_{\text{cor}}+\varepsilon_{\text{sec}}$
against collective (Gaussian) attacks.

For the heterodyne version of the protocol, we can again extend the security
from collective to general attacks by performing a similar modification as in
Eq.~(\ref{RhetgenFIN}). In this case, the secret-key rate
becomes~\cite{FreeSpacePaper}%
\begin{align}
R_{\text{gen}}^{\text{het}}  &  \geq\frac{np_{\text{th}}p_{\text{ec}}}%
{N}\left[  R_{\text{LB}}^{\text{het}}-\frac{\Delta_{\text{aep}}}%
{\sqrt{np_{\text{th}}}}\right. \nonumber\\
&  \left.  +\frac{\Theta-2\left\lceil \log_{2}\binom{K_{np_{\text{th}}}+4}%
{4}\right\rceil }{np_{\text{th}}}\right]  , \label{ratePOSTSgen}%
\end{align}
where $K_{n}$ is given in Eq.~(\ref{keyP2}) and the total number of
key-generation pulses is now given by
\begin{equation}
n=N-(m+m_{\text{PL}}+m_{\text{et}})=\frac{N-(m+m_{\text{PL}})}%
{1+f_{\mathrm{et}}}.
\end{equation}
Final epsilon security is equal to $\varepsilon^{\prime}=K_{np_{\text{th}}%
}^{4}\varepsilon/50$.

\subsection{Setup noise and observations about the LO\label{SECqkd5}}

The setup noise $\bar{n}_{\text{ex}}$ strictly depends on what type of LO is
used. In fact, in the case of a TLO, there is no phase error, but the
electronic noise of the detector becomes non-trivial due to the attenuation
that the LO undergoes during its transmission. By contrast, in the case of an
LLO, the electronic noise is lower due to the fact that the LO\ is always
bright, independently from the transmissivity; however, there will be some
non-trivial phase noise that comes from the imperfect digital reconstruction
of the rotating reference frame at the receiver.

The electronic noise can be described by an additive Gaussian channel with
associated variance $\nu_{\text{el}}$ or equivalent number of photons $\bar
{n}_{\text{el}}=\nu_{\text{el}}/2$. Its value depends on various quantities,
namely the noise equivalent power (NEP) of the amplifiers/photodiodes in the
homodyne detectors, the detection bandwidth $W$, the duration of the
LO\ pulses $\Delta t_{\text{LO}}$, the LO\ power at the detector
$P_{\text{LO}}^{\text{det}}$, and the frequency of the light $\nu$. Let us
consider the power $P_{\text{LO}}$ at which the LO\ is generated, so that we
have $P_{\text{LO}}^{\text{det}}=\tau P_{\text{LO}}$ for the TLO and
$P_{\text{LO}}^{\text{det}}=P_{\text{LO}}$ for the LLO. Then, let us introduce
the parameter (see also Ref.~\cite{FreeSpacePaper})
\begin{equation}
\Theta_{\text{el}}:=\frac{\nu_{\text{det}}\mathrm{NEP}^{2}W\Delta
t_{\text{LO}}}{2h\nu P_{\text{LO}}}. \label{eleCONTRIBUT}%
\end{equation}
In our notation, we may write
\begin{equation}
\bar{n}_{\text{el}}^{\text{TLO}}=\Theta_{\text{el}}/\tau,~~\bar{n}_{\text{el}%
}^{\text{LLO}}=\Theta_{\text{el}}.
\end{equation}

For a detector with bandwidth $W=100~$MHz, we may consider $\mathrm{NEP}%
=6~$pW/$\sqrt{\text{Hz}}$. Then, let us take an LO\ power $P_{\text{LO}}%
=100~$mW. At $800~$nm and using LO pulses of duration $\Delta t_{\text{LO}%
}=10~$ns, we get $\Theta_{\text{el}}\simeq1.45\times10^{-3}$ for the case of
an heterodyne setup ($\nu_{\text{det}}=2$). It is clear the advantage of the
LLO in reducing the electronic noise.

Such a reduction of the electronic noise induced by the LLO comes at the price
of introducing phase errors. In our notation, we quantify this contribution as
follows
\begin{equation}
\bar{n}_{\text{phase}}^{\text{LLO}}=\frac{\pi\sigma_{x}^{2}l_{\text{W}}}%
{C}\tau,
\end{equation}
where $C$ is the clock of the system and $l_{\text{W}}$ is the laser
linewidth~\cite{FreeSpacePaper}. In our investigations we have $\sigma_{x}%
^{2}\lesssim10$ and $C=10~$MHz, so a low value $\bar{n}_{\text{phase}%
}^{\text{LLO}}$ can be reached with a linewidth $l_{\text{W}}\simeq1.6$~KHz.

\begin{figure}[t]
\vspace{0.2cm}
\par
\begin{center}
\includegraphics[width=0.40\textwidth] {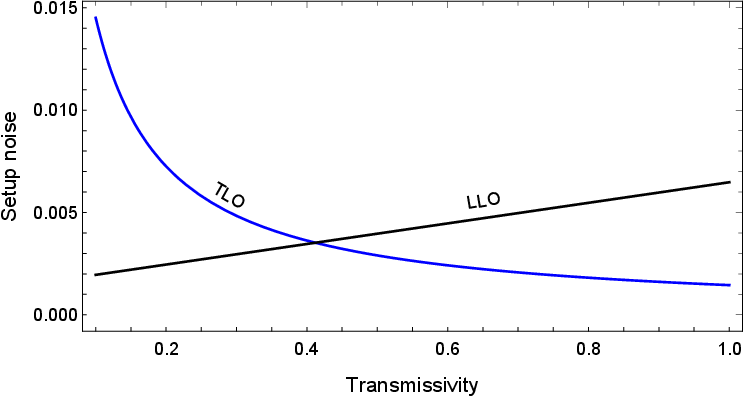}
\end{center}
\par
\vspace{-0.3cm}\caption{Comparison of the setup noise (in terms of equivalent
number of thermal photons $\bar{n}_{\text{el}}$) versus channel transmissivity
$\tau$, for the LLO (black line) and the TLO (blue line). These are computed
using Eq.~(\ref{TLOLLOex}). Parameters are chosen as in the text, i.e.,
$\lambda=800~$nm, $l_{\text{W}}=1.6$~KHz, $W=100~$MHz, $\mathrm{NEP}%
=6~$pW/$\sqrt{\text{Hz}}$, $P_{\text{LO}}=100~$mW, $\Delta t_{\text{LO}}%
=10~$ns,$\ C=10~$MHz, $\sigma_{x}^{2}=10$, and $\nu_{\text{det}}=2$
(heterodyne detection).}%
\label{TLOLLOpic}%
\end{figure}

Overall, we have that the setup noise for the TLO\ and LLO takes the following
expressions
\begin{equation}
\bar{n}_{\text{ex}}^{\text{TLO}}(\tau)=\frac{\Theta_{\text{el}}}{\tau}%
,~~\bar{n}_{\text{ex}}^{\text{LLO}}(\tau)=\Theta_{\text{el}}+\frac{\pi
\sigma_{x}^{2}l_{\text{W}}}{C}\tau,\label{TLOLLOex}%
\end{equation}
where we see the different monotonic behavior of these quantities versus the
transmissivity $\tau$. In particular, the TLO\ seems to be preferable for
short distances (high values of $\tau$), while the LLO is better for long
distances (low value of $\tau$). See also Fig.~\ref{TLOLLOpic}%
.

In order to maximize $\bar{n}_{\text{ex}}$ over the post-selection interval
$\Delta$, we therefore compute $\bar{n}_{\text{ex}}^{\text{TLO}}(\tau)$ at the
minimum border value $\tau=\eta_{\text{th}}$, and $\bar{n}_{\text{ex}%
}^{\text{LLO}}(\tau)$ at the maximum border value $\tau=\eta$. Thus, the
worst-case value for the setup noise will be given by%
\begin{equation}
\bar{n}_{\text{wc}}=\eta_{\text{eff}}\bar{n}_{B}+\bar{n}_{\text{ex,wc}},
\label{UB1e}%
\end{equation}
where we have set%
\begin{equation}
\bar{n}_{\text{ex,wc}}^{\text{TLO}}=\Theta_{\text{el}}/\eta_{\text{th}}%
,~~\bar{n}_{\text{ex,wc}}^{\text{LLO}}=\Theta_{\text{el}}+\frac{\pi\sigma
_{x}^{2}l_{\text{W}}}{C}\eta. \label{exwc2}%
\end{equation}

As already mentioned, an LLO is more convenient for long-distance quantum
communications (low values of the transmissivity).\ In fact, in such a case,
the associated phase error decreases to zero, and the total setup noise tends
to the constant term $\Theta_{\text{el}}$. For this reason, in the following
we specifically investigate the performances achievable with an LLO, but we
stress that both approaches of TLO\ and LLO are encompassed by our theory. We
also stress that an LLO requires a more sophisticated hardware than that
needed by a TLO.

As already mentioned in Sec.~\ref{NoiseFILTERsection}, an important aspect in
the use of an LLO is that each pulse (signal or pilot) must be transmitted in
the middle of two reference pulses, so that there is a regular interleaving
between these bright references and all the other pulses. From the detection
of the references, Bob reconstructs Alice's rotating frame and therefore
suitably rotates the outcomes he has obtained from the measurements of the
signals and the pilots (for which a locally-generated LO was employed for the
homodyne/heterodyne detection). It is clear that, in this process, the
free-space link is half of the time used by the phase-synchronizing
references. This means that, in terms of actual throughput (bits/second),
there is a factor of $1/2$ to account for the LLO. However, it is also true
that such a time-multiplexing of the LO enables Alice to use both
polarizations for the transmission of the signals, which therefore leads to a
compensation of the $1/2$\ extra factor. We assume this scenario.

\subsection{Key rate analysis with orbital dynamics\label{orbitalSEC}}

With all the ingredients in our hands, we now study the numerical behavior of
the composable key rate, accounting not only for the fading process but also
for the variability associated with the orbital dynamics of the satellite. In
fact, there are two basic reasons why the channel is not stable in satellite
communications: One is the inevitable fading process affecting the free-space
transmission, which occurs even when the satellite is assumed to be `frozen'
at some fixed altitude and zenith angle; the other is the temporal variation
of the zenith angle (and altitude) due to the specific orbit. Even if we
assume the satellite to be at some fixed (or approximately fixed) altitude,
the variation of the zenith angle is inevitable for all orbits in the LEO and
MEO regions.

Assume that the satellite is on an orbit with a constant altitude $h$, which
is an assumption valid for polar orbits and approximately valid for
sun-synchronous orbits. Then, during the quantum communication of a block of
size $N$, the satellite performs a corresponding slice of its orbit, spanning
a range of different zenith angles. The angular size of this slice depends on
various factors, including the clock of the system $C$ and the speed of the
satellite (which depends on its altitude $h$). As a result, we have that the
value of the maximum transmissivity $\eta$ (and that of the threshold
transmissivity $\eta_{\text{th}}$) sensibly changes during the quantum
communication. It is clear that this problem also affects the thermal noise
$\bar{n}$, due to the general dependence of the setup noise $\bar
{n}_{\text{ex}}$ on the instantaneous transmissivity.

There are various strategies to deal with this situation. One strategy is to
minimize $\eta_{\text{LB}}$ and maximize $\bar{n}_{\text{UB}}$ over the slice,
and then use these values to lower bound the achievable key rate (for an LLO,
one could also use a constant maximum value of $\bar{n}_{\text{UB}}$, as
computed at the K\'{a}rm\'{a}n line or at the lower LEO boundary). Another
method, which is more practical, is to directly minimize the key rate over the
orbital slice. Under typical conditions, the fixed-geometry rate $R(h,\theta)$
decreases for increasing $\theta$, i.e., from the zenith towards the horizon.
For this reason, we can lowerbound the actual rate by taking the value of
$R(h,\theta)$ at the largest zenith angle $\theta$ (for any fixed altitude
$h$). In particular, assume that the slice associated with the block has
zenith angles $\theta\lesssim1$. We can therefore lower bound the key rate by
taking the value $R(h,\theta=1)$. We call this the `$1$-radiant' key
rate~\cite{NoteSlice}.

Let us analyze the $1$-radiant key rates that are achievable by a pilot-guided
post-selected heterodyne protocol with LLO at various satellite altitudes for
the various configurations. In particular, we choose the protocol parameters
specified in Table~\ref{tablePARAMETERS}. We assume a $1\%$ quota for the
pilots~\cite{PilotNOTA}. The values of the input modulation $\mu$ (i.e.,
$\bar{n}_{T}$) and the threshold transmissivity $f_{\text{th}}$ (i.e.,
$\eta_{\text{th}}$) are implicitly optimized at each altitude. This is the
case for rates under collective attacks (plotted for each configuration). For
the study of the general attacks (only done in the best configuration of
night-time downlink), we have made the sub-optimal choices of $\mu=7$ and
$f_{\text{th}}=0.75$. Also note that the epsilon security $\varepsilon
^{\prime}$ versus general attacks depends on the altitude; its maximum value
shown in Table~\ref{tablePARAMETERS} is achieved for the minimum altitude of
$h=100~$km.

\begin{table}[t]
\vspace{0.2cm}
\begin{tabular}
[c]{|l|l|l|l|}\hline
$%
\begin{array}
[c]{l}%
\text{Protocol}\\
\text{parameter}%
\end{array}
$ & Symbol & $%
\begin{array}
[c]{l}%
\text{Collective}\\
\text{attacks}%
\end{array}
$ & $%
\begin{array}
[c]{l}%
\text{General}\\
\text{attacks}%
\end{array}
$\\\hline\hline
Total pulses & $N$ & $10^{8}$ & $10^{8}$\\\hline
Pilot pulses & $m_{\text{PL}}$ & $0.01\times N$ & $0.01\times N$\\\hline
PE signals & $m$ & $0.1\times N$ & $0.1\times N$\\\hline
Energy tests & $f_{\text{et}}$ & $-$ & $0.2$\\\hline
KG signals & $n$ & $0.89\times N$ & $\simeq7.4\times10^{7}$\\\hline
Digitalization & $d$ & $2^{5}$ & $2^{5}$\\\hline
Rec. efficiency & $\beta$ & $0.96$ & $0.96$\\\hline
EC success prob & $p_{\text{ec}}$ & $0.9$ & $0.1$\\\hline
Epsilons & $\varepsilon_{\text{h,s,\ldots}}$ & $2^{-33}\simeq10^{-10}$ &
$10^{-43}$\\\hline
Confidence & $w$ & $\simeq6.34$ & $\simeq14.07$\\\hline
Security & $\varepsilon,\varepsilon^{\prime}$ & $\simeq5.6\times10^{-10}$ &
$\lesssim2.6\times10^{-10}$\\\hline
Modulation & $\mu$ & optimized & $7$\\\hline
Threshold & $f_{\text{th}}~$ & optimized & $0.75$\\\hline
\end{tabular}
\caption{Protocol parameters adopted with respect to collective attacks and
general attacks.}%
\label{tablePARAMETERS}%
\end{table}

\begin{table}[t]
\vspace{0.2cm}
\begin{tabular}
[c]{|l|l|l|}\hline
Physical parameter & Symbol & Value\\\hline\hline
Beam curvature & $R_{0}$ & $\infty$\\\hline
Wavelength & $\lambda$ & $800~$nm\\\hline
Beam spot size & $w_{0}$ & $%
\begin{array}
[c]{c}%
20~\text{cm (setup 1)}\\
40~\text{cm (setup 2)}\\
60~\text{cm (setup 3)}%
\end{array}
$\\\hline
Receiver aperture & $a_{R}$ & $%
\begin{array}
[c]{l}%
40~\text{cm (setup 1)}\\
1~\text{m (setup 2)}\\
2~\text{m (setup 3)}%
\end{array}
$\\\hline
Receiver field of view & $\Omega_{\text{fov}}$ & $10^{-10}~$sr\\\hline
Homodyne filter & $\Delta\lambda$ & $0.1~\text{pm}$\\\hline
Detector shot-noise & $\nu_{\text{det}}$ & $2$~~(heterodyne)\\\hline
Detector efficiency & $\eta_{\text{eff}}$ & $0.4$\\\hline
Detector bandwidth & $W$ & $100~$MHz\\\hline
Noise equivalent power & NEP & $6~$pW/$\sqrt{\text{Hz}}$\\\hline
Linewidth & $l_{\text{W}}$ & $1.6$~KHz\\\hline
LO power & $P_{\text{LO}}$ & $100~$mW\\\hline
Clock & $C$ & $10~$MHz\\\hline
Pulse duration & $\Delta t,\Delta t_{\text{LO}}$ & $10~$ns\\\hline
Extinction (at $1$ rad) & $\eta_{\text{atm}}$ & $\simeq0.94$\\\hline
Pointing error & $\sigma_{\text{P}}^{2}$ & $\simeq(10^{-6}z)^{2}~~$($1~\mu
$rad)\\\hline
Structure constant & $C_{n}^{2}$ & night/day H-V model\\\hline
Turbulence parameters & $w_{\text{st}}$,$~\sigma_{\text{TB}}^{2}$ &
Appendix~\ref{TurbSECTION}\\\hline
Background noise & $\bar{n}_{B}$ & Eqs.~(\ref{UPthermalNOISE}),
(\ref{DOWNthermalNOISE})\\\hline
\end{tabular}
\caption{Physical parameters and theoretical models.}%
\label{tableMODELS}%
\end{table}

Then, we follow all the physical parameters and theoretical models considered
so far, explicitly listed in Table~\ref{tableMODELS}. However, we allow for
the possibilities of different setups on the basis of the spot-size $w_{0}$ and
receiver aperture $a_{R}$. With setup$~1$, we reproduce the physical
conditions adopted for the study of the ultimate bounds presented in
Fig.~\ref{Trittico} (under the assumption of a narrow filter $\Delta
\lambda=0.1~$pm). The other setups offer a more generous hardware that allows
us to improve the key rates in the various configurations.

\begin{figure}[t]
\vspace{0.2cm}
\par
\begin{center}
\includegraphics[width=0.48\textwidth] {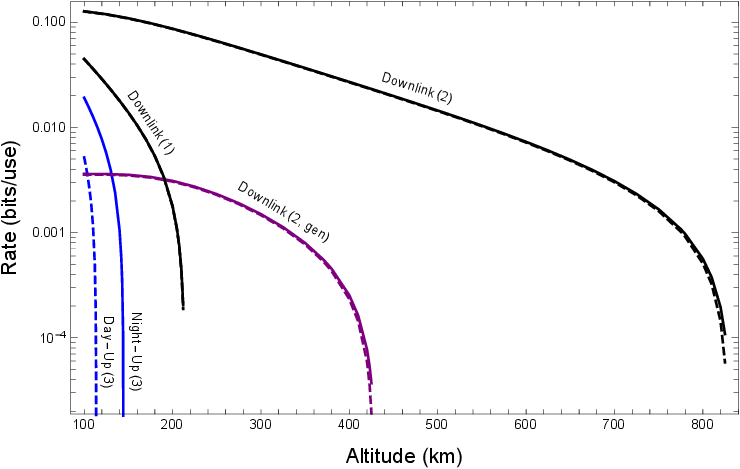}
\end{center}
\par
\vspace{-0.3cm}\caption{$1$-radiant composable key rates (bits/use) achievable
by a pilot-guided post-selected heterodyne protocol with LLO at various
satellite altitudes $h$ (km). We consider downlink with setups$~1$ and~$2$
from Table~\ref{tableMODELS}. Black lines refer to the key rate of
Eq.~(\ref{ratePOSTS}) while the purple line is the rate of
Eq.~(\ref{ratePOSTSgen}) against general attacks. Solid lines refer to night
time, while the overlapping/almost-overlapping dashed lines refer to day time
(difference becomes appreciable only for longer distances). We also show the
performance in uplink (blue lines), considering Eq.~(\ref{ratePOSTS}) and
setup$~3$ from Table~\ref{tableMODELS}. In particular, we compare the
night-time uplink (solid blue) with day-time uplink (dashed blue).}%
\label{RatesConfigsPic}%
\end{figure}

In Fig.~\ref{RatesConfigsPic}, we show the performances that are achievable in
downlink with setup$~1$, which clearly improve when the better setup$~2$ is
considered, with positive key rates in the LEO region. By contrast, we need to
assume the expensive setup$~3$ for enabling uplink, whose key rates appear to
be restricted to the sub-LEO region (between $100~$km and $160~$km). Note
that, thanks to the narrow homodyne filter, we may achieve positive key rates
for day-time operation. In particular, in downlink, the suppression of the
background is such that the day-time rate coincides with the night-time rate.
For uplink, there is still a gap though, which is due to the effects of the
ground-level turbulence on the beam. These effects reduce the transmissivity
and are higher during the day.

According to our investigation, all the configurations allow for secure
quantum communications with a satellite in the LEO/sub-LEO region, even though
with different hardware requirements. The homodyne filter largely suppresses
the background noise, paving the way for day-light implementations under
various weather conditions.

\subsection{Orbital slicing\label{SECqkd7}}

The $1$-radiant key rates are pessimistic estimates of what we can actually
achieve with a satellite orbiting at an approximately constant altitude. For
this reason, we now consider a more accurate treatment where we explicitly
account for the fact that different blocks of data correspond to different
slices of the orbit within the $1$-radiant sector. For each slice we consider
the corresponding minimum rate, which is achieved at the largest zenith angle
along that particular slice. The overall orbital rate is given by an average
over the slices.

Before presenting the improved results, we need to make some preliminary
considerations about an ideal modus operandi for the ground-satellite link.

In fact, we identify the following ideal conditions:

\begin{enumerate}
\item[(i)] The transit time of the satellite should allow the parties to
distribute many quantum data points;

\item[(ii)] There should be additional time for classical communication and
data-processing, in such a way that a secret key is generated before the end of the fly-by;

\item[(iii)] There should be time for an encrypted communication, exploiting
part of the key already distributed.
\end{enumerate}

To satisfy these ideal conditions, it is better to have a satellite that is
able to reach small zenith angles. Clearly, an optimal solution is a satellite
crossing the zenith point above the ground station (zenith-crossing orbit).
For simplicity, assume that its orbit is circular, with constant radius
$R_{\text{S}}=R_{\text{E}}+h$ from the centre of the Earth. Examples of
circular orbits are polar and near-polar sun-synchronous orbits.

\begin{figure}[t]
\vspace{-0.8cm}
\par
\begin{center}
\includegraphics[width=0.48\textwidth] {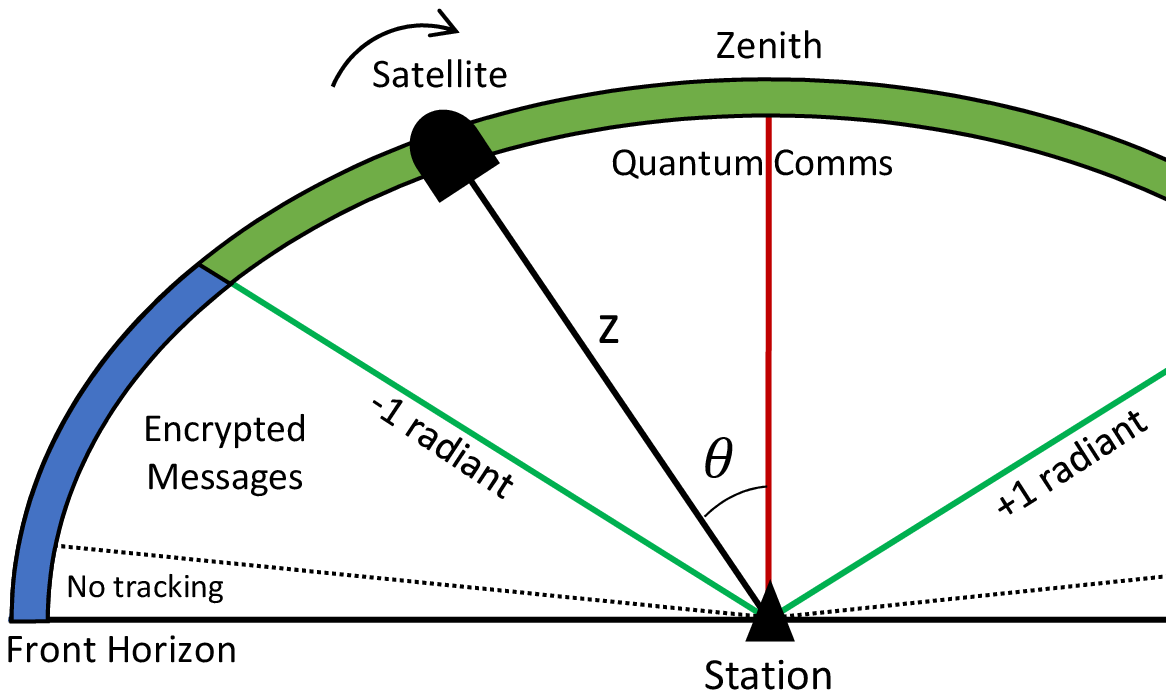}
\end{center}
\par
\vspace{-1.5cm}\caption{Orbital sectors for a zenith-crossing circular orbit.
Zenith angle $\theta$ has an associated sign and we identify the front and the
back horizons. Quantum communication occurs within the (green) good sector
associated with the angular window $-1\leq\theta\leq1$ (where the angle is
negative on the left, i.e., for a rising satellite). There is a transit time
$t_{\text{Q}}$ associated with this sector. Total transit time $t_{\text{T}}%
$\ is associated with the entire flyby from the front to the back horizon. The
right blue sector $1<\theta<\pi/2$ can be used for data processing and key
generation. The left blue sector $-\pi/2<\theta<-1$ can be used for encrypted
communication using a previously generated key (e.g., the satellite may
download the key exchanged with another station). In practical scenarios, the
satellite is not tracked within $5^{\circ}-10^{\circ}$ of the horizon
(depending on the urban setting etc.). In other words, there is an effective
\textquotedblleft mask\textquotedblright\ angle $\theta_{\text{m}}$ (or
minimum acceptable elevation above the horizon) for the satellite. In the text
we assume $\theta_{\text{m}}=10^{\circ}$.}%
\label{orbSECT2}%
\end{figure}

For a zenith-crossing orbit, it is useful to introduce a sign for the zenith
angle $\theta$, so that a negative $\theta$ corresponds to the zenith angle
with respect to a satellite which is arising from the \textquotedblleft
front\textquotedblright\ horizon and moving towards the zenith, while a
positive $\theta$ corresponds to a satellite that has passed the zenith point
and it is descending towards the \textquotedblleft back\textquotedblright%
\ horizon (see Fig.~\ref{orbSECT2}).

For a zenith-crossing circular orbit, the slant distance $z=z(R_{\text{S}%
},\alpha)$ can be expressed as $z(R_{\text{S}},\alpha)=\sqrt{R_{\text{E}}%
^{2}+R_{\text{S}}^{2}-2R_{\text{E}}R_{\text{S}}\cos\alpha}$ in terms of
$R_{\text{S}}=R_{\text{E}}+h$ and the orbital angle $\alpha$ (which may also
be negative). See also Eq.~(\ref{zOrbital}) and Fig.~\ref{geomPIC} in
Appendix~\ref{SecGEOMETRY}. Then, we may write the orbital period (in
seconds)
\begin{equation}
T_{\mathrm{S}}=2\pi\sqrt{\frac{R_{\text{S}}^{3}}{\mu_{\text{G}}}},
\end{equation}
where $\mu_{\text{G}}=GM_{\text{E}}$ is the standard gravitational parameter,
with $G=6.674\times10^{-11}$~N~m$^{2}~$kg$^{-2}$\ being the gravitational
constant and $M_{\text{E}}\simeq5.972\times10^{24}$~kg the approximate Earth's
mass. As a result, the orbital angle $\alpha$ varies over time $t$ according
to the law
\begin{equation}
\alpha(t,h)=\frac{2\pi t}{T_{\mathrm{S}}}=t\sqrt{\frac{\mu_{\text{G}}}{R_{\text{S}}%
^{3}}},\label{alphatheta}%
\end{equation}
where we have implicitly set $\alpha=0$ (satellite at the zenith) for
$t=0$.\ Correspondingly, the time-varying zenith angle $\theta=\theta(t,h)$
can be computed from%
\begin{equation}
\sin\theta=\frac{R_{\text{S}}\sin\alpha}{z(R_{\text{S}},\alpha)}%
.\label{sintheta}%
\end{equation}

\begin{figure}[t]
\vspace{0.2cm}
\par
\begin{center}
\includegraphics[width=0.44\textwidth] {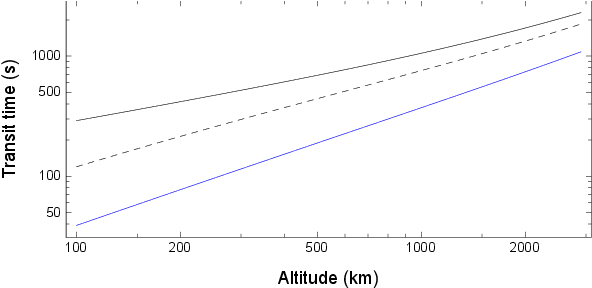}
\end{center}
\par
\vspace{-0.3cm}\caption{Transit times (seconds) versus altitudes (kms) for a
satellite passing through the zenith in a circular orbit. We plot the total
transit time $t_{\mathrm{T}}$ from horizon to horizon (solid black), the
effective transit time $t_{\mathrm{E}}$ accounting for the mask angle (dashed
black) and the quantum transit time $t_{\mathrm{Q}}$ within the angular window
$-1\leq\theta\leq1$ (solid blue).}%
\label{TransitPIC}%
\end{figure}

We can divide the orbit in different sectors as depicted in
Fig.~\ref{orbSECT2}. Assuming that the quantum communication occurs within $1$
radiant from the zenith (good sector), we compute a corresponding
\textit{quantum transit time} $t_{\text{Q}}$. Within $|\theta|\leq\pi/2$, we
may certainly invert $\theta=\theta(t,h)$ into $t=t(\theta,h)$. In fact, we
may write%
\begin{equation}
t(\theta,h)=\sqrt{\frac{(R_{\text{E}}+h)^{3}}{GM_{\text{E}}}}\arccos\left[
\frac{R_{\text{E}}+z(h,\theta)\cos\theta}{R_{\text{E}}+h}\right]  ,
\end{equation}
where $z(h,\theta)$\ is given in Eq.~(\ref{slantANALYTICAL}) and the initial
condition is $t(0,h)=0$. We then compute the quantum transit time
$t_{\text{Q}}(h)=2t(1,h)$, the total transit time $t_{\text{T}}(h):=2t(\pi
/2,h)$ of the satellite from horizon to horizon, and the effective transit
time $t_{\text{E}}(h):=2t(\pi/2-\theta_{\text{m}},h)$, where $\theta
_{\text{m}}$ is the mask angle (here assumed to be of $10^{\circ}$). Their
behaviors are plotted in Fig.~\ref{TransitPIC}.

As we can see, at $530$~km, the total transit time is about $716$ seconds, of
which $200$ seconds are within $1$ radiant. Assuming $t_{\text{Q}}\simeq200$
and a clock of $C=10~$MHz, we have the quantum communication of $Ct_{\text{Q}%
}\simeq2\times10^{9}$ pulses. Assuming blocks of size $N=10^{8}$ (each
corresponding to $2N$ pairs of data points for the heterodyne protocol), we
have $20$ blocks distributed within the $1$-radiant window of each passage.
Since we assume that the protocol runs with an LLO, we also assume that both
polarizations are used for the signals (so that we compensate for the time
multiplexing associated with the transmission of the reference pulses). Thus,
condition~(i) above can be met.

In order to realize condition~(ii), we exploit the orbital sector after the
$1$-radiant window (see Fig.~\ref{orbSECT2}) where a $530$~km satellite spends
$(t_{\text{T}}-t_{\text{Q}})/2\simeq258$~seconds. In particular, accounting
for the mask angle $\theta_{\text{m}}=10^{\circ}$, we have that the satellite
is visible for $(t_{\text{E}}-t_{\text{Q}})/2\simeq131$~seconds. During this
time, the parties can implement the classical procedures of error correction
and privacy amplification, e.g., using the high-speed methods of
Refs.~\cite{Xwang1,Xwang2}. Ideally, by using optimal LDPC codes over a $1$GHz
GPU, the processing of $\simeq2\times10^{9}$ data points may take $\simeq120$
seconds~\cite{LeoEXP,LeoComms}. Such performances for data processing require
an highly-performant computing hardware. High speed data processing is
achievable because of the relatively-high signal-to-noise ratio (so that the
number of iterations for syndrome extraction in the error correcting procedure
is low). As a matter of fact, for downlink from about $500$~km, the total loss
is less than $10$~dB. Note that there is also a latency time in communications
with the satellite, of the order of $1.6$~ms at that altitude. However,
because the procedures can be implemented with limited sessions of one-way CC,
this is negligible.

Finally, there is the condition~(iii) which is about having enough time for an
encrypted communication between the satellite and the ground station. This
session can be used for authentication and/or for downloading the key of
another ground station\ via one-time pad. This step can be implemented during
the first sector of the orbit, i.e., at zenith angles $-\pi/2<\theta<-1$. In
the basic scenario of Fig.~\ref{orbSECT2}, this phase is symmetric to that
discussed above and the satellite is visible for about $131$~seconds, clearly
sufficient for the encrypted communication.

Let us now slice the good sector for quantum communication ($-1\leq\theta
\leq1$) for a zenith-crossing circular orbit. Assume that $n_{\text{bks}}$
blocks are transmitted during the flyby. Dividing the quantum transit time
$t_{\text{Q}}$ by $n_{\text{bks}}$, we get the time $\delta t$ that is needed
for each block to be transmitted. These time intervals identify corresponding
angular slices $\{\theta_{i},\theta_{i+1}\}$ along the orbit, that can be
computed using Eqs.~(\ref{alphatheta}) and~(\ref{sintheta}) for any fixed
altitude $h$. For slice $i=1,\ldots,n_{\text{bks}}$ we consider the minimum
value%
\begin{equation}
R_{i}=\min_{\theta\in\lbrack\theta_{i},\theta_{i+1}]}R(\theta
),\label{RiSLICED}%
\end{equation}
where $R(\theta)$ is the $\theta$-dependent key rate for the considered
altitude. Thus, the average orbital rate is equal to%
\begin{equation}
R_{\text{orb}}=\frac{1}{n_{\text{bks}}}\sum_{i=1}^{n_{\text{bks}}}%
\max\{0,R_{i}\}\geq R_{1\text{-rad}},\label{orbitalRATE}%
\end{equation}
where the bound $R_{1\text{-rad}}:=\max\{0,R(\theta=\pm1)\}$ is the rate
associated with the border values (considered in the previous subsection and
Fig.~\ref{RatesConfigsPic}).

\begin{figure}[t]
\vspace{0.2cm}
\par
\begin{center}
\includegraphics[width=0.44\textwidth] {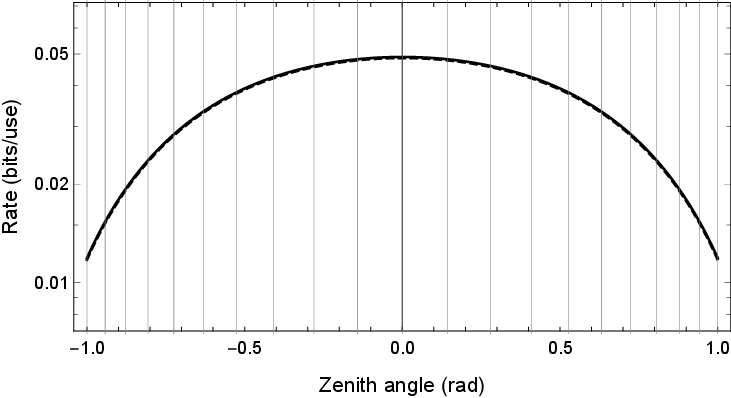}
\end{center}
\par
\vspace{-0.3cm}\caption{Composable finite-size key rate $R$ of
Eq.~(\ref{ratePOSTS}) for a pilot-guided post-selected heterodyne protocol
with LLO. This is plotted versus the zenith angle $\theta$ for downlink from a
satellite which is in a zenith-crossing circular orbit at $h=530~$km. We plot
the performance for night-time (solid) and day-time (dashed, overlapping).
Parameters are specified in Table~\ref{tablePARAMETERS} for collective
attacks, and Table~\ref{tableMODELS} for setup~$2$. In particular, we choose
$\mu=7.18$ and $f_{\text{th}}=0.76$. Besides the rates we explicitly show the
angular lattice in Eq.~(\ref{SlicesAA}).}%
\label{OrbitalDOWN}%
\end{figure}

As previously said, for a zenith-crossing circular orbit at $h=530~$km, the
quantum transit time is about $200$~seconds. Therefore we may consider
$n_{\text{bks}}=20$ blocks, where each block of size $N=10^{8}$ corresponds to
a transmission time of $\delta t=10$~seconds for a $C=10~$MHz clock. This
configuration identifies angular slices%
\begin{equation}
\{\theta_{i},\theta_{i+1}\}\simeq\{-1,-0.942\},\ldots
,\{0.942,1\},\label{SlicesAA}%
\end{equation}
that are shown in Fig.~\ref{OrbitalDOWN} together with the rate $R(\theta)$ of
Eq.~(\ref{ratePOSTS}) specified for night-time and day-time downlink.

For the angle-dependent rate $R(\theta)$, we assume the protocol parameters in
Table~\ref{tablePARAMETERS} for collective attacks, and the physical
parameters in Table~\ref{tableMODELS}, by choosing spot size and receiver
aperture according to setup~$2$. The values for the modulation and the
threshold are chosen in such a way to maximize the lowest rate in the
ensemble $\{R_{i}\}_{i=1}^{n_{\text{bks}}}$ which coincides with
$R_{1\text{-rad}}$. Thus, in the figure, we have chosen $\mu=7.18$ and
$f_{\text{th}}=0.76$.

Using Eqs.~(\ref{ratePOSTS}) and~(\ref{SlicesAA}) in Eq.~(\ref{RiSLICED}) and
then Eq.~(\ref{orbitalRATE}), we compute the average orbital rate for
downlink, which is approximately the same for night and day, i.e.,%
\begin{equation}
R_{\text{orb}}^{\text{down}}\simeq\left\{
\begin{array}
[c]{l}%
3.066\times10^{-2}~\text{bits/use (night-time)}\\
3.041\times10^{-2}~\text{bits/use~(day-time).}%
\end{array}
\right.  \label{LBprimo}%
\end{equation}
Here \textquotedblleft per use\textquotedblright\ means per use of the quantum
communication channel, occurring within $1$ radiant. When we plug a clock
$C=10~$MHz, we have a rate of $R_{\text{orb}}^{\text{down}}\simeq307$ kbits/s
during night time, and $R_{\text{orb}}^{\text{down}}\simeq304$ kbits/s during
day time. Accounting for the time of the quantum communication ($200~$s), each
night-time zenith-crossing passage distributes $\simeq6.13\times10^{7}$ secret
bits, while a day-time zenith-crossing passage distributes $\simeq
6.08\times10^{7}$ secret bits. Considering that, within $24$~hours, there will
also be \textit{non}-zenith-crossing passages (exploitable for QKD), the above
estimates lowerbound the number of bits that can be distributed per day via
night- and day-time operation. (Alternatively, in a less-performant hardware,
the other passages can be used for data processing by the parties).

Now consider uplink to a satellite in the sub-LEO region. We take $h=103$~km,
just after the K\'{a}rm\'{a}n line. As we have already mentioned, the main
reason for the inferior performance in uplink is the non-trivial effect of the
atmospheric turbulence, with bigger impact during the day. We consider the
rate $R(\theta)$ of Eq.~(\ref{ratePOSTS}) for the case of a pilot-guided
post-selected heterodyne protocol with LLO, specified for night-time and
day-time uplink. We assume the protocol parameters given in
Table~\ref{tablePARAMETERS} for collective attacks, and the physical
parameters in Table~\ref{tableMODELS}, where we assume the more demanding
setup~$3$. The values for the modulation $\mu$ and the threshold
$f_{\text{th}}$ are chosen to maximize the lowest rate ($R_{1\text{-rad}}$).
In particular, we choose $\mu=6.5$ and $f_{\text{th}}=0.74$.

\begin{figure}[t]
\vspace{0.2cm}
\par
\begin{center}
\includegraphics[width=0.44\textwidth] {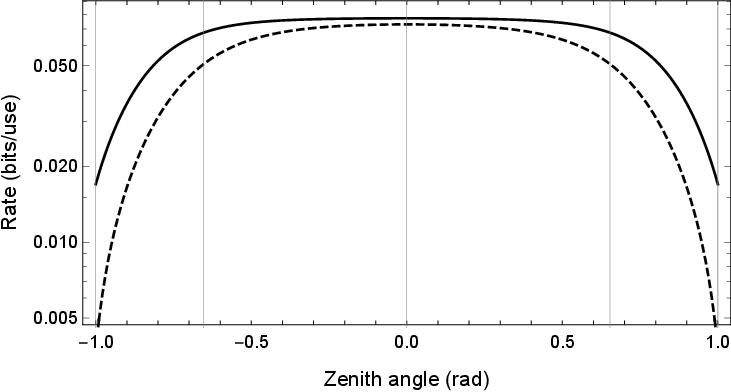}
\end{center}
\par
\vspace{-0.3cm}\caption{Composable finite-size key rate $R$ of
Eq.~(\ref{ratePOSTS}) for a pilot-guided post-selected heterodyne protocol
with LLO. This is plotted versus the zenith angle $\theta$ for uplink to a
satellite which is in a zenith-crossing circular orbit at $h=103~$km. We plot
the performance for night-time (solid) and day-time (dashed). Parameters are
specified in Table~\ref{tablePARAMETERS} for collective attacks, and
Table~\ref{tableMODELS} for setup~$3$. In particular, we choose $\mu=6.5$ and
$f_{\text{th}}=0.74$. Besides the rates we show the angular lattice generated
by the slices of Eq.~(\ref{slicesBB}).}%
\label{UPPP}%
\end{figure}

Because of the lower altitude, we have a total transit time of just
$t_{\text{T}}\simeq295$~s, an effective transit time of $t_{\text{E}}%
\simeq123$~s, and a quantum transit time of $t_{\text{Q}}\simeq40$~s. The
latter allows the parties to distribute 4 blocks of size $10^{8}$ with a clock
of $C=10~$MHz (and these blocks may be data-processed in $\simeq24$~s, which
is enough for the last sector of the orbit by assuming highly-performant
error-correcting codes). The orbital slices associated with the $10$~s-long
blocks are
\begin{equation}
\{\theta_{i},\theta_{i+1}\}\simeq
\{-1,-0.65\},\{-0.65,0\},\{0,0.65\},\{0.65,1\}.\label{slicesBB}%
\end{equation}
Both the rate $R(\theta)$ and the slices are shown in Fig.~\ref{UPPP}.

Using Eqs.~(\ref{ratePOSTS}) and~(\ref{slicesBB}) in Eq.~(\ref{RiSLICED}) and
then Eq.~(\ref{orbitalRATE}), we compute the average orbital rate in uplink,
which is equal to%
\begin{equation}
R_{\text{orb}}^{\text{up}}\simeq\left\{
\begin{array}
[c]{l}%
4.244\times10^{-2}~\text{bits/use (night-time)}\\
2.737\times10^{-2}~\text{bits/use~(day-time).}%
\end{array}
\right.  \label{LBsecondo}%
\end{equation}
At $10~$MHz this rate corresponds to about $424$~kbit/s at night and
$273$~kbit/s during the day. Accounting for a quantum transit time of $40$~s,
we have that each zenith-crossing passage distributes $\simeq1.69\times10^{7}%
$~secret bits for night uplink, and $\simeq1.09\times10^{7}$~secret bits for
day uplink.

According to our analysis, it is indeed feasible to use CV-QKD protocols to
distribute keys with a satellite orbiting in the LEO/sub-LEO region,
accounting not only for the fading process due to turbulence and pointing
errors, but also for the fast orbital dynamics (which creates additional
problems for the transmission of reasonably-large block sizes). Our analysis
shows this possibility not only for the best-considered scenario of night-time
downlink, but also for day-time downlink and for the more challenging
scenarios of night- and day-time uplink.

\subsection{Satellites versus ground-based repeaters\label{satcompsec}}

Once we have computed the secret key rates that are achievable in the most
relevant configurations for satellite and ground station, we now show that
satellite quantum communications are able to provide a non-trivial advantage
with respect to the use of quantum repeaters on the ground, when the aim is to
connect two remote end-users that are sufficiently far apart on Earth's surface.

An example of circular orbit is a polar orbit, i.e., with orbital inclination
$\iota=90^{\circ}$. However, a more practical scenario is considering a
near-polar satellite in sun-synchronous orbit (which is approximately
circular). This type of orbit guarantees that the satellite passes over any
point on Earth's surface at the same local mean solar time. This clearly
implies the possibility of stable conditions for night-time or day-time
operation, so that the quantum communication with the satellite occurs at
roughly the same time of the night or the day.

For a sun-synchronous satellite at altitude $h$, the orbital inclination is
given by
\begin{equation}
\iota=\frac{360}{2\pi}\arccos\left[  -\left(  \frac{R_{\text{E}}+h}%
{12352}\right)  ^{7/2}\right],
\end{equation}
where $h\leq5980$ is expressed in km. At $h=530~$km, this corresponds to
$\iota\simeq97.5^{\circ}$. Because the orbital period is $T_{\mathrm{S}}\simeq95$
minutes, the satellite performs $15$ orbits a day, before returning above the
initial point. Note that the Micius satellite is sun-synchronous with
$\iota\simeq97.4^{\circ}$ and with altitude between $488$ and $584~$km. At
$h=103~$km, we have $\iota\simeq96^{\circ}$, $T_{\mathrm{S}}\simeq86$ minutes and $16$
orbits a day.

Assume that two ground stations are along the orbital path, so that the
satellite crosses both their zenith positions, which happens once per day. We
assume the worst-case scenario in which the stations interact with the
satellite only during the sections of the orbit where the zenith positions are
crossed (of course this assumption can be relaxed and the ground stations
could also use other passages that are not zenith-crossing). Also assume that
the stations are in similar operational conditions, so that we can
simultaneously adopt the results for night time or day time for both of them.
Finally, assume that the satellite may have the option to communicate with two
stations simultaneously (e.g., using two quantum transmitters or receivers);
this is assumed to address the particular case where the stations are close,
so that the satellite appears within their $1$-radiant angular windows roughly
at the same time (clearly this is not the case for very distant ground stations).

Start with the satellite at the zenith of the first station ($t=0$) and assume
that it reaches the zenith of the second station after time $\Delta t$. For
$\Delta t\leq T_{\mathrm{S}}/2$, the distance between the two stations is equal to
\begin{equation}
d_{\text{st}}=\alpha(\Delta t,h)R_{\text{E}}=\frac{2\pi\Delta tR_{\text{E}}%
}{T_{\mathrm{S}}}\in\lbrack0,\pi R_{\text{E}}],
\end{equation}
where we have used Eq.~(\ref{alphatheta}) and accounted for Earth's curvature.
Then, assume that the two stations are also connected by an optical fiber with
standard loss-rate of $\alpha_{\text{fib}}=0.2$dB/km, so that we have a total
fiber transmissivity of $\eta_{\text{fib}}=10^{-\alpha_{\text{fib}%
}d_{\text{st}}/10}$. The maximum fiber-based repeater-less rate (bits per use)
is given by the PLOB bound $R_{\text{fib}}=-\log_{2}(1-\eta_{\text{fib}}%
)$~\cite{QKDpaper}. Multiplying by the clock $C=10~$MHz and the number of
seconds in one day $\#_{\text{day}}\simeq8.6\times10^{4}$, we may compute the
maximum number of secret bits that can the distributed in one day
$CR_{\text{fib}}\#_{\text{day}}$ as a function of the station-to-station
ground distance. We also assume the situation where a number $N_{\text{rep}%
}\geq1$ of ideal repeaters are inserted along the fiber line, so that we have
the fiber-based rate becomes $R_{\text{fib}}^{\text{rep}}=-\log_{2}%
(1-\sqrt[N_{\text{rep}}+1]{\eta_{\text{fib}}})$~\cite{netpaper}. We have a
corresponding number of secret bits $CR_{\text{fib}}^{\text{rep}%
}\#_{\text{day}}$ per day.\begin{figure}[t]
\vspace{0.2cm}
\par
\begin{center}
\includegraphics[width=0.44\textwidth] {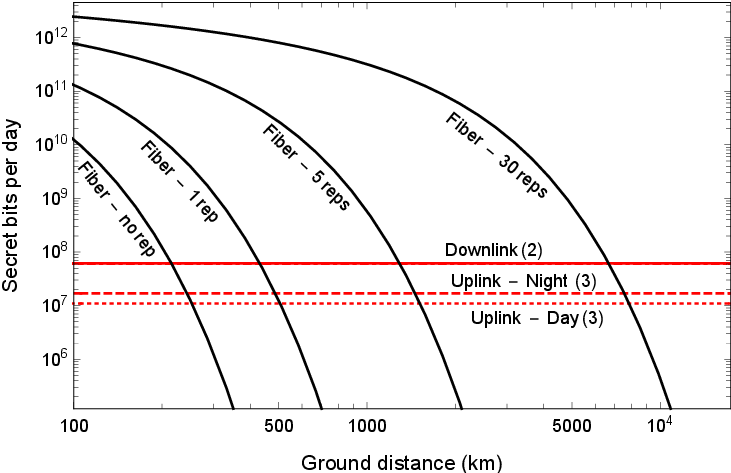}
\end{center}
\par
\vspace{-0.3cm}\caption{Secret-key bits per day versus ground distance (km)
between two stations, assuming a clock of $10$~MHz. We consider the maximum
performances achievable by a repeaterless fiber-connection (PLOB bound) and
repeater-based fiber-connections assisted by $1$, $5$ and $30$ ideal quantum
repeaters (solid lines). These are compared with the constant performances
achievable by connecting the two ground stations via a near-polar
sun-synchronous satellite. We consider $h=530$~km in downlink
[Eq.~(\ref{LBprimo}) based on setup~2 in Table~\ref{tableMODELS}] and
$h=103~$km in uplink [Eq.~(\ref{LBsecondo}) based on setup~3 in
Table~\ref{tableMODELS}]. In particular, from the top to the bottom, we show
downlink at night-time (solid red) and day-time (dot-dashed red, overlapping
with the solid line). Then we show uplink at night-time (dashed red) and
day-time (dotted red).}%
\label{SatNetPic}%
\end{figure}

In Fig.~\ref{SatNetPic}, we consider the maximum number of secret bits per day
(versus ground distance) that can be distributed by a repeaterless fiber link
between the stations and also by fiber-links assisted by ideal quantum
repeaters. Assuming the same clock, we compare these ground-based performances
with the secret-bits per day that can be distributed by using a satellite
moving between the two stations, the latter operating in the same way with
respect to the satellite, i.e., via night/day-time downlink
[Eq.~(\ref{LBprimo})] or night/day-time uplink [Eq.~(\ref{LBsecondo})].

We can see how a satellite can beat the fiber-based repeaterless bound when
the stations are separated by more than $215$~km, and how it can achieve the
same rate of $30$ ideal quantum repeaters when the station-to-station ground
separation is about $6675~$km. This performance is achievable in downlink no
matter if during the day or the night (solid red line in Fig.~\ref{SatNetPic}%
). As expected, in uplink, the performances are worse than downlink (despite
the better setup). That being said, via uplink to the satellite, the remote
users are still able to beat\ ground chains of quantum repeaters after similar distances.

\section{Conclusions\label{SECconclusions}}

In this work we have established the ultimate limits and the practical rates
that can be achieved in secure quantum communications with satellites,
assuming various configurations (downlink/uplink) and operational settings
(night- or day-time). While our study is based on ideas and tools developed in
Ref.~\cite{FreeSpacePaper} for free-space quantum communications, it also
required a number of non-trivial generalizations in order to account for the
slant propagation at variable altitudes and beyond the atmosphere. As a matter
of fact, the underlying physical models for atmospheric turbulence and
background noise have crucial differences with respect to the models adopted
for free-space communications on the ground.

We have started our work by establishing information-theoretic upper limits to
the maximum number of secret bits (and ebits) that can be achieved per use of
the satellite link in all scenarios. Our theoretical analysis considers
all relevant effects, as due to diffraction, atmospheric extinction, limited
efficiency, pointing error, turbulence, thermal background, and setup noise.

In uplink, turbulence is very important because it affects the beam close to
the transmitter where the spot size is small. In this case, both pointing
error and turbulence effects must be accounted for. By contrast, in downlink,
the spot size is already very large when it enters the atmosphere, compared to
the typical size of the turbulent eddies. For this reason, turbulence is
negligible and the only relevant effect is the pointing error.

A further asymmetry is introduced by the background noise that is induced by
sky brightness and planetary albedos (Moon and Earth), even though this
background can be greatly suppressed by using narrow frequency filters. Such
filters are indeed created by the interferometric process occurring in
homodyne-like setups where the signals are mode-matched with a strong local oscillator.

For all configurations, we have studied the numerical behavior of the ultimate
key rates when the satellite is at the zenith position or at $1$ radiant from
the zenith, showing that a large range of altitudes are possible for secure
key generation (and entanglement distribution) when we adopt optimal
protocols. For the same configurations, we have then studied secret key rates
that are achievable in practice by accounting for finite-size effects and
composable security. Our analysis therefore addresses the problem of block
size, which is particularly relevant for satellite quantum communications (see
also Ref.~\cite{Lim} for the setting of discrete-variable QKD).

In our paper, the use of a pilot-guided post-selected heterodyne protocol,
combined with a careful consideration for the orbital dynamics, enables the
implementation of CV-QKD between station and satellite in all configurations
of night-time downlink/uplink and day-time downlink/uplink. It is interesting
that all these scenarios represent indeed a viable option for secure quantum
communications, whereas only the setting of night-time downlink has been
considered in other works~\cite{LeverrierSAT}.

As a further analysis beyond this work, it would be interesting to compare the QKD rates that are achievable by CV protocols with those of discrete-variable protocols in the various configurations of communication with the satellite. Since discrete-variable QKD is more robust for long-distance implementations on the ground, we would expect this approach to be particularly suitable for the far-LEO and MEO regions. By contrast, CV-QKD is generally better for high-rate implementations at shorter ground distances, so it appears an approach specifically suitable for the LEO and sub-LEO regions.

Finally, we have shown that a sun-synchronous satellite, exchanging keys with
ground stations, is able to distribute more secret bits per day than a direct
fiber-connection between the stations, even if the latter communicate at the
ultimate PLOB bound. Remarkably, the satellite is also able to outperform a
chain of many ideal quantum repeaters operating between the remote stations.
These results are obtained when the distance between the remote stations
surpasses certain thresholds, which depend on the hardware available for the
satellite together with the adopted configuration (downlink/uplink) and
operational setting (night/day time).

\bigskip

\textit{Acknowledgements}.--~The author acknowledges funding from the European
Union's Horizon 2020 research and innovation programme under grant agreement
No 820466 (Quantum-Flagship Project CiViQ: \textquotedblleft Continuous
Variable Quantum Communications\textquotedblright).

\appendix

\section{Basic geometry for satellite communications\label{SecGEOMETRY}}

Here we discuss some geometrical elements about communications with
satellites. The slant distance $z$ between a sea-level ground station and a
satellite can be connected with other two important parameters. The first one
is the (positive) zenith angle $\theta$ which is defined as the angle between
the vertical direction (zenith) and the pointing direction from the ground
station to the satellite. The second one is the altitude $h$ of the satellite,
which is defined as the distance from the satellite to the ground (sea-level)
orthogonally to the surface of the Earth. Slant distance $z$, altitude $h$,
and positive zenith angle $\theta\in\lbrack0,\pi/2]$ can be easily connected
by means of simple trigonometric observations. (Note that the following
geometric formulas do not change if we include a sign in the zenith angle, so
that $\theta\in\lbrack-\pi/2,\pi/2]$, as considered for the study with the
zenith-crossing orbit in the main text).

\begin{figure}[t]
\vspace{-0.0cm}
\par
\begin{center}
\includegraphics[width=0.4\textwidth] {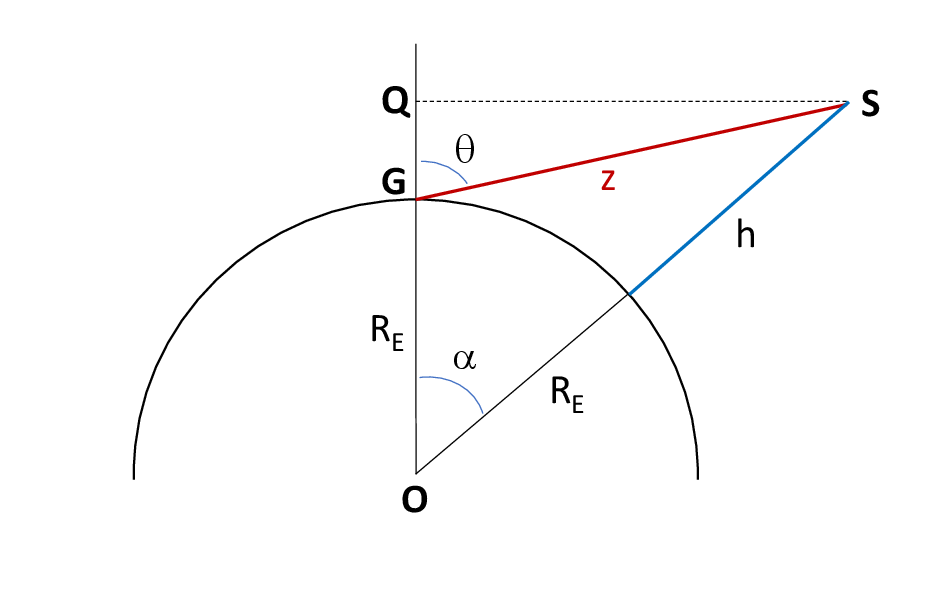}
\end{center}
\par
\vspace{-0.8cm}\caption{Basic geometry for satellite communications. A
satellite (S) is at slant distance $z$ from a sea-level ground station (G).
The satellite is at zenith angle $\theta$ and altitude $h$ over the Earth's
surface. Clearly, $z=h$ only at the zenith ($\theta=0$). The satellite is also
at the orbital angle $\alpha$\ and orbital radius $R_{\text{S}}=h+R_{\text{E}%
}$, where $R_{\text{E}}$ is the radius of the Earth. Note that $\overline
{\text{OQ}}=R_{\text{E}}+\overline{\text{QG}}=\overline{\text{OS}}\cos\alpha$
leads to $(1)~R_{\text{E}}+z\cos\theta=R_{\text{S}}\cos\alpha$. Now, from
$\overline{\text{QS}}$ have: $(2)~z\sin\theta=R_{\text{S}}\sin\alpha$. By
taking the squared in $(1)$ and $(2)$ and using $\cos^{2}\alpha=1-\sin
^{2}\alpha$, we derive $\ h^{2}+2R_{\text{E}}h-z(z+2R_{\text{E}}\cos\theta)=0$
whose positive solution gives Eq.~(\ref{hztheta}). In a similar way, one can
use $\cos^{2}\theta=1-\sin^{2}\theta$ and derive Eq.~(\ref{zOrbital}).}%
\label{geomPIC}%
\end{figure}

It is easy to express the altitude $h$ as a function of $z$ and $\theta$. A
simple trigonometric calculus provides
\begin{equation}
h(z,\theta)=\sqrt{R_{\text{E}}^{2}+z^{2}+2zR_{\text{E}}\cos\theta}%
-R_{\text{E}}, \label{hztheta}%
\end{equation}
where $R_{\text{E}}\simeq6371$~km is the approximate radius of the Earth (see
Fig.~\ref{geomPIC}). In particular, at small angles $\theta\simeq0$, we can
write the simplified \textquotedblleft flat-Earth\textquotedblright%
\ approximation
\begin{equation}
h(z,\theta)\simeq z\cos\theta_{z}+\mathcal{O}(\theta^{3}),~\theta_{z}%
:=\theta\sqrt{\frac{R_{\text{E}}}{R_{\text{E}}+z}}. \label{hAPPROX}%
\end{equation}
For $\theta\leq1$ and relatively small $h$ (of the order of the atmospheric
thickness, i.e., $20~$km), one finds that $h\simeq z\cos\theta$ and $z\simeq
h\sec\theta$ represent excellent approximations.

It is easy to see that Eq.~(\ref{hztheta}) can be inverted into
\begin{equation}
\cos\theta=\frac{h}{z}+\frac{h^{2}-z^{2}}{2zR_{\text{E}}},
\end{equation}
which gives the zenith angle $\theta$\ in terms of $z$ and $h$ (see also
Ref.~\cite[Ch.~13]{ThomsonBOOK}). Similarly, the slant range $z$ can be
expressed as a simple function of $h$ and $\theta$, i.e., we may write the
slant distance functional (see also Ref.~\cite{Vasy19})%
\begin{equation}
z(h,\theta)=\sqrt{h^{2}+2hR_{\text{E}}+R_{\text{E}}^{2}\cos^{2}\theta
}-R_{\text{E}}\cos\theta. \label{slantHT}%
\end{equation}
Note that $z(h,\theta)\leq h\sec\theta$ for any $h$ and $\theta\in\lbrack
0,\pi/2]$.

It is easy to verify that the previous formulas can immediately be generalized
to the scenario where the ground station is located at some non-zero altitude
$h_{0}$. Setting $R_{\text{G}}:=R_{\text{E}}+h_{0}$ and $R_{\text{S}%
}=R_{\text{E}}+h$, we may in fact write%
\begin{align}
h(z,\theta)  &  =\sqrt{R_{\text{G}}^{2}+z^{2}+2zR_{\text{G}}\cos\theta
}-R_{\text{E}},\label{hzT2}\\
z(h,\theta)  &  =\sqrt{R_{\text{S}}^{2}+R_{\text{G}}^{2}(\cos^{2}\theta
-1)}-R_{\text{G}}\cos\theta. \label{zT2}%
\end{align}

Another parametrization is in terms of orbital radius $R_{\text{S}}$\ and the
orbital angle $\alpha$, i.e., the angle between the position of the ground
station and the position of the satellite as seen from the centre of the
Earth. It is immediate to see that%
\begin{equation}
z(R_{\text{S}},\alpha)=\sqrt{R_{\text{E}}^{2}+R_{\text{S}}^{2}-2R_{\text{E}%
}R_{\text{S}}\cos\alpha}. \label{zOrbital}%
\end{equation}
This parametrization is useful for circular orbits, where $R_{\text{S}}$ is
constant. In this case, we can write $\alpha=2\pi t/T_{\mathrm{S}}$ where $t$ is time
and $T_{\mathrm{S}}$ is the orbital period, i.e., the time needed for a complete
revolution around the Earth.

\section{Refraction effects\label{SECrefraction}}

In a more refined description, we need to consider atmospheric refraction.
Assuming the atmosphere to be modeled as a set of thin uniform slabs provides
the same result of an atmosphere modeled as a single uniform slab with surface
refractive index $n_{0}$~\cite[Ch.~13, Fig.~13.4]{ThomsonBOOK}. Therefore,
atmospheric refraction creates an apparent zenith angle $\theta_{\text{app}}$
satisfying Snell's law%
\begin{equation}
\sin\theta_{\text{app}}=n_{0}^{-1}\sin\theta,
\end{equation}
where $n_{0}\simeq1.00027$ is the surface value of the refractive index. We
see that the angle of refraction $\Delta\theta:=\theta-\theta_{\text{app}}%
$\ exceeds one degree when the satellite is at the horizon, where $\theta
=\pi/2$ corresponds to $\theta_{\text{app}}^{\text{max}}\simeq1.548$
($\simeq88.7^{\circ}$). Besides the apparent angle, refraction also increases
the optical path by an elongation factor $\varepsilon_{\text{elo}}%
=\varepsilon_{\text{elo}}(\theta_{\text{app}})$.

Replacing $\theta=\theta(\theta_{\text{app}}):=\arcsin(n_{0}\sin
\theta_{\text{app}})$ in $z(h,\theta)$ and multiplying by $\varepsilon
_{\text{elo}}$, one gets the refracted slant range
\begin{equation}
z_{\text{ref}}(h,\theta_{\text{app}})=\varepsilon_{\text{elo}}(\theta
_{\text{app}})z[h,\theta(\theta_{\text{app}})], \label{refractedslant}%
\end{equation}
as a function of the altitude $h$ and the apparent angle $\theta_{\text{app}}%
$. Replacing $\theta=\theta(\theta_{\text{app}})$ and $z=z_{\text{ref}%
}/\varepsilon_{\text{elo}}(\theta_{\text{app}})$\ in $h(z,\theta)$, we get the
altitude in terms of the refracted parameters, i.e.,
\begin{equation}
h=h_{\text{ref}}(z_{\text{ref}},\theta_{\text{app}}):=h[z_{\text{ref}%
}/\varepsilon_{\text{elo}}(\theta_{\text{app}}),\theta(\theta_{\text{app}})].
\end{equation}

With these modifications in hand, we can formulate the refracted version of
the atmospheric extinction in Eq.~(\ref{atmEQ2}) of the main text. We need to
integrate $\alpha(h)=\alpha_{0}\exp(-h/\tilde{h})$ (with $\alpha_{0}$ and
$\tilde{h}$ given in the main text)\ using the modified expression for the
slant distance $z_{\text{ref}}(h,\theta_{\text{app}})$ and the expression of
the altitude $h_{\text{ref}}(z_{\text{ref}},\theta_{\text{app}})$. We get%
\begin{align}
\eta_{\text{atm}}^{\text{ref}}(h,\theta_{\text{app}}) &  =\exp\left\{
-\int_{0}^{z_{\text{ref}}(h,\theta_{\text{app}})}dy~\alpha\lbrack
h_{\text{ref}}(y,\theta_{\text{app}})]\right\}  \nonumber\\
&  =e^{-\alpha_{0}g_{\text{ref}}(h,\theta_{\text{app}})},\label{atmLOSSref}%
\end{align}
where we have defined
\begin{equation}
g_{\text{ref}}(h,\theta_{\text{app}}):=\int_{0}^{z_{\text{ref}}(h,\theta
_{\text{app}})}dy\exp\left[  -\frac{h_{\text{ref}}(y,\theta_{\text{app}}%
)}{\tilde{h}}\right]  .
\end{equation}
Correspondingly, we can modify the bound in Eq.~(\ref{PLOBder1}) of the main
text to account for refraction.\ We obtain%
\begin{align}
\mathcal{B}_{\text{ref}}(h,\theta_{\text{app}}) &  =-\log_{2}\left[
1-\eta_{\text{eff}}e^{-\alpha_{0}g_{\text{ref}}(h,\theta_{\text{app}}%
)}\right.  \nonumber\\
&  \times\left.  \left(  1-e^{-\frac{2a_{R}^{2}}{w_{\text{d}}[z_{\text{ref}%
}(h,\theta_{\text{app}})]^{2}}}\right)  \right]  .\label{UBrefrac}%
\end{align}
For low transmissivity, which is certainly the case in the far field regime
$z\gg z_{R}$, we can approximate%
\begin{equation}
\mathcal{B}_{\text{ref}}(h,\theta_{\text{app}})\simeq\frac{2}{\ln2}\frac
{a_{R}^{2}\eta_{\text{eff}}e^{-\alpha_{0}g_{\text{ref}}(h,\theta_{\text{app}%
})}}{w_{\text{d}}[z_{\text{ref}}(h,\theta_{\text{app}})]^{2}}.
\end{equation}

\begin{figure}[t]
\vspace{0.2cm}
\par
\begin{center}
\includegraphics[width=0.4\textwidth] {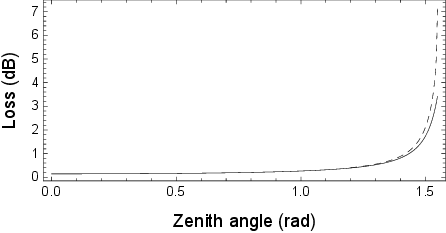}
\end{center}
\caption{Atmospheric loss (decibels) versus zenith angle (radiants) in the
link between a sea-level ground station and a satellite at $h=780$~km, for
$\lambda=800~$nm. We plot the refraction-free model for atmospheric loss
$-10\log_{10}\eta_{\text{atm}}$ [given by Eq.~(\ref{atmEQ2}) of the main text]
with respect to the zenith angle $\theta$ (solid line), and the
refraction-based model $-10\log_{10}\eta_{\text{atm}}^{\text{ref}}$ of
Eq.~(\ref{atmLOSSref}) with respect to the apparent zenith angle
$\theta_{\text{app}}$ (dashed line).}%
\label{REFPICs}%
\end{figure}

In Fig.~\ref{REFPICs}, we investigate the effects of refraction on the channel
loss. For typical parameters, we see that refraction is negligible within
$\simeq1$ radiant from the zenith, while it becomes more and more relevant in
the proximity of the horizon. For a sea-level ground station communicating
with a satellite at $h=780~$km and apparent zenith angle $\theta_{\text{app}%
}^{\text{max}}$, we compute $\eta_{\text{atm}}^{\text{ref}}\simeq7.1~$dB from
Eq.~(\ref{atmLOSSref}) instead of $\eta_{\text{atm}}\simeq3.4~$dB from
Eq.~(\ref{atmEQ2}) of the main text (setting $\theta=\theta_{\text{app}%
}^{\text{max}}$). This discrepancy leads to differences between $\mathcal{B}$
and its refraction-based version $\mathcal{B}_{\text{ref}}$ for large angles,
i.e., close to the horizon.

Finally, it is worth mentioning that the formula in Eq.~(\ref{UBrefrac}) can
also be applied to the case where the ground station is at some non-negligible
altitude $h_{0}$. In fact, it is sufficient to use Eqs.~(\ref{hzT2})
and~(\ref{zT2}) in all the previous expressions that lead to
Eq.~(\ref{UBrefrac}).

\section{Atmospheric turbulence\label{TurbSECTION}}

\subsection{Weak turbulence}

Atmospheric turbulence leads different treatments depending on its strength.
From a physical point of view, turbulence effects are due to eddies affecting
the travelling beam. In a regime of weak turbulence, one can clearly
distinguish the action of small and large turbulent eddies. Those smaller than
the beam waist tend to broaden the beam (on a fast time scale), while those
larger than the beam waist tend to deflect the beam, randomly but on a slower
time scale~\cite{Fante75}. As a result, the broadening of the beam can be
decomposed into a sum of two contributions, the short-term spot size
$w_{\text{st}}^{2}$ plus the random wandering of the beam centroid with
variance $\sigma_{\text{TB}}^{2}$, so that there is a long-term spot
size~\cite[Eq.~(32)]{Fante75}
\begin{equation}
w_{\text{lt}}^{2}=w_{\text{st}}^{2}+\sigma_{\text{TB}}^{2}. \label{turbmodel}%
\end{equation}
The slower time scale is of the order of $10-100~$ms~\cite{Burgoin}, which
means that this dynamics can be resolved by a fast-enough detector.

For long-distance communication, if turbulence becomes stronger, the motion of
the centroid becomes negligible ($\sigma_{\text{TB}}^{2}\simeq0$). At some
point, strong beam deformation comes into place as a major effect, and the
beam will eventually break up in multiple patches~\cite{Fante75,Kon}.

The first step is therefore the correct characterization of the relevant
regime of turbulence, which requires the introduction of parameters from the
theory of optical propagation through turbulent media. The most important of
these parameters is the refraction index structure constant $C_{n}^{2}%
$~\cite{AndrewsBook,Hemani}, since this is at the basis of the others and, in
particular, the scintillation index~\cite{AndrewsBook}, that characterizes the
strength of turbulence, and the spherical-wave coherence length~\cite{Fante75}%
, that directly enters in the expressions of the spot sizes of
Eq.~(\ref{turbmodel}).

\subsection{Refraction index structure constant\label{HVmodelAPP}}

The structure constant $C_{n}^{2}$ measures the strength of the fluctuations
in the refraction index, due to spatial variations of temperature and
pressure. We need to consider an adequate model for the structure constant
$C_{n}^{2}(h)$, so that this quantity can be suitably averaged over different
altitudes for up- and down-link communication.

Assuming the Hufnagel-Valley (H-V) model of atmospheric
turbulence~\cite{Stanley,Valley}, the structure constant reads
\begin{align}
C_{n}^{2}(h)  &  =5.94\times10^{-53}\left(  \frac{v}{27}\right)  ^{2}%
h^{10}e^{-h/1000}\nonumber\\
&  +2.7\times10^{-16}e^{-h/1500}+Ae^{-h/100}, \label{HFmodel}%
\end{align}
where $h>0$ is expressed in meters, $v$ is the windspeed (m/s) and $A$ is the
nominal value of $C_{n}^{2}(0)$ at the ground in units m$^{-2/3}$ (MKS units
are implicitly assumed in all these formulas). These parameters depend on the
atmospheric conditions and the time of the day.

Similarly to Ref.~\cite{Burgoin} one can assume the typical night-time value
$A=1.7\times10^{-14}~$m$^{-2/3}$ and low-wind $v=21~$m/s~\cite{Tofsted}. This
is also known as the H-V$_{5/7}$ model~\cite[Sec.~12.2.1]{AndrewsBook}.
However, during the day, parameter $A$ can be of the order of $2.75\times
10^{-14}~$m$^{-2/3}$~\cite{Vasy17} and, for high-wind conditions, $v$ can be
of the order of $v=57~$m/s~\cite{BrussSAT}. In our work, we adopt H-V$_{5/7}$
as night-time model, and H-V with parameters $A=2.75\times10^{-14}~$m$^{-2/3}$
and $v=21~$m/s as day-time model. Finally, we may also consider H-V with
parameters $A=2.75\times10^{-14}~$m$^{-2/3}$ and $v=57~$m/s as the worst-case
day-time model.

It is important to remark that there are other models for $C_{n}^{2}(h)$.
These include the VanZandt model~\cite{Vanzdant}, with a simplified version
proposed by Dewan et al.~\cite{Dewan}, and the Walters and Kunkel
model~\cite{Walters}. For instance, they have been used in
Ref.~\cite[Appendix~D]{Vasy19}. These other approaches and the H-V model are
in good agreement with thermosonde data in Ref.~\cite{Thermosonde} (see also
Ref.~\cite[Fig.~13]{Vasy19}). Here we also consider a simplified version of
this model, as originally proposed by Hufnagel and Stanley~\cite[Fig.~6]%
{Stanley}. This is given by~\cite[Eq.~(3.1)]{Fried}
\begin{align}
C_{n}^{2}(h)  &  \simeq c_{1}h^{-1/3}\exp\left(  -\frac{h}{c_{2}}\right)
,\label{Cn2}\\
c_{1}  &  =4.2\times10^{-14},~c_{2}=3200,
\end{align}
so that $C_{n}^{2}\simeq10^{-14}~$m$^{-2/3}$ a few meters high (see also
Ref.~\cite[Ch.~3]{Majumdar} and Ref.~\cite[Ch.~8]{Goodman}).

\begin{figure}[t]
\vspace{0.2cm}
\par
\begin{center}
\includegraphics[width=0.48\textwidth] {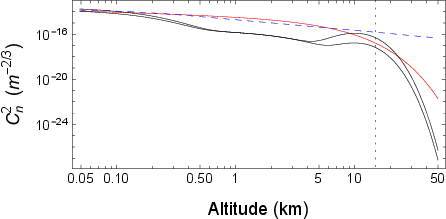}
\end{center}
\par
\vspace{-0.3cm}\caption{Optical turbulence profile in the atmosphere,
quantified by the refraction index structure constant $C_{n}^{2}$ as a
function of the altitude $h$. More precisely, we plot the predictions of the
Hufnagel-Valley model of Eq.~(\ref{HFmodel}) (black curves) considering the
typical night-time parameters (H-V$_{5/7}$, lower black curve) and the
worst-case day-time parameters (i.e., the worst-case day-time model, upper
black curve). We compare these predictions with the simplified model of
Eq.~(\ref{Cn2}) (red curve), which is an approximate upper-bound at most
altitudes. We can clearly see the exponential fall of $C_{n}^{2}(h)$ after
$\simeq15$~km (vertical dotted line), compared with the single-layer average
value $\bar{C}_{n}^{2}(h)$ of Eq.~(\ref{avCn}) (dashed blue line).}%
\label{ConstantPic}%
\end{figure}

In Fig.~\ref{ConstantPic}, we show the H-V model of Eq.~(\ref{HFmodel}) and
the simplified Hufnagel-Stanley model in Eq.~(\ref{Cn2}) which are in good
agreement. We can see that, at higher altitudes, $C_{n}^{2}$ starts to
decrease exponentially. As a matter of fact, it can be considered to be
negligible beyond $h_{\text{max}}\simeq20~$km. This altitude corresponds to
the upper edge of the tropopause, below which most of the mass of the
atmosphere is contained. Taking $h_{\text{max}}$ as effective thickness of the
atmosphere can also be justified by the following argument. Let us treat the
atmosphere as a single layer of thickness $h$ and structure constant given by
the average
\begin{equation}
\bar{C}_{n}^{2}(h)=h^{-1}\int_{0}^{h}d\xi C_{n}^{2}(\xi), \label{avCn}%
\end{equation}
computed according to the standard H-V$_{5/7}$ model. From
Fig.~\ref{ConstantPic}, we can see how this quantity exponentially departs
from the previous models after $15~$km. At $\simeq20~$km the difference is
about two orders of magnitude.

For a satellite at altitude $h$ and zenith angle $\theta\lesssim1$
communicating with a sea-level ground station, the effective section of the
atmosphere which is crossed by the beam is given by $z_{\text{atm}}%
(\theta)=z(h_{\text{max}},\theta)$ using the slant functional in
Eq.~(\ref{slantHT}). At one radiant, we have $z_{\text{atm}}(1)\simeq37~$km,
which is of the same order of magnitude of $h_{\text{max}}$. For larger
angles, refraction comes into place and one needs to use the elongated slant
distance in Eq.~(\ref{refractedslant}). At the horizon, the section becomes
large even neglecting the elongation by refraction. In fact, we have
\begin{equation}
z_{\text{atm}}(\pi/2)=\sqrt{h_{\text{max}}^{2}+2h_{\text{max}}R_{\text{E}}%
}\simeq505~\text{km.}%
\end{equation}

An important observation is that, for $\theta\lesssim1$ and $h\leq
h_{\text{max}}$, we may certainly use the approximations
\begin{equation}
z(h,\theta)\simeq h\sec\theta,~~h(z,\theta)\simeq z\cos\theta,
\end{equation}
since the relative error $[z(h,\theta)-h\sec\theta]/z(h,\theta)$ remains less
than $0.4\%$. In the integral of Eq.~(\ref{avCn}) the structure constant
$C_{n}^{2}$ is non-negligible only for values $\xi\leq h_{\text{max}}$, so we may write
\begin{equation}
\bar{C}_{n}^{2}(h)\simeq h^{-1}\int_{0}^{h_{\text{max}}}d\xi C_{n}^{2}%
(\xi)\simeq h^{-1}\int_{0}^{\infty}d\xi C_{n}^{2}(\xi).
\end{equation}
This observation leads to a simplification when we write $\bar{C}_{n}^{2}$ and
similar integrals in terms of the slant distance $z=z(h,\theta)$. In fact, for
zenith angles $\theta\lesssim1$, we may write the approximation%
\begin{align}
\bar{C}_{n}^{2}(z,\theta)  &  :=z^{-1}\int_{0}^{z}dz^{\prime}C_{n}%
^{2}[h(z^{\prime},\theta)]\\
&  \simeq z^{-1}\sec\theta\int_{0}^{h}d\xi C_{n}^{2}(\xi).
\label{simplificationCNN}%
\end{align}

\subsection{Scintillation index and Rytov variance}

An important issue in free-space communication with turbulence is the
evaluation of the scintillation effects. In general, scintillation corresponds
to irradiance fluctuations, causing variations of the field intensity across
the aperture of the receiver, both temporally (twinkles) and spatially
(speckles). As a result, for an input Gaussian beam, the intensity profile at
the receiver will not be simply given by%
\begin{equation}
I(z,r)=\frac{w_{0}^{2}}{w_{\text{d}}(z)^{2}}\exp[-2r^{2}/w_{\text{d}}(z)^{2}],
\label{Irz}%
\end{equation}
but there will be some instantaneous random profile $I(z,\mathbf{r})$, where
$\mathbf{r}=(x,y)$ is the radial coordinate at the receiver and $z$ the
longitudinal coordinate.

Mathematically, one defines the scintillation index as the normalized variance
of the field intensity~\cite{AndrewsYu95}
\begin{equation}
\sigma_{I}^{2}(z,\mathbf{r}):=\frac{\left\langle I(z,\mathbf{r})^{2}%
\right\rangle }{\left\langle I(z,\mathbf{r})\right\rangle ^{2}}-1,
\end{equation}
where the average is taken over the random fluctuations. This index is usually
decomposed into a longitudinal (on-axis) and transverse (off-axis)
parts~\cite{AndrewsYu95,AndrewsBook}
\begin{equation}
\sigma_{I}^{2}(z,\mathbf{r})=\sigma_{I}^{2}(z,\mathbf{0})+\sigma_{I,r}%
^{2}(z,\mathbf{r}).
\end{equation}
The condition of weak fluctuation (weak turbulence) corresponds to $\sigma
_{I}^{2}(z,\mathbf{r})<1$ throughout the beam profile. If this is the case,
the mean intensity can be closely approximated by a Gaussian spatial
profile~\cite{Gprofile1,Gprofile2,Gprofile3}.

According to previous studies~\cite{AndrewsYu95,Andrews2000}, the regime of
weak turbulence holds for zenith angles smaller than $1$ radiant (i.e., about
$60^{\circ}$) assuming the standard H-V$_{5/7}$ atmospheric model. For
downlink, this is true for any beam waist $w_{0}$. For uplink, the off-axis
scintillation index $\sigma_{I,r}^{2}(z,\mathbf{r})$ increases with $w_{0}$,
but still remains reasonably small over the receiver's aperture if this is not
too large (condition which is typically satisfied at the satellite). Under the
assumption of weak fluctuations, one can use Rytov approximation for the beam
field~\cite{RytovAPPROX}, together with the Kolmogorov power-law
spectrum~\cite{Kolmogorov}, and develop a simple formalism for the theory of turbulence.

In a weak-fluctuation theory, the longitudinal scintillation index $\sigma
_{L}^{2}:=\sigma_{I}^{2}(z,\mathbf{0})$ can be easily written for both
downlink and uplink. For a downlink path from a satellite at altitude $h$ and
zenith angle $\theta\lesssim1$, this index equals the Rytov variance for a
plane wave $\sigma_{R,\text{plane}}^{2}$, i.e.~\cite{Andrews2000,AndrewsBook}%
\begin{align}
\sigma_{L\text{,down}}^{2}  &  =\sigma_{\text{Rytov}}^{2}:=2.25k^{7/6}%
h^{5/6}(\sec\theta)^{11/6}\mu(h),\label{rytovDOWN}\\
\mu(h)  &  :=\int_{0}^{h}d\xi C_{n}^{2}(\xi)\left(  \frac{\xi}{h}\right)
^{5/6}.\label{integralmuH}%
\end{align}
Note that, if we impose $C_{n}^{2}$ to be constant in the integral of
Eq~(\ref{integralmuH}), then $\sigma_{\text{Rytov}}^{2}$ becomes
$1.23C_{n}^{2}k^{7/6}z^{11/6}$, which is the expression for the Rytov variance
that is valid for fixed-altitude $z$-long propagation.

For an uplink path, we may instead write~\cite{YuraMcKinley83}
\begin{align}
\sigma_{L\text{,up}}^{2}  &  =\sigma_{L\text{,down}}^{2}\frac{\tilde{\mu}%
(h)}{\mu(h)},\\
\tilde{\mu}(h)  &  :=\int_{0}^{h}d\xi C_{n}^{2}(\xi)\left(  \frac{\xi}%
{h}\right)  ^{5/6}\left(  1-\frac{\xi}{h}\right)  ^{5/6}.
\end{align}
As noted in Ref.~\cite{Andrews2000}, one can approximate $\tilde{\mu}%
(h)\simeq\mu(h)$, so that $\sigma_{L\text{,up}}^{2}\simeq\sigma_{\text{Rytov}%
}^{2}$ also for uplink.

\begin{figure}[t]
\vspace{0.2cm}
\par
\begin{center}
\includegraphics[width=0.48\textwidth] {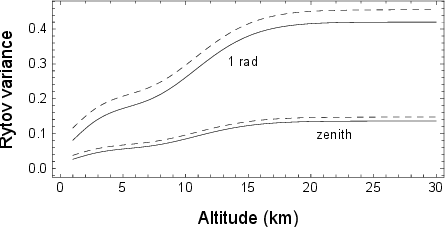}
\end{center}
\par
\vspace{-0.3cm}\caption{Rytov variance versus altitude $h$ (km)\ for
$\theta=0$ (zenith) and $\theta=1$, considering $\lambda=800~$nm. We plot the
Rytov variance assuming the H-V model with night-time parameters (H-V$_{5/7}$
model, solid lines) and the H-V model with typical day-time parameters (dashed
lines). In all cases, the Rytov variance saturates at values that are $<1$.}%
\label{RytovPic}%
\end{figure}

For these reasons the Rytov variance for a plane wave in Eq.~(\ref{rytovDOWN})
can be used as a measure of the scintillation (in the weak fluctuation regime)
and, most importantly, as a parameter to check if the condition of weak
turbulence is indeed satisfied, corresponding to $\sigma_{\text{Rytov}}^{2}%
<1$. As we can see from Fig.~\ref{RytovPic}, the value of the Rytov variance
quickly saturates within the atmosphere and its values are below unity for
zenith angles within $1$ radiant. It is easy to check that the Rytov variance
exceeds $1$ for larger zenith angles; for instance, at $h=20$~km, we have that
$\sigma_{\text{Rytov}}^{2}>1$ for $\theta\gtrsim1.2$, i.e., beyond $69^{\circ
}$. One can also check that, if we assume not typical but worst-case day-time
parameters for the H-V model (i.e., high-wind conditions, see
Appendix~\ref{HVmodelAPP}), the Rytov variance tends to values that are below
the unity at the zenith ($\simeq0.6$), but quickly violate the unity for
increasing zenith angle, e.g., $\simeq2$ already at $\theta=1$.

\subsection{Coherence length}

Once we have clarified the working regime of weak turbulence, we introduce the
spherical-wave coherence length $\rho_{0}$, which directly enters in the
explicit expressions of the spot sizes of Eq.~(\ref{turbmodel}). This is
related to the well known Fried's parameter $r_{\text{F}}$%
~\cite{Fried,BelandBook}, that can be written as $r_{\text{F}}=2.088\rho_{0}%
$~\cite{Fante80}, and describes the transverse spatial separation at the
receiver over which the field phase correlations decay by $1/e$. At the
optical frequencies, typical values of $\rho_{0}$ or $r_{\text{F}}$ are in cm.
When this value is particularly large, e.g., of the order of meters, then the
effects of turbulence are completely negligible from the point of view of the
receiver. In this regard, we will see a stark difference between uplink and downlink.

For wavenumber $k$ and propagation distance $z$, the spherical-wave coherence
length is given by~\cite[Eq.~(38)]{Fante75}
\begin{align}
\rho_{0}  &  =\left[  1.46k^{2}I_{0}(z)\right]  ^{-3/5},\label{spericalwave}\\
I_{0}(z)  &  :=\int_{0}^{z}d\xi\left(  1-\frac{\xi}{z}\right)  ^{5/3}C_{n}%
^{2}(\xi), \label{cohL}%
\end{align}
where the explicit functional dependence of $C_{n}^{2}(\xi)$ needs to be
specified and depends on the type of propagation.
For free-space propagation at a fixed altitude, the value of $C_{n}^{2}$ is
constant and we have the simple form
\begin{equation}
\rho_{0}^{\text{fix}}=(0.548k^{2}C_{n}^{2}z)^{-3/5}. \label{fixCLeq}%
\end{equation}

For uplink communications where the altitude $h$ increases with the beam
propagation, we assume the H-V$_{5/7}$ model for $C_{n}^{2}(h)$ and we write
$\rho_{0}$ in terms of the slant distance $z$ and the zenith angle $\theta$,
by replacing $I_{0}(z)$ with the following integral%
\begin{equation}
I_{0}^{\text{up}}(z,\theta):=\int_{0}^{z}d\xi\left(  1-\frac{\xi}{z}\right)
^{5/3}C_{n}^{2}[h(\xi,\theta)], \label{coHL2}%
\end{equation}
where $h(z,\theta)$ is given in Eq.~(\ref{hztheta}). For downlink, the
altitude decreases with the propagation. This is accounted by replacing
$\xi\rightarrow z-\xi$ in the structure constant, so that we replace
$I_{0}(z)$ with the following integral%
\begin{equation}
I_{0}^{\text{down}}(z,\theta):=\int_{0}^{z}d\xi\left(  1-\frac{\xi}{z}\right)
^{5/3}C_{n}^{2}[h(z-\xi,\theta)].
\end{equation}

In downlink, the term $\left(  1-\xi/z\right)  ^{5/3}$ goes to zero in the
region where $C_{n}^{2}$\ has the higher values (close to the ground). For
this reason, the downlink coherence length $\rho_{0}^{\text{down}}$ becomes
large very quickly, for any object beyond the tropopause ($\simeq20~$km). For
instance, consider an object at the slant distance of $z=100~$km sending down
a beam with wavelength $\lambda=800~$nm. We compute $\rho_{0}^{\text{down}%
}\simeq1.8~$m at the zenith (compared to the uplink value $\rho_{0}%
^{\text{up}}\simeq4.2$~cm) and $\rho_{0}^{\text{down}}\simeq68~$cm at
$\theta=1$ (compared to $\simeq2.9$~cm in uplink). At $\lambda=1~\mu$m, we compute
$\rho_{0}^{\text{down}}\simeq2.4~$m at the zenith, and $\rho_{0}^{\text{down}%
}\simeq0.9~$m at $\theta=1$. Note that these values increase both in distance
$z$ and wavelength. In particular, one has $\rho_{0}\propto\lambda^{6/5}$.

It is clear that, within $1$ radiant from the zenith (weak scintillation
regime), the effect of the atmospheric turbulence is practically negligible in
downlink paths. More precisely, this is true as long as the receiver's
aperture $a_{R}$ does not become too large (e.g., of the order of $2$ meters).
In fact, recall that the number of turbulence-induced short-term speckles from
a point source is of the order of $N_{s}=1+(a_{R}/\rho_{0})^{2}$%
~\cite{AndrewsBook}. Assuming $a_{R}=40$~cm and an object at altitude
$h=100~$km communicating at $\lambda=800~$nm, we compute $N_{s}\simeq1.05$ at
the zenith and $\simeq1.35$ at $\theta=1$ radiant. These values are very close
to the perfect coherent limit ($N_{s}=1$).

Then, consider an increased aperture $a_{R}=1$~m and a satellite in the LEO
region, so that $h\geq160~$km, we compute $N_{s}\lesssim1.11$ at the zenith
and $\lesssim1.82$ at $\theta=1$ radiant. In particular, for a satellite at
$h=530$~km (as the one studied in the main text), we get $N_{s}\lesssim1.01$
at the zenith and $\lesssim1.07$ at $\theta=1$ radiant. For these reasons,
turbulence-induced beam spreading and wandering are negligible in downlink.
This means that long- and short-term spot sizes are both equal to the
diffraction-limited spot size, i.e., we can set $w_{\text{lt}}\simeq
w_{\text{st}}\simeq w_{\text{d}}$.

For uplink the situation is completely different, and turbulence effects
cannot be neglected even at the zenith position. Before proceeding further, it
is important to note some simplifications which can be enforced for zenith
angles $\theta\lesssim1$ (that are useful for the spot sizes in uplink
discussed in the next subsection). First of all, we may simplify the
expression of the slant distance as\ in Eq.~(\ref{simplificationCNN}) and
write%
\begin{equation}
\rho_{0}^{\text{up}}\simeq\left[  1.46k^{2}\sec\theta\int_{0}^{h}d\xi\left(
1-\frac{\xi}{h}\right)  ^{5/3}C_{n}^{2}(\xi)\right]  ^{-3/5}. \label{upSLLL}%
\end{equation}
Then, we observe that, for $\theta\lesssim1$, any satellite slant distance $z$
is much larger than the thickness of the atmosphere ($20-37~$km). As a result,
the term $\left(  1-\xi/z\right)  ^{5/3}$ in Eq.~(\ref{upSLLL}) is $\simeq1$
for all values of $\xi$ falling in the atmosphere, where the quantity
$C_{n}^{2}$ is non-negligible.

\begin{figure}[t]
\vspace{0.2cm}
\par
\begin{center}
\includegraphics[width=0.42\textwidth] {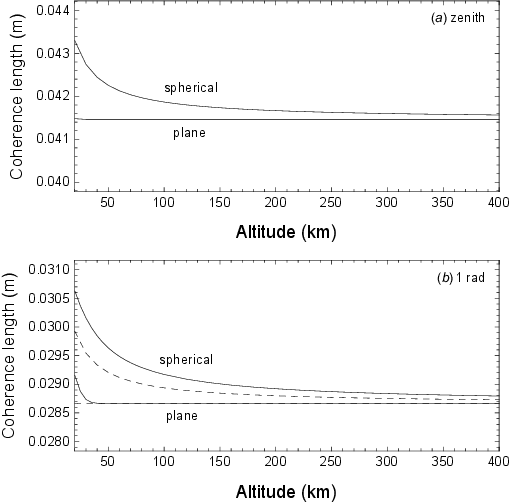}
\end{center}
\par
\vspace{-0.3cm}\caption{Coherence length (m) versus altitude $h$ (km)\ for
uplink communication with a collimated Gaussian beam at $\lambda=800~$nm.
Turbulence is modelled by H-V$_{5/7}$. We compare the spherical-wave coherence
length $\rho_{0}^{\text{up}}$ (upper curve) and its plane-wave approximation
$\rho_{p}^{\text{up}}$ (lower curve). In panel (a), we consider the zenith
position, while in panel (b) we consider $\theta=1$. In both panels the solid
curves are computed with the exact integrals in Eqs.~(\ref{coHL2})
and~(\ref{planarIntegral}). The dashed curves are associated with the secant
approximations in Eqs.~(\ref{upSLLL}) and~(\ref{planarIntegral2}). The latter
are not shown in panel (a) because they overlap with the solid curves. Note
that $\rho_{0}^{\text{up}}$ rapidly converges to $\rho_{p}^{\text{up}}$, and
the asymptotic (lower bound) value in Eq.~(\ref{finalAPP}) is approximately
achieved already in the LEO region, i.e., for $h\geq h_{\text{LEO}}=160~$km.}%
\label{compareCLpic}%
\end{figure}

For this reason, we can approximate the spherical-wave coherence length
$\rho_{0}^{\text{up}}$ with a plane-wave coherence length~\cite[Eq.~(51)]%
{Fante75}, which is given by%
\begin{align}
\rho_{p}^{\text{up}}  &  =\left[  1.46k^{2}I_{p}(z,\theta)\right]
^{-3/5},\label{planewavephase}\\
I_{p}(z,\theta)  &  =\int_{0}^{z}d\xi C_{n}^{2}[h(\xi,\theta
)]\label{planarIntegral}\\
&  \simeq\sec\theta\int_{0}^{h}d\xi C_{n}^{2}(\xi). \label{planarIntegral2}%
\end{align}
This planar approximation is numerically investigated in
Fig.~\ref{compareCLpic}, where we see how $\rho_{0}^{\text{up}}$ rapidly
approaches $\rho_{p}^{\text{up}}$ already in the LEO\ region. These coherence
lengths are computed using the exact integrals in Eqs.~(\ref{coHL2})
and~(\ref{planarIntegral}). The secant approximations in Eqs.~(\ref{upSLLL})
and~(\ref{planarIntegral2}) provide curves that are very close to those based
on the exact integrals. As a matter of fact, they practically overlap with
them at the zenith position. (Note that, while Fig.~\ref{compareCLpic} is
plotted for the night-time H-V model, an equivalent behavior is found for the
day-time H-V model, but with different asymptotic values). 

It is therefore clear that we can set $\rho_{0}^{\text{up}}\simeq\rho
_{p}^{\text{up}}$ and use the integral in Eq.~(\ref{planarIntegral2}). We can
further simplify the formulas above by directly replacing the asymptotic value
of the coherence lengths. In other words, we may extend the integral in
Eq.~(\ref{planarIntegral2}) to infinity, and write%
\begin{align}
\rho_{0}^{\text{up}} &  \simeq\rho_{p}^{\text{up}}\simeq\left[  1.46k^{2}%
(\sec\theta)I_{\infty}\right]  ^{-3/5},\label{finalAPP}\\
I_{\infty} &  :=\int_{0}^{\infty}d\xi C_{n}^{2}(\xi).
\end{align}
Assuming the H-V$_{5/7}$ model of atmosphere ($A=1.7\times10^{-14}~$m$^{-2/3}$
and $v=21~$m/s ), particularly appropriate for night-time operation, we
compute $I_{\infty}\simeq2.2354\times10^{-12}$~m$^{1/3}$, leading to
\begin{equation}
\rho_{0}^{\text{up}}\simeq\rho_{p}^{\text{up}}\simeq8.59\times10^{5}%
\lambda^{6/5}\left(  \sec\theta\right)  ^{-3/5},\label{approxSIMPLE}%
\end{equation}
which is an excellent approximation for any slant distance $z\geq
h_{\text{LEO}}=160~$km and zenith angle $\theta\lesssim1$. For the day-time H-V
model ($A=2.75\times10^{-14}~$m$^{-2/3}$ and $v=21~$m/s), we compute the
different value $I_{\infty}\simeq3.2854\times10^{-12}$~m$^{1/3}$, so the
approximation in Eq.~(\ref{approxSIMPLE}) holds with a different prefactor.

\subsection{Spot sizes for uplink}

Consider a Gaussian beam with wavelength $\lambda$, spot size $w_{0}$ and
curvature radius ${R_{0}}$, which freely propagates in uplink for a distance
$z$ with a zenith angle $\theta\lesssim1$, so that we are in the regime of
weak turbulence. In particular, we may assume a collimated beam (${R_{0}%
=+\infty}$), even though this assumption is not necessary for the following
theory. The associated spherical-wave coherence length $\rho_{0}^{\text{up}}$
is based on the integral in Eq.~(\ref{coHL2}) which can be closely
approximated by Eq.~(\ref{finalAPP}).

We now impose Yura's condition~\cite{Yura73,Fante75}
\begin{equation}
0.33\left(  \frac{\rho_{0}^{\text{up}}}{w_{0}}\right)  ^{\frac{1}{3}}\ll1.
\label{condsYURA}%
\end{equation}
Using Eq.~(\ref{approxSIMPLE}), it is easy to show that Eq.~(\ref{condsYURA})
is implied by $w_{0}^{1/3}\gg31\lambda^{2/5}$, which is compatible with
typical satellite communications. For instance, at $\lambda=800~$nm, it means
to considering spot sizes $w_{0}\gg1.4$~mm. Furthermore, the condition in
Eq.~(\ref{condsYURA}) could also be imposed more weakly as $\rho
_{0}^{\text{up}}/w_{0}<1$, in which case the resulting expressions (discussed
below) will be valid with a higher degree of approximation.

The satisfaction of Yura's condition in Eq.~(\ref{condsYURA}) allows us to
write the decomposition in Eq.~(\ref{turbmodel}) with analytical expressions
for the long- and short-term spot sizes. Specifically, we have the
formulas~\cite{Yura73,Fante75}%
\begin{align}
w_{\text{lt}}^{2}  &  \simeq w_{\text{d}}^{2}+2\left(  \frac{\lambda z}%
{\pi\rho_{0}^{\text{up}}}\right)  ^{2},\label{LTexpression}\\
w_{\text{st}}^{2}  &  \simeq w_{\text{d}}^{2}+2\left(  \frac{\lambda z}%
{\pi\rho_{0}^{\text{up}}}\right)  ^{2}\Psi, \label{STexpression}%
\end{align}
where $w_{\text{d}}$ is the diffraction-limited field spot size and $\Psi$ is
given by~\cite{Yura73}%
\begin{equation}
\Psi=\left[  1-0.33\left(  \frac{\rho_{0}^{\text{up}}}{w_{0}}\right)
^{1/3}\right]  ^{2}\simeq1-0.66\left(  \frac{\rho_{0}^{\text{up}}}{w_{0}%
}\right)  ^{1/3}. \label{GammaFF}%
\end{equation}
As a consequence, the variance associated to the centroid wandering is given
by~\cite{Yura73}%
\begin{equation}
\sigma_{\text{TB}}^{2}=w_{\text{lt}}^{2}-w_{\text{st}}^{2}\simeq
\frac{0.1337\lambda^{2}z^{2}}{w_{0}^{1/3}\left(  \rho_{0}^{\text{up}}\right)
^{5/3}}. \label{SigmaCwander}%
\end{equation}

Note that the expressions in Eqs.~(\ref{LTexpression}) and~(\ref{STexpression}%
) are derived from Ref.~\cite[Eqs.~(16-18)]{Yura73} and Ref.~\cite[Eq.~(37)]%
{Fante75}, changing their notation from intensity spot size ($w_{\text{int}}$)
to field spot size ($w=\sqrt{2}w_{\text{int}}$). See also
Refs.~\cite{Poirier72,Bunkin70,Dios04,Belmonte} for related derivations.

The formulas above undergo a great simplification by explicitly accounting for
the asymptotic expression of the coherence length $\rho_{0}^{\text{up}}$ given
in Eq.~(\ref{finalAPP}). Thus, for uplink satellite communications with zenith
angle $\theta\lesssim1$, we may write the following approximations%
\begin{align}
w_{\text{lt}}^{2}  &  \simeq w_{\text{d}}^{2}+a\lambda^{-2/5}z^{2}(\sec
\theta)^{6/5},\label{appp1}\\
\Psi &  \simeq1-bw_{0}^{-1/3}(\lambda^{2}\cos\theta)^{1/5},\label{appp2}\\
\sigma_{\text{TB}}^{2}  &  \simeq cw_{0}^{-1/3}z^{2}\sec\theta, \label{appp2b}%
\end{align}
where we set $a=26.28I_{\infty}^{6/5}$, $b=0.2934I_{\infty}^{-1/5}$ and
$c=ab\simeq7.71I_{\infty}$, whose numerical values depend on the specific
atmospheric profile (e.g., for the H-V$_{5/7}$ model, they become
$a\simeq2.75\times10^{-13}$, $b\simeq63$, and $c\simeq1.72\times10^{-11}$).
Then, for the short-term spot size, we may write the following simple
expression%
\begin{align}
w_{\text{st}}^{2}  &  =w_{\text{lt}}^{2}-\sigma_{\text{TB}}^{2}\simeq
w_{\text{d}}^{2}+z^{2}\Delta(\theta),\label{STapproxxx}\\
\Delta(\theta)  &  :=a\lambda^{-2/5}(\sec\theta)^{6/5}-cw_{0}^{-1/3}\sec
\theta. \label{overheadST}%
\end{align}

By using the geometric expression of the slant distance $z=z(h,\theta)$ from
Eq.~(\ref{slantHT}) in Eqs.~(\ref{appp1}), (\ref{appp2b})
and~(\ref{STapproxxx}), we can study the behavior of the spot sizes and that
of the centroid wandering as a function of the altitude $h$ and zenith angles
$\theta\lesssim1$. For a typical optical frequency, it is easy to see that
their values practically coincide with those that can be computed from
Eqs.~(\ref{LTexpression}), (\ref{STexpression}) and~(\ref{SigmaCwander}),
while providing much simpler analytical expressions.

\begin{figure}[t]
\vspace{0.2cm}
\par
\begin{center}
\includegraphics[width=0.48\textwidth] {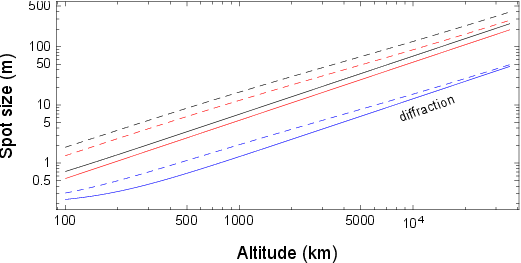}
\end{center}
\par
\vspace{-0.3cm}\caption{Spot sizes (m) versus altitude $h$ (km)\ for uplink
communication by means of a collimated Gaussian beam with $\lambda=800~$nm and
$w_{0}=20~$cm. Here, turbulence is described by the H-V$_{5/7}$ model (night
time). At the zenith position, we plot the short-term spot size $w_{\text{st}%
}$ (black solid) and the standard deviation of the centroid wandering
$\sigma_{\text{TB}}$ (red solid), to be compared with the diffraction-limited
spot size $w_{\text{d}}$ (blue solid). Dashed lines refer to a zenith angle of
$\theta=1$.}%
\label{stPic}%
\end{figure}

For uplink communication with a collimated beam with $w_{0}=20~$cm and
$\lambda=800$~nm, we perform a numerical study in Fig.~\ref{stPic}. Here we
note that the short-term spot size $w_{\text{st}}$ becomes considerably larger
than the diffraction-limited spot size $w_{\text{d}}$, and the standard
deviation of the centroid wandering $\sigma_{\text{TB}}$ increases from about
$0.5-1~$m at the K\'{a}rm\'{a}n line to about $200-300~$m at the GEO altitude
($\simeq36000~$km), depending on the value of the zenith
angle.

\section{Background noise in satellite communications\label{APPnoiseSAT}}

Let us discuss the basic theoretical models which describe the background
noise affecting satellite communications. With good approximation, both the
Moon and the Earth can be considered to be Lambertian
disks~\cite{LambertScatterer,LBbook}. This means that the scattering from
their surfaces can be approximated to be uniform (radiance independent from
the angle), which in turn implies that the intensity detected by the
satellite's receiver strictly depends on its angular field of view
$\Omega_{\text{fov}}$.

First consider uplink. In day-time operation, the main source of noise comes
from the sunlight directly reflected by the Earth towards the satellite. The
total amount depends on the solar spectral irradiance $H_{\lambda}%
^{\text{Sun}}$ at the relevant wavelength $\lambda$ and the albedo of the
Earth ($A_{\text{E}}\simeq0.3$). During night-time operation, the noise is
mainly due to the sunlight reflected by the Moon towards the Earth, and then
from the Earth towards the satellite. Therefore, this noise also depends on
the albedo of the Moon ($A_{\text{M}}\simeq0.12$), the radius of the Moon
($R_{\text{M}}\simeq1.737\times10^{6}$~m), and the average Earth-Moon distance
($d_{\text{EM}}\simeq3.84\times10^{8}$~m).

Considering these parameters, the mean number of environmental thermal photons
impinging on the satellite's receiver with aperture $a_{R}$ and (solid)
angular field of view $\Omega_{\text{fov}}$, within the time window $\Delta t$
and the spectral filter $\Delta\lambda$, is given by~\cite{Bonato} $\bar
{n}_{B}^{\text{up}}=\kappa H_{\lambda}^{\text{sun}}\Gamma_{R}$. Here the
parameter $\Gamma_{R}=\Delta\lambda\Delta t\Omega_{\text{fov}}a_{R}^{2}$ only
depends on the specific features of the receiver, while the dimensionless
factor $\kappa$ is equal to $\kappa_{\text{day}}=A_{\text{E}}\simeq0.3$ for
day-time and to $\kappa_{\text{night}}=A_{\text{E}}A_{\text{M}}R_{\text{M}%
}^{2}d_{\text{EM}}^{-2}\simeq7.36\times10^{-7}$ for full-Moon night-time
(roughly $10^{-6}$ of the day-time value).

At the optical regime, the typical values of $\bar{n}_{B}^{\text{up}}$ are
orders of magnitude higher than the photon numbers due to the black-body
thermal radiation emitted by the Earth. Recall that the spectral radiance of a
black body at temperature $T$ and wavelength $\lambda$\ is given by
\begin{equation}
N(\lambda,T)=2c\lambda^{-4}\left[  e^{hc/(\lambda k_{\text{B}}T)}-1\right]
^{-1},
\end{equation}
in terms of number of photons per unit area, time, wavelength, and solid angle
(photons m$^{-2}$ s$^{-1}$ nm$^{-1}$ sr$^{-1}$). In the formula above, it is
understood that $c$ is the speed of light and $k_{\text{B}}$ is the Boltzmann
constant. Therefore, the total number of photons impinging on the receiver is
given by $\bar{n}_{\text{body}}^{\text{up}}=N(\lambda,T)\Gamma_{R}$.
Considering the optical wavelength $\lambda=800~$nm and assuming the average
surface temperature of the Earth ($T\simeq288~$K), one has $N(\lambda
,T)\simeq3\times10^{6}$~photons m$^{-2}$ s$^{-1}$ nm$^{-1}$ sr$^{-1}$. For a
receiver with $\Gamma_{R}=1.6\times10^{-19}~$m$^{2}$ s nm sr, we compute
$\bar{n}_{\text{body}}^{\text{up}}\simeq4.8\times10^{-13}$ mean photons, which
is clearly negligible with respect to the values of $\bar{n}_{B}^{\text{up}}$
given in Table~\ref{TableDDDD} of the main text.

In downlink, the transmitter is the satellite and the receiver is a ground
station with aperture $a_{R}$ and angular field of view $\Omega_{\text{fov}}$.
In this case, the number of environmental photons reaching the receiver within
the time window $\Delta t$ and the spectral filter $\Delta\lambda$ is given
by~\cite{Miao,BrussSAT} $\bar{n}_{B}^{\text{down}}=H_{\lambda}^{\text{sky}%
}\Gamma_{R}$, where $H_{\lambda}^{\text{sky}}:=\pi\tilde{H}_{\lambda
}^{\text{sky}}/(\hbar\omega)$ and $\tilde{H}_{\lambda}^{\text{sky}}$ is the
background spectral irradiance of the sky in units W m$^{-2}$ nm$^{-1}$
sr$^{-1}$. In these units, its value ranges between $1.5\times10^{-6}$
(full-Moon clear night)\ to $1.5\times10^{-3}$ (clear day-time) and
$1.5\times10^{-1}$ (cloudy day-time)~\cite[Table~1]{Miao}, assuming that the
field of view of the receiver does not include the Moon or the
Sun~\cite{sky1,sky2}. At $\lambda=800~$nm, we have $\pi/(\hbar\omega
)\simeq1.27\times10^{19}$ W$^{-1}$ s$^{-1}$ sr, which means that $H_{\lambda
}^{\text{sky}}$ ranges between $1.9\times10^{13}$ and $1.9\times10^{18}$
photons m$^{-2}$ s$^{-1}$ nm$^{-1}$ sr$^{-1}$. Using the same parameters for
the receiver as above ($\Gamma_{R}=1.6\times10^{-19}~$m$^{2}$ s nm sr), we
find that $\bar{n}_{B}^{\text{down}}$ ranges between $\simeq3\times10^{-6}$
and $\simeq0.3$ mean photons, which are the values for downlink reported in
Table~\ref{TableDDDD} of the main text.

Let us conclude by noticing that these estimates for the background thermal
noise can be made more precise by employing dedicated programs. For instance,
a more detailed calculation of sky brightness can be achieved by using
software such as MODTRAN~\cite{Modtran}
LibRadtran~\cite{LibRadtran1,LibRadtran2}, or 6SV~\cite{6SV}.


\begin{thebibliography}{99}                                                                                               %


\bibitem {QKDreview}S. Pirandola, U. L. Andersen, L. Banchi, M. Berta, D.
Bunandar, R. Colbeck, D. Englund, T. Gehring, C. Lupo, C. Ottaviani, J.
Pereira, M. Razavi, J. S. Shaari, M. Tomamichel, V. C. Usenko, G. Vallone, P.
Villoresi, and P. Wallden, \textit{Advances in Quantum Cryptography}, Adv. Opt. Photon. textbf{12}, 1012 (2020).

\bibitem {satexp1}G. Vallone, D. Bacco, D. Dequal, S. Gaiarin, V. Luceri, G.
Bianco, and P. Villoresi, \textit{Experimental Satellite Quantum
Communications}, Phys. Rev. Lett. \textbf{115}, 040502 (2015).

\bibitem {satexp1b}S.-K. Liao, W.-Q. Cai, W.-Y. Liu, L. Zhang, Y. Li, J.-G.
Ren, J. Yin, Q. Shen, Y. Cao, and Z.-P. Li et al., \textit{Satellite-to-ground
quantum key distribution},\ Nature \textbf{549}, 43 (2017).

\bibitem {satexp1c}S.-K. Liao, W.-Q. Cai, J. Handsteiner, B. Liu, J. Yin, L.
Zhang, D. Rauch, M. Fink, J.-G. Ren, and W.-Y. Liu et al.,
\textit{Satellite-Relayed Intercontinental Quantum Network}, Phys. Rev. Lett.
\textbf{120}, 030501 (2018).

\bibitem {satexp2}J. Yin, Y. Cao, Y.-H. Li, S.-K. Liao, L. Zhang, J.-G. Ren,
W.-Q. Cai, W.-Y. Liu, B. Li, and H. Dai et al., \textit{Satellite-based
entanglement distribution over 1200 kilometers}, Science \textbf{356}, 1140 (2017).

\bibitem {satexp2b}J. Yin, Y. Cao, Y. H. Li, J. G. Ren, S. K. Liao, L. Zhang,
W. Q. Cai,W. Y. Liu, B. Li, and H. Dai et al., \textit{Satellite-to-Ground
Entanglement-Based Quantum Key Distribution}, Phys. Rev. Lett. \textbf{119},
200501 (2017).

\bibitem {satexp2c}J. Yin, Y.-H. Li, S.-K. Liao, M. Yang, Y. Cao, L. Zhang,
J.-G. Ren, W.-Q. Cai, W.-Y. Liu, S.-L. Li, et al., \textit{Entanglement-based
secure quantum cryptography over 1,120 kilometres}, Nature \textbf{582}, 501 (2020).

\bibitem {satexp3}J. G. Ren, P. Xu, H. L. Yong, L. Zhang, S. K. Liao, J. Yin,
W. Y. Liu, W. Q. Cai, M. Yang, and L. Li et al., \textit{Ground-to-satellite
quantum teleportation}, Nature \textbf{549}, 70 (2017).

\bibitem {satexp4}S. K. Liao, J. Lin, J. G. Ren, W. Y. Liu, J. Qiang, J. Yin,
Y. Li, Q. Shen, L. Zhang, and X. F. Liang et al., \textit{Space-to-Ground
Quantum Key Distribution Using a Small-Sized Payload on Tiangong-2 Space Lab},
Chin. Phys. Lett. \textbf{34}, 090302 (2017).

\bibitem {satexp4b}H. Takenaka, A. Carrasco-Casado, M. Fujiwara, M. Kitamura,
M. Sasaki, and M. Toyoshima, \textit{Satellite-to ground quantum-limited
communication using a 50-kgclass microsatellite}, Nat. Photon. \textbf{11},
502 (2017).

\bibitem {satexp4c}A. Villar, A. Lohrmann, X. Bai, T. Vergoossen, R.
Bedington, C. Perumangatt, H. Y. Lim, T. Islam, A. Reezwana, Z. Tang, et al.,
\textit{Entanglement demonstration on board a nano-satellite}, Optica
\textbf{7}, 734 (2020).

\bibitem {QKDpaper}S. Pirandola, R. Laurenza, C. Ottaviani, and L. Banchi,
\textit{Fundamental limits of repeaterless quantum communications}, Nat.
Commun. \textbf{8}, 15043 (2017). See also arXiv:1510.08863\ (2015).

\bibitem {RCI}S. Pirandola, R. Garc\'{\i}a-Patr\'{o}n, S. L. Braunstein, and
S. Lloyd,\ \textit{Direct and reverse secret-key capacities of a quantum
channel}, Phys. Rev. Lett. \textbf{102}, 050503 (2009).

\bibitem {FreeSpacePaper}S. Pirandola, \textit{Limits and Security of
Free-Space Quantum Communications}, Phys. Rev. Res. \textbf{3}, 013279 (2021).

\bibitem {RMP}C. Weedbrook, S. Pirandola, R. Garc\'{\i}a-Patr\'{o}n, N. J.
Cerf, T. C. Ralph, J. H. Shapiro, and S. Lloyd, \textit{Gaussian Quantum
Information}, Rev. Mod. Phys. \textbf{84}, 621 (2012).

\bibitem {netpaper}S. Pirandola, \textit{End-to-end capacities of a quantum
communication network}, Commun. Phys. \textbf{2}, 51 (2019). See also
arXiv:1601.00966 (2016).

\bibitem {svelto}O. Svelto, \textit{Principles of Lasers}, 5th edn. (Springer,
New York 2010).

\bibitem {Andrews93}L. C. Andrews, W. B. Miller, and J. C. Ricklin,
\textit{Geometrical representation of Gaussian beams propagating through
complex paraxial optical systems}, Appl. Opt. \textbf{32}, 5918-5929 (1993).

\bibitem {Andrews94} M. Born and E. Wolf, \textit{Principles of Optics} (Cambridge University Press, Cambridge, 2013).

\bibitem {Siegman}A. Siegman, \textit{Lasers} (University Science Books, 1986).

\bibitem {Huffman}C. F. Bohren and D. R. Huffman, \textit{Absorption and
scattering of light by small particles} (John Wiley \& Sons, 2008).

\bibitem {Vasy19}D. Vasylyev, W. Vogel, and F. Moll,
\textit{Satellite-mediated quantum atmospheric links},\ Phys. Rev. A
\textbf{99}, 053830 (2019).

\bibitem {BrussSAT}C. Liorni, H. Kampermann, and D. Bru\ss ,
\textit{Satellite-based links for quantum key distribution: beam effects and
weather dependence}, New J. Phys. \textbf{21}, 093055 (2019).

\bibitem {Esposito}R. Esposito, \textit{Power Scintillations Due to the
Wandering of the. Laser Beam}, Proc. IEEE \textbf{55}, 1533-1534 (1967).

\bibitem {Fried73}D. Fried, \textit{Statistics of laser beam fade induced by
pointing jitter}, App. Opt. \textbf{12}, 422-423 (1973).

\bibitem {Titterton}P. Titterton, \textit{Power reduction and fluctuations
caused by narrow laser beam motion in the far field},\ Appl. Opt. \textbf{12},
423-425 (1973).

\bibitem {Vasy12}D. Yu. Vasylyev, A. A. Semenov, W. Vogel, \textit{Toward
Global Quantum Communication: Beam Wandering Preserves Nonclassicality}, Phys.
Rev. Lett. \textbf{108}, 220501 (2012).

\bibitem {Fante75}R. L. Fante, \textit{Electromagnetic Beam Propagation in
Turbulent Media}, Proc. IEEE \textbf{63}, 1669 (1975).

\bibitem {Burgoin}J.-P. Bourgoin, E. Meyer-Scott, B. L. Higgins, B. Helou, C.
Erven, H. H\"{u}bel, B. Kumar, D. Hudson, I. D'Souza, R. Girard, R. Laflamme,
and T. Jennewein, \textit{A comprehensive design and performance analysis of
low Earth orbit satellite quantum communication}, New J. Phys. \textbf{15},
023006 (2013).

\bibitem {Stanley}R. E. Hufnagel and N. R. Stanley, \textit{Modulation
transfer function associated with image transmission through turbulent media},
J. Opt. Soc. Am. \textbf{54}, 52-61 (1964).

\bibitem {Valley}G. C. Valley, \textit{Isoplanatic degradation of tilt
correction and short-term imaging systems}, Appl. Opt. \textbf{19}, 574-577 (1980).

\bibitem {AndrewsBook}L. C. Andrews and R. L. Phillips, \textit{Laser Beam
Propagation Through Random Medium}, 2nd edn. (SPIE, Bellinghan, 2005).

\bibitem {Fante80}R.L. Fante, \textit{Electromagnetic beam propagation in
turbulent media: An update}, Proc. IEEE \textbf{68}, 1424 (1980).

\bibitem {Yura73}H. Yura, \textit{Short term average optical-beam spread in a
turbulent medium}, J. Opt. Soc. Am. \textbf{63}, 567-572 (1973).

\bibitem {Dowling}J. Dowling and P. Livington, \textit{Behavior of focused
beams in atmospheric turbulence: Measurements and comments on the theory}, J.
Opt. Soc. Am. \textbf{63}, 846-858 (1973).

\bibitem {Gruneisen}M. T. Gruneisen, M. L. Eickhoff, S. C. Newey, K. E.
Stoltenberg, J. F. Morris, M. Bareian, M. A. Harris, D. W. Oesch, M. D.
Oliker, M. B. Flanagan, B. T. Kay, J. D. Schiller, and R. N. Lanning,
\textit{Adaptive-optics-enabled daytime field experiment for satellite-Earth
quantum networking}, arXiv:2006.07745 (2020).

\bibitem {LLO}B. Qi, P. Lougovski, R. Pooser, W. Grice, and M. Bobrek,
\textit{Generating the local oscillator \textquotedblleft
locally\textquotedblright\ in continuous-variable quantum key distribution
based on coherent detection}, Phys. Rev. X \textbf{5}, 041009 (2015).

\bibitem {CollectiveATT}S. Pirandola, S. L. Braunstein, S. Lloyd,
\textit{Characterization of collective Gaussian attacks and security of
coherent-state quantum cryptography}, Phys. Rev. Lett. \textbf{101}, 200504 (2008).

\bibitem {NoteTRUSTED}Note that, under the hypothesis that the environmental
noise is assumed to be trusted, then Eqs.~(\ref{ThermalLOSSub})
and~(\ref{ThermalLOSSlb}) do not apply. In that case, the optimal QKD\ rate is
still bounded by the pure-loss bound in Eq.~(\ref{mainOTHER}).

\bibitem {NoteCapacity}It is understood that we refer to an \textquotedblleft
extended\textquotedblright\ quantum channel which includes imperfections of
the transmitter (pointing error $\sigma_{\text{P}}^{2}$) and the receiver
(quantum efficiency $\eta_{\text{eff}}$). By assuming an optimal value for
these setup parameters ($\sigma_{\text{P}}^{2}=0$ and $\eta_{\text{eff}}=1$),
one retrieves the bounds/capacity that are associated with the
\textquotedblleft external\textquotedblright\ quantum channel (accounting for
diffraction, extinction and turbulence).

\bibitem {RuppertCluster}L. Ruppert, C. Peuntinger, B. Heim, K. G\"{u}nthner,
V. C. Usenko, D. Elser, G. Leuchs, R. Filip, and C. Marquardt, \textit{Fading
channel estimation for free-space continuous-variable secure quantum
communication}, New J. Phys. \textbf{21}, 123036 (2019).

\bibitem {GG02}F. Grosshans and P. Grangier, \textit{Continuous Variable
Quantum Cryptography Using Coherent States}, Phys. Rev. Lett. \textbf{88},
057902 (2002).

\bibitem {Noswitch}C. Weedbrook, A. M. Lance, W. P. Bowen, T. Symul, T. C.
Ralph, and P. K. Lam, \textit{Quantum Cryptography Without Switching}, Phys.
Rev. Lett. \textbf{93}, 170504 (2004).

\bibitem {Lev2017}A. Leverrier, \textit{Security of Continuous-Variable
Quantum Key Distribution via a Gaussian de Finetti Reduction}, Phys. Rev.
Lett. \textbf{118}, 200501 (2017).

\bibitem {LeoEXP}Y.-C. Zhang, Z. Chen, S. Pirandola, X. Wang, C. Zhou, B. Chu,
Y. Zhao, B. Xu, S. Yu, and H. Guo, \textit{Long-Distance Continuous-Variable
Quantum Key Distribution over 202.81 km of Fiber}, Phys. Rev. Lett.
\textbf{125}, 010502 (2020).

\bibitem {NoteSlice}From an operational point of view, the parties may
approximately reach the $1$-radiant rate by applying a de-fading procedure
which maps all their data points (collected along the slice) into the
threshold transmissivity $\eta_{\text{th}}$ associated with the angle
$\theta=1$.

\bibitem {PilotNOTA}At reasonably high values of the clock, just a small
percentage of the pulses needs to be employed for the pilots. Indeed, for a
$10~$MHz clock, we have $10^{7}$ pulses per second. Assume that $1\%$ of the
pulses are pilots. Then, within the typical timescale associated with
beam-wandering ($10-100$ ~ms), we have $10^{3}-10^{4}$ pilots. On average
these allow the parties to create a very fine lattice in transmissivity, with
a small step $\simeq10^{-3}-10^{-4}$. Also note that the dynamics of the
pilots is fast with respect to the orbital dynamics. In fact, within a
timescale of $10$~ms, a fast object in circular orbit at $h=100~$km travels a
zenith angle of about $8\times10^{-4}$ rad, so that the post-selection
interval $\Delta$ can be considered to be approximately constant.

\bibitem {Xwang1}X. Wang, Y. Zhang, S. Yu, and H. Guo, \textit{High speed
error correction for continuous-variable quantum key distribution with
multi-edge type LDPC code}, Sci. Rep. \textbf{8}, 10543 (2018).

\bibitem {Xwang2}X. Wang, Y. Zhang, S. Yu, and H. Guo, \textit{High-speed
implementation of length-compatible privacy amplification in
continuous-variable quantum key distribution,} IEEE Photon. Journal
\textbf{10}, 7600309 (2018).

\bibitem {LeoComms}Yichen Zhang, private communication.

\bibitem {Lim}C. C.-W. Lim, F. Xu, J.-W. Pan, and A. Ekert, \textit{Security
analysis of quantum key distribution with small block length and its
application to quantum space communications}, arXiv:2009.04882 (2020).

\bibitem {LeverrierSAT}D. Dequal, L. T. Vidarte, V. R. Rodriguez, G. Vallone,
P. Villoresi, A. Leverrier, and E. Diamanti, \textit{Feasibility of
satellite-to-ground continuous-variable quantum key distribution},
arXiv:2002.02002 (2020).

\bibitem {ThomsonBOOK}A. R. Thompson, J. M. Moran, and G. W. Jr. Swenson,
\textit{Interferometry and Synthesis in Radio Astronomy} (3rd ed., Springer
Open, 2017).

\bibitem {Kon}V. Klyatskin and A. Kon, \textit{On the displacement of
spatially bounded light beams in a turbulent medium in the Markovian
random-process approximation},\ Radiophys. Quantum Electron \textbf{15},
1056-1061 (1972).

\bibitem {Hemani}H. Kaushal, V. K. Jain, and S. Kar, \textit{Free Space
Optical Communication} (Springer, New York, 2017).

\bibitem {Tofsted}D. H. Tofsted, S. G. O'Brien, and G. T. Vaucher, \textit{An
atmospheric turbulence profile model for use in army war gaming applications
I}, Technical Report ARL-TR-3748 US Army Research Laboratory (2006).

\bibitem {Vasy17}D. Vasylyev, A. A. Semenov, W. Vogel, K. G\"{u}nthner, A.
Thurn, \"{O}. Bayraktar, and Ch. Marquardt, \textit{Free-space quantum links
under diverse weather conditions}, Phys. Rev. A \textbf{96}, 043856 (2017).

\bibitem {Vanzdant}T. W. VanZandt, J. L. Green, K. S. Gage, and W. L. Clark,
\textit{Vertical profiles of refractivity turbulence structure constant:
Comparison of observations by the Sunset Radar with a new theoretical model},
Radio Sci. \textbf{5}, 819 (1978).

\bibitem {Dewan}E. M. Dewan, R. E. Good, R. Beland, and J. Brown, \textit{A
model for }$C_{n}^{2}$\textit{ (optical turbulence) profiles using radiosonde
data}, Environmental Research Papers, Report No. \textbf{1121}, PL-TR-93-2043,
1993 (unpublished).

\bibitem {Walters}D. L. Walters and K. E. Kunkel, \textit{Atmospheric
modulation transfer function for desert and mountain locations: The
atmospheric effects on }$r_{0}$, J. Opt. Soc. Am. \textbf{71}, 397 (1981).

\bibitem {Thermosonde}R. Frehlich, R. Sharman, F. Vandenberghe, W. Yu, Y. Liu,
and J. Knievel, \textit{Estimates of }$C_{n}^{2}$\textit{ from numerical
weather prediction model output and comparison with thermosonde data}, J.
Appl. Meteor. Climatol. \textbf{49}, 1742 (2010).

\bibitem {Fried}D. L. Fried, \textit{Limiting Resolution Looking Down Through
the Atmosphere}, J. Opt. Soc. Am. \textbf{56}, 1380-1384 (1966).

\bibitem {Majumdar}A. K. Majumdar and J. C. Ricklin, \textit{Free-Space Laser
Communications} (Springer New York, 2008).

\bibitem {Goodman}J. W. Goodman, \textit{Statistical Optics} (John Wiley \&
Sons, Inc., 1985).

\bibitem {AndrewsYu95}L. C. Andrews, R. L. Phillips, and P. T. Yu,
\textit{Optical scintillations and fade statistics for a
satellite-communication system}, Appl. Opt. \textbf{34}, 7742-7751 (1995).

\bibitem {Gprofile1}P. A. Lightsey, \textit{Scintillation in ground-to-space
and retroreflected laser beams},\ Opt. Eng. \textbf{33}, 2535-2543 (1994).

\bibitem {Gprofile2}A. M. Prokhorov, F. V. Bunkin, K. S. Gochelashvily, and V.
I. Shishov, \textit{Laser irradiance propagation in turbulent media},\ Proc.
IEEE \textbf{63}, 790-809 (1975).

\bibitem {Gprofile3}L. C. Andrews, W. B. Miller, and J. C. Ricklin,
\textit{Spatial coherence of a Gaussian-beam wave in weak and strong optical
turbulence},\ J. Opt. Soc. Am. A \textbf{11}, 1653-1660 (1994).

\bibitem {Andrews2000}L. C. Andrews, R. L. Phillips, and C. Y. Young,
\textit{Scintillation model for a satellite communication link at large zenith
angles}, Opt. Eng. \textbf{39}, 3272--3280 (2000).

\bibitem {RytovAPPROX}S. M. Rytov, \textit{Diffraction of light by ultrasonic
waves}, Izvestiya Akademii Nauk SSSR, Seriya Fizicheskaya (Bulletin of the
Academy of Sciences of the USSR, Physical Series) \textbf{2}, 223--259 (1937).

\bibitem {Kolmogorov}A. N. Kolmogorov, \textit{The local structure of
turbulence in an incompressible viscous fluid for very large Reynolds
numbers}, C. R. (Doki) Acad. Sci. U.S.S.R. \textbf{30}, 301--305 (1941).

\bibitem {YuraMcKinley83}H. T. Yura and W. G. McKinley, \textit{Optical
scintillation statistics for IR ground-to-space laser communication systems},
Appl. Opt. \textbf{22}, 3353-3358 (1983).

\bibitem {BelandBook}B. Beland, \textit{The Infrared and Electro-Optical
System Handbook}, vol. 2 (SPIE Press, 1993).

\bibitem {Poirier72}J. Poirier and D. Korff, \textit{Beam spreading in a
turbulent medium}, J. Opt. Soc. Am. \textbf{62}, 893-898 (1972).

\bibitem {Bunkin70}F. Bunkin and K. Gochelashvily, \textit{Spreading of a
light beam in a turbulent medium}, Radiophys. Quantum Electron. \textbf{13},
811-821 (1970).

\bibitem {Dios04}F. Dios, J. A. Rubio, A. Rodr\'{\i}guez, and A. Comer\'{o}n,
\textit{Scintillation and beam-wander analysis in an optical ground
station-satellite} \textit{uplink}, Appl. Opt. \textbf{43}, 3866-3873 (2004).

\bibitem {Belmonte}A. Belmonte, \textit{Feasibility study for the simulation
of beam propagation: consideration of coherent lidar performance}, Appl. Opt.
\textbf{39}, 5426-5445 (2000).

\bibitem {LambertScatterer}J. H. Lambert, \textit{Photometria, sive de Mensura
et gradibus luminis, colorum et umbrae} (1760).

\bibitem {LBbook}F. L. Pedrotti and L. S. Pedrotti, \textit{Introduction to
Optics} (Prentice Hall, 1993).

\bibitem {Bonato}C. Bonato, A. Tomaello, V. Da Deppo, G. Naletto, and P.
Villoresi, F\textit{easibility of satellite quantum key distribution}, New J.
Phys \textbf{11}, 045017 (2009).

\bibitem {Miao}E.-L. Miao, Z.-F. Han, S.-S. Gong, T. Zhang, D.-S. Diao, and
G.-C. Guo, \textit{Background noise of satellite-to-ground quantum key
distribution}, New J. Phys. \textbf{7}, 215 (2005).

\bibitem {sky1}C. Leinert, S. Bowyer, L. K. Haikala, M. S. Hanner, M. G.
Hauser, A.-C. Levasseur-Regourd, I. Mann, K. Mattila, W. T. Reach, W.
Schlosser, H. J. Staude, G. N. Toller, J. L. Weiland, J. L. Weinberg, and A.
N. Witt, \textit{The 1997 reference of diffuse night sky brightness}, Astron.
Astrophys. Suppl. Ser. \textbf{127}, 1-99 (1998).

\bibitem {sky2}V. Hansen, \textit{Spectral Distribution of Solar Radiation on
Clear Days: A Comparison Between Measurements and Model Estimates}, Journal of
Climate and Applied Meteorology \textbf{23}, 772-780 (1984).



\bibitem {Modtran}A. Berk, P. Conforti, R. Kennett, T. Perkins, F. Hawes, and
J. van den Bosch, \textit{MODTRAN6: a major upgrade of the MODTRAN radiative
transfer code}, Proc. SPIE 9088, Algorithms and Technologies for
Multispectral, Hyperspectral, and Ultraspectral Imagery XX, 90880H (June 13, 2014).

\bibitem {LibRadtran1}B. Mayer and A. Kylling, \textit{Technical note: The
libRadtran software package for radiative transfer calculations - description
and examples of use}, Atmos. Chem. Phys. \textbf{5}, 1855-1877 (2005).

\bibitem {LibRadtran2}C. Emde, R. Buras-Schnell, A. Kylling, B. Mayer, J.
Gasteiger, U. Hamann, J. Kylling, B. Richter, C. Pause, T. Dowling, and L.
Bugliaro. \textit{The libradtran software package for radiative transfer
calculations (version 2.0.1)}, Geoscientific Model Development \textbf{9},
1647-1672 (2016).

\bibitem {6SV}E. F. Vermote, D. Tanr\'{e}, J. L. Deuz\'{e}, M. Herman, J.-J.
Morcrette, Second \textit{Simulation of the Satellite Signal in the Solar
Spectrum, 6S: An Overview}, IEEE Transactions on Geoscience and Remote Sensing
\textbf{35}, 675-686 (1997).
\end{thebibliography}
\end{document}